\documentclass{eas}
\usepackage{graphicx}
\usepackage{journaux}
\usepackage{natbib}
\usepackage{amsfonts}
\usepackage{amsmath}
\usepackage{txfonts}
\usepackage{hyperref} 

\begin{document}
\title{Observational techniques to measure solar and stellar oscillations} 
\author{Rafael A. Garc\'\i a}
\address{Laboratoire AIM, CEA/DSM -- CNRS - Univ. Paris Diderot -- IRFU/SAp, Centre de Saclay, 91191 Gif-sur-Yvette Cedex, France}
%
%
\runningtitle{Garc\'\i a, R.A.: Observational techniques for stellar oscillations}
\begin{abstract}
As said by Sir A. Eddington in 1925: ``Our telescopes may probe farther and farther into the depths of space.
At first sight it would seem that the deep interior of the sun and stars is less accessible to scientific investigation than any other region of the universe. What appliance can pierce through the outer layers of a star and test the conditions within?''
\citet{1926ics..book.....E}. \\ Nowadays, asteroseismology has proven its ability to pierce below stellar photospheres and allow us to ``see'' inside the interior of thousands of stars down to the stellar cores, answering the question asked by Eddington ninety years ago. In this chapter we review the general properties of the spectral analysis which is the base of any asteroseismic investigation. After describing the stellar power spectrum, we will describe in details the characterization of the modal spectrum. This chapter will end by a brief description of the instrumentation in both helio and asteroseismology.

\end{abstract}
\maketitle

\tableofcontents

\section{Introduction}
For a long time, investigations of stellar interiors have been restricted to theoretical studies only constrained by observations of their global properties and external characteristics. However, in the last few decades the field has been revolutionized by the ability to perform seismic investigations of the internal properties of stars. Not surprisingly, it started with the Sun, where helioseismology \citep[the seismic study of the Sun, e.g.][]{1996Sci...272.1296G,JCD2002} has been yielding information competing with what can be inferred about the Earth's interior from geoseismology. 

The last few years have witnessed the advent of asteroseismology \citep[the seismic study of stars, e.g.][]{2003PASA...20..203B}, in particular for solar-like pulsators, thanks to a dramatic development of new observing facilities providing the first reliable results on the interiors of distant stars similar to the Sun. The coming years will see a huge development in this field.

Helio- and asteroseismology provide unique tools to infer the fundamental stellar properties (e.g., mass, radius, sound speed...) and to probe the internal conditions inside the Sun and stars \citep[e.g.][]{2009ApJ...700.1589S}. Today, asteroseismology also provides invaluable information to other scientific communities. As an example, it can give a good estimate of the masses, radii, and ages of the stars hosting planets \citep[e.g.][]{2004A&A...427..965B,2005A&A...440..615B,2010A&A...513A..49S,2010MNRAS.406..566M,2010ApJ...713L.164C,2010A&A...524A..47G,2011ApJ...729...27B,2011IAUS..276...30V,2012ApJ...745..120B,2012ApJ...746..123H,2013ApJ...767..127H,2015ApJ...799..170C,2015MNRAS.452.2127S}, as it has been demonstrated for example comparing with stars for which Hipparcos parallaxes, spectroscopy and asteroseismology is available \citep{2012ApJ...757...99S}.  This is a key-information to understand the formation of these planetary systems and their evolution, and to constrain the habitable zones of the surrounding exoplanets that can also be influenced by the magnetic activity of the host stars \citep[e.g.][]{2005A&A...431L..13M,2009A&A...506..245M,2009A&A...506...33M,2009MNRAS.399..914K,2010Sci...329.1032G,2010ApJ...723L.213M,2011ApJ...735...59P,2013ApJ...769...37B,2013A&A...550A..32M,2014A&A...562A.124M,2014JSWSC...4A..15M}. 
Asteroseismology can also be used  to determine if a star belongs to a cluster \citep{2011ApJ...739...13S} and to verify  cluster's properties \citep{2011ApJ...729L..10B,2012ApJ...757..190C}.
Asteroseismology leads to the testing and revision of our theories of stellar structure, dynamical processes, and evolution. Helio- and asteroseismology are today in a blooming phase both in their goals and in impact. Helioseismology has shown the way to asteroseismology, which is reaching its maturity thanks to the path opened by the CNES CoRoT satellite \citep[Convection, Rotation and planetary Transits,][]{2006cosp...36.3749B}, the NASA's {\it Kepler} spacecraft  \citep{2010Sci...327..977B,2010ApJ...713L..79K} --and its extended mission K2 \citep{2014PASP..126..398H}--, the BRITE-Constellation of nanosatellites for precision photometry of Bright Stars \citep{2014PASP..126..573W}, and the future missions such as NASA's TESS \citep[Transiting Exoplanet Survey Satellite,][]{2014SPIE.9143E..20R} and the ESA's M3 PLATO mission \citep{2014ExA....38..249R}. It is important to remember the pioneers of this research done thanks to some episodic ground-based campaigns \citep[e.g.][]{2007MNRAS.377..584S,2010ApJ...713..935B} and some solar-like oscillating stars observed from space using both, the American satellite WIRE \citep[Wide-Field Infrared Explorer,][]{2000ApJ...532L.133B} and the Canadian MOST satellite \citep[Microvariability and Oscillations of Stars,][]{2000ASPC..203...74M}.

\subsection{Helio and asteroseismology}

Helio and asteroseismology aim at studying the internal structure and dynamics of the Sun and other stars by means of their resonant oscillations \citep[e.g.][and references therein]{Gou1985,STCDap1993,JCD2002}. These vibrations manifest themselves in small motions of the visible surface of the star and in the associated small variations of stellar luminosity. Variable stars can be found all across the Hertzsprung-Russell, H-R, diagram.

During the last 30 years, helioseismology has proven its ability to study the structure and dynamics of the solar interior in a stratified way. These seismic tools allow us to infer some physical quantities as a function of the radius and latitude: the sound-speed profile \citep[e.g.][]{BasJCD1997,STCBas1997}, the density profile \citep[e.g.][]{2009ApJ...699.1403B}, the internal rotation profile in the convective  zone \citep[e.g.][]{ThoToo1996} and the radiative zone \citep[e.g.][]{1999MNRAS.308..405C,CouGar2003,2008ApJ...679.1636E,2013SoPh..287...43E,ElsHow1995,GarCor2004,2008SoPh..251..119G} or the conditions and properties of the solar core  \citep[e.g.][]{2010A&ARv..18..197A,BasJCD1997,2009ApJ...699.1403B,2007Sci...316.1591G,2008AN....329..476G,2008SoPh..251..135G,STCCou2001,STCGar2004} are some well-known examples. Moreover, thanks to the detailed study of these variables, the position of the base of the convection zone \citep[e.g.][]{JCDGou1985} or the Helium abundances \citep[e.g.][]{1991Natur.349...49V} are some examples of what has been inferred. These observational constraints have significantly improved the standard solar models. 


The Sun, because of its proximity, has been a fundamental calibrator of stellar evolution but observations of many other stars \citep[e.g.][]{2011Sci...332..213C,2011ApJ...743..143H} --covering a larger region of the H-R diagram through asteroseismology-- will allow testing stellar structure, evolution, and dynamo theories under many different conditions  \cite[e.g.][]{2010Ap&SS.328...51C}. In this case, due to the absence of spatial resolution in the observations, only low-degree modes (those with a small number of nodal lines on the surface of the star, see Fig.~\ref{spher_harm}) will be accessible. Therefore compared to the Sun, less detailed information will be available on stellar interiors. 


Stars are also known to be magnetic rotating objects. Such dynamical factors, magnetism and rotation, affect the internal structure and evolution of stars \citep[e.g.][]{2004ApJ...614.1073B,2008sf2a.conf..341Z,2010MNRAS.402..271D,2010A&A...519A.116E}, and modifies the observed spectrum. High precision observations provided by modern facilities (from ground-based or spaceborne instruments) allow to constraint these dynamical process with a precision never achieved before.

\subsection{Type of oscillation modes}

The quest to improve our knowledge of the structure and dynamics of the solar interior has been possible thanks to the study of the resonant modes that are trapped in its interior and their comparison with stellar models.  The difference between the two provides valuable information on the errors and omissions of the theoretical models. Before describing and interpreting the observed spectra, it is important to have some basic knowledge of the properties of the waves we want to characterize.

The theory behind stellar oscillations is very well known and it has been longly described by several authors \citep[e.g.][]{1980tsp..book.....C,1989nos..book.....Ub,2002RvMP...74.1073C}. Without going into deep details, a brief review of some magnitudes and theoretical concepts that will be used in the rest of the manuscript are given here. 

\begin{figure}[!htb]
\begin{center}
\begin{tabular}{cc}
	\includegraphics[width=0.4\textwidth]{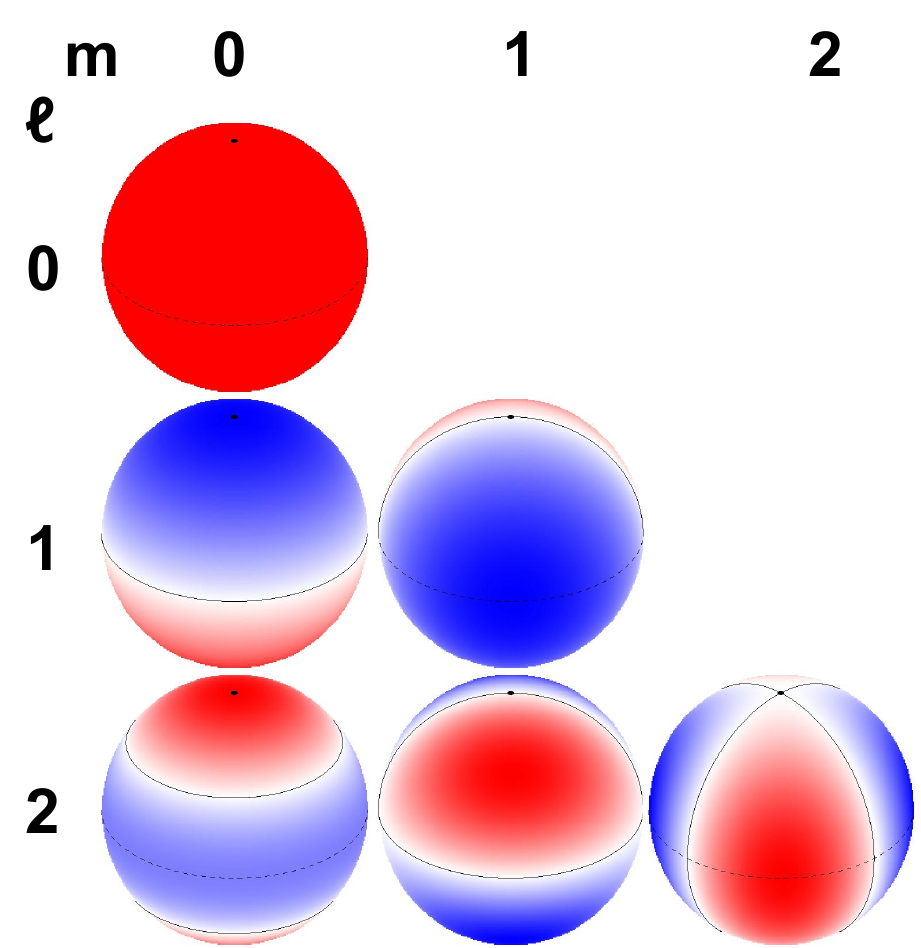} &
	\includegraphics[width=0.4\textwidth]{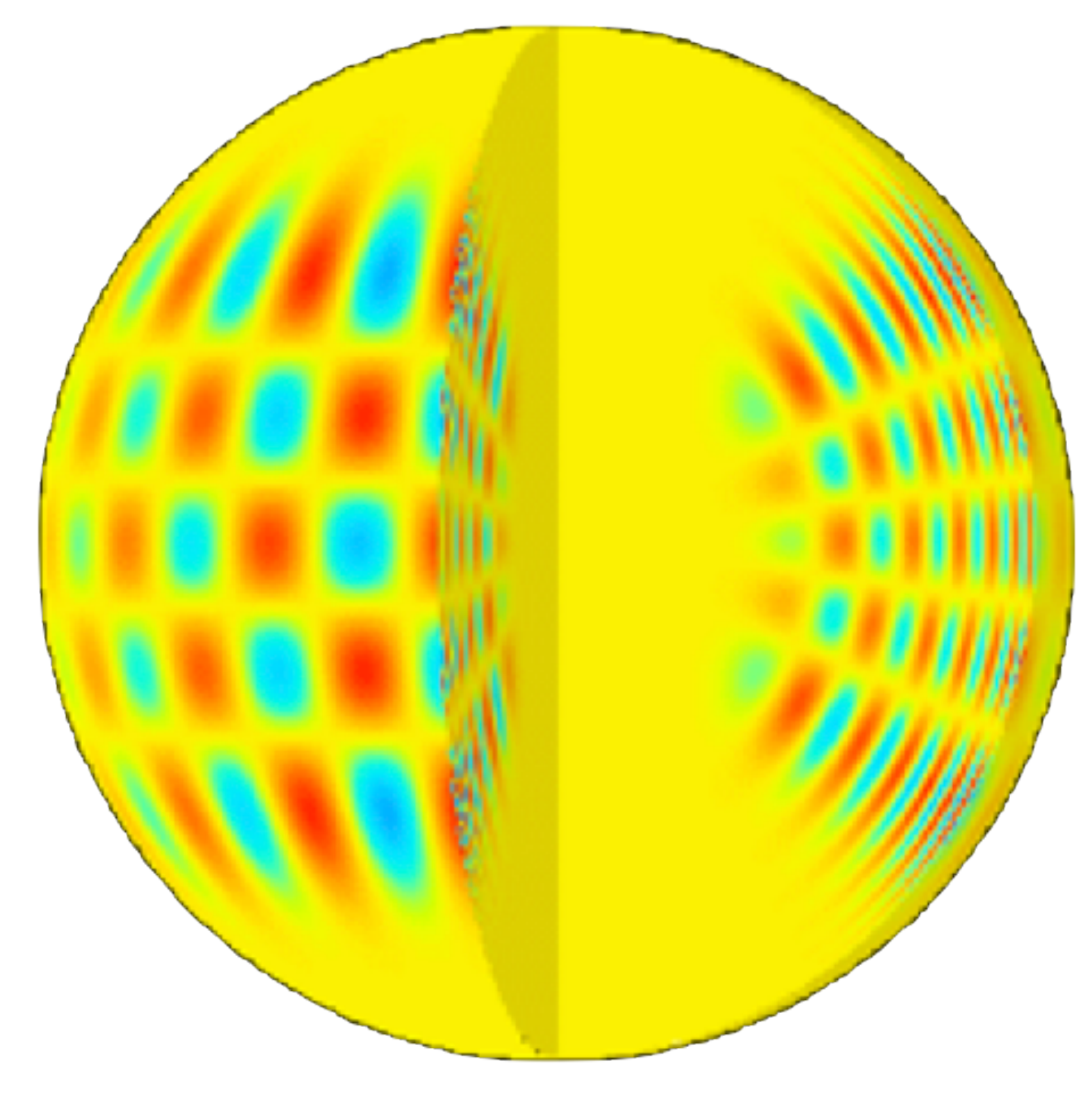}
\end{tabular}
\end{center}
\caption[Example of spherical harmonics of degree $\ell$=0,1,2 and azimuthal order $m$=0, 1, 2]{\label{spher_harm} Left: Example of spherical harmonics of degree $\ell$=0,1,2 and azimuthal order $m$=0, 1, 2. The blue regions are those coming close to the observer, while the red regions represent those that are moving away. Right: Mode $\ell$=20, $m$=16 and $n$=14.
}
\end{figure}

Solar-like oscillations are standing waves characterized by three integers: $n, \ell$, and $m$ (see Fig.~\ref{spher_harm}). $n$ is the radial order indicating the number of nodal shells along the radius, and it is an integer greater than zero. By convention, we denote the p modes by positive numbers and g modes by negative ones. The angular degree, $\ell$, is an integer greater or equal zero that denotes the number of nodes in the surface of the sphere. The first ones usually receive special names. Thus, modes with $\ell$=0 are called radial modes while those with $\ell \ge 1$ are the non-radial modes. Moreover, those modes with $\ell$=1 are called dipole modes, those with $\ell$=2 are the quadrupole modes and finally the $\ell$=3 are the octupole modes. Finally, $m$ is the azimuthal order and gives the number of nodal lines passing through the poles. It can take values from $-\ell$ to $+\ell$ passing by zero. For each eigenmode we can define a characteristic frequency $\nu_{n,\ell,m}=\omega_{n,\ell,m} / 2 \pi$. 

Since the solar rotation lifts the azimuthal degeneracy of the resonant modes, their eigenfrequencies, $\nu_{n \ell m}$,  are split into their $2\ell+1$ $m$-components that are usually called the mode multiplet or simply multiplet. The frequency separation between two consecutive components is usually called rotational splitting (or just splitting) and it depends on the rotation rate in the region sampled by the mode. In the same way, the precise frequency of a mode depends on the physical properties of the cavity where the mode propagates. Using inversion techniques the rotation rate, the sound speed or the density profile at different locations inside the Sun can be inferred from a suitable lineal combination of the measured modes. 

When we observe the Sun or the stars without any spatial resolution only low-degree modes ($\ell \le5$) can be observed. This is because for higher degree modes the regions with positive and negative velocities cancel out. 

In the interior of solar-like stars we can define two main types of oscillations modes: the acoustic and the gravity modes. 

\subsubsection{Acoustic modes}

Pressure driven modes (or p modes) are acoustic waves for which the restoring force arises from the pressure gradient. In the case of the Sun, the modes that are excited with the highest amplitudes are around the 3300 $\mu$Hz producing the so-called 5-minutes oscillation of the Sun. They were first detected by \citet{1962ApJ...135..474L}, but interpreted as being part of the turbulent motions of the Sun. Tracing back the history of helioseismology \citep[for a full review on the history of helioseismology see e.g.][]{chaplin2006}, the same year \citet{1962ApJ...136..493E} confirmed the existence of the previous detection but it was not until 1970 when these oscillations were explained as standing waves trapped between the photosphere and the solar interior \citep{1970ApJ...162..993U,1971ApL.....7..191L}, with a particular relationship between the frequency and horizontal wave number. This explained the peaks or bands that appeared in a ``diagnostic diagram''  around 3 mHz, or 5 minutes \citep{1969SoPh....9..328T,1975A&A....44..371D}. Subsequently, \citep{1974SoPh...36..313M} concluded that there was no spatial correlation between turbulent convection and the oscillatory waves, giving final independence to both events. \citet{1975A&A....44..371D} confirmed experimentally the existence of eigenmodes, finding a relationship between the period and horizontal wavelength consistent with the predictions done by \citep{1970ApJ...162..993U}.  While previous observations showed evidences of spatially-localized oscillations in the solar atmosphere, \citet{1975ApJ...200..484H} announced the detection of oscillations in the solar diameter, suggesting the existence of global oscillations and, consequently, the possibility to use these pulsations to probe the solar interior \citep[e.g.][]{1975A&A....45...15S,1976Natur.259...89C}. This theoretical developments were confirmed by the detection of the p-mode power spectrum of the 5 minutes oscillations reported by \citet{1979Natur.282..591C} confirming the existence of global modes. These observations were then improved by \citet{1980Natur.288..541G}  thanks to the measurements obtained during 120 continuous hours from the South Pole, which sensibly improved the overall quality of the spectrum by reducing the daily aliases. The helioseismology, as we know it today was officially born.

Without going into the details of the theory behind the stellar oscillations, we can describe some useful characteristics. Therefore, while p modes propagate\footnote{Rigorously, waves propagate while eigenmodes do not.} inside the stellar interior, the sound speed increases and the waves are refracted. The deepest layer reached by the modes has a radius usually called the internal turning point, $r_t$ which is defined by:
\[
r_t = c_t\, L/( 2 \pi \nu_{n \ell})
\]
where $L=\ell+1/2$, $\nu_{n \ell}$ is the frequency of the mode, and $c_t = c(r_t)$ the sound-speed at the radius $r_t$ \citep[see for example][]{1994A&A...290..845L}. 

\begin{figure}[!htb]
\begin{center}
\begin{tabular}{cc}
\includegraphics[width=0.4\textwidth]{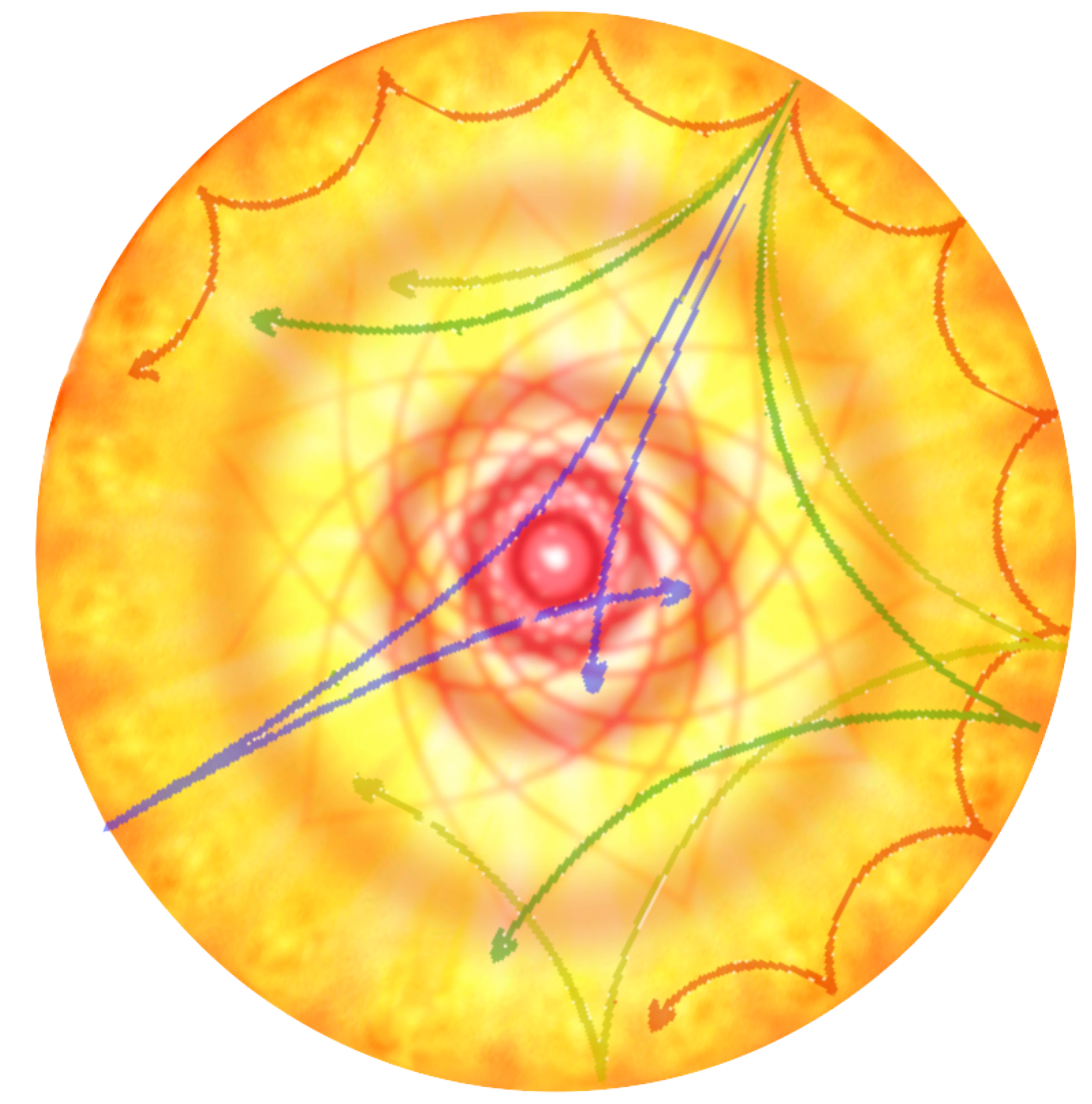}
\includegraphics[width=0.55\textwidth]{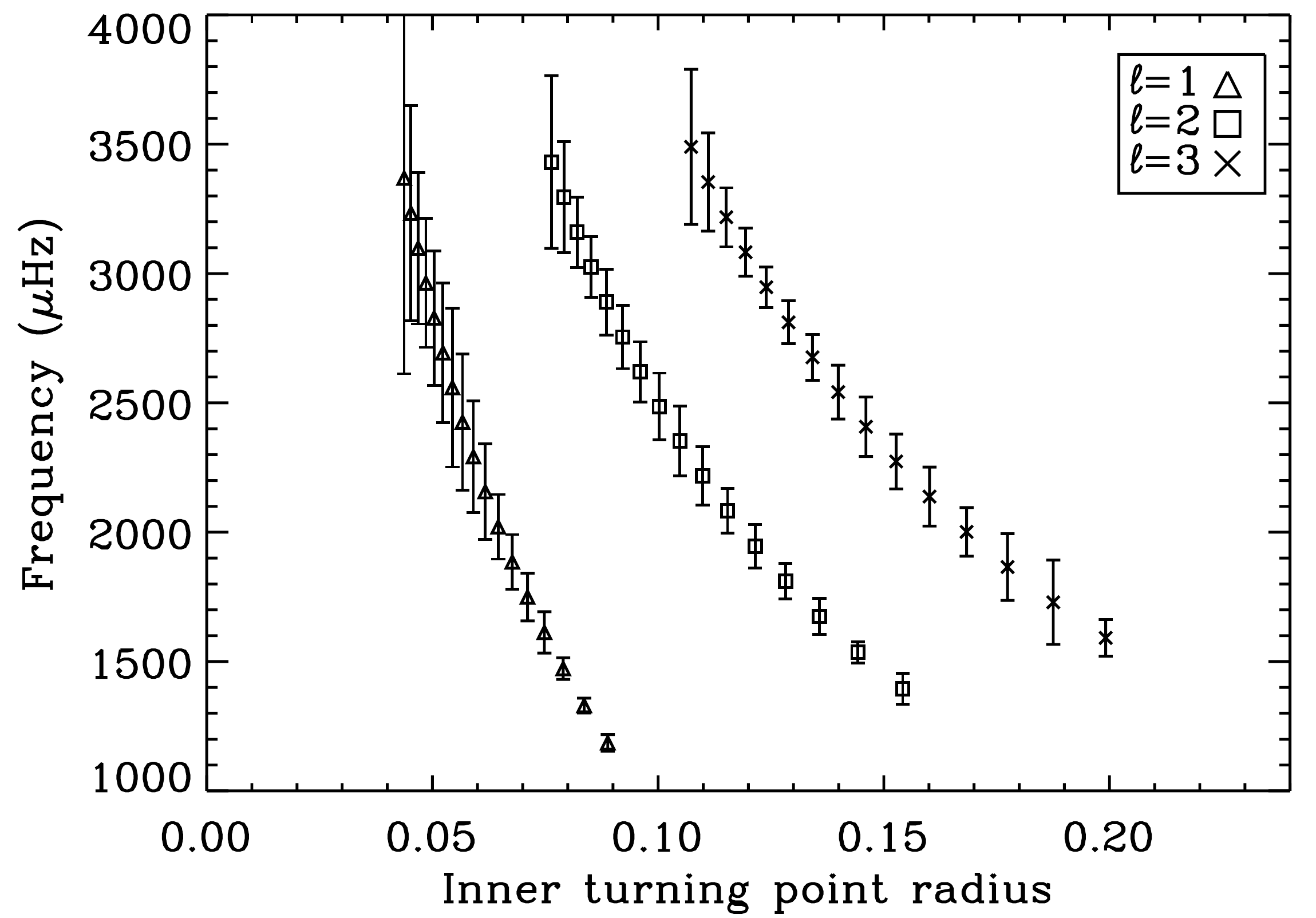}
\end{tabular}
\end{center}
\caption{\label{Fig_prop_modes} Left: Representation of the solar interior with the radiative and the convective zones. Gravity modes can only propagate inside the radiative zone. Low-degree acoustic modes can propagate through most of the Sun (blue and green lines). High-degree modes only propagates in the outer layers of the Sun (orange lines). Right: Fractional radii of the Inner turning point for low-degree acoustic modes obtained from GOLF \citep{2008SoPh..251..119G}. The error bars are the splitting error bars in nanohertz magnified by a factor of $10^4$.}
\end{figure}

Therefore, the internal turning point rises when the degree $\ell$ of the modes increases (see Fig.~\ref{Fig_prop_modes}). For example, radial modes in the Sun will propagate all along the radius and cross the center. Dipole modes will have internal turning points in a range 0.04 to 0.1 $R_\odot$, while octupole modes will propagate above 0.1 $R_\odot$. Moreover, for a fixed $\ell$, the modes with increasing frequencies -- higher radial order $n$ -- penetrate deeper inside the Sun (see right panel in Fig.~\ref{Fig_prop_modes}). 


Acoustic modes of the same degree are equidistant in frequency (in the asymptotic regime) which allows to define some global parameters of the p-mode pattern, such as the large and the small frequency separations (a detailed explanation of these quantities can be found in Mosser's chapter in this volume).

\subsubsection{Gravity modes}
\label{Sect_gm_def}
For gravity modes (g modes) the restoring force is buoyancy. These modes propagate in the radiative interiors but they become evanescent in the convective zones reaching the surface of the Sun --and other stars with a thick outer convective zone-- with very small amplitudes. This complicates their detection (see left panel in Fig.~\ref{Fig_prop_modes}). These modes are located at lower frequencies compared to the p modes. In the case of the Sun, they have frequencies below $\sim$ 470 $\mu$Hz and there are no $\ell$=0 g modes. The frequency of g modes decreases with n. In the asymptotic regime, g modes are equidistant in period and not in frequency, with a very dense spectrum when going to higher periods (lower frequencies).

\begin{figure}[!htb]
\begin{center}
\includegraphics[width=0.6\textwidth]{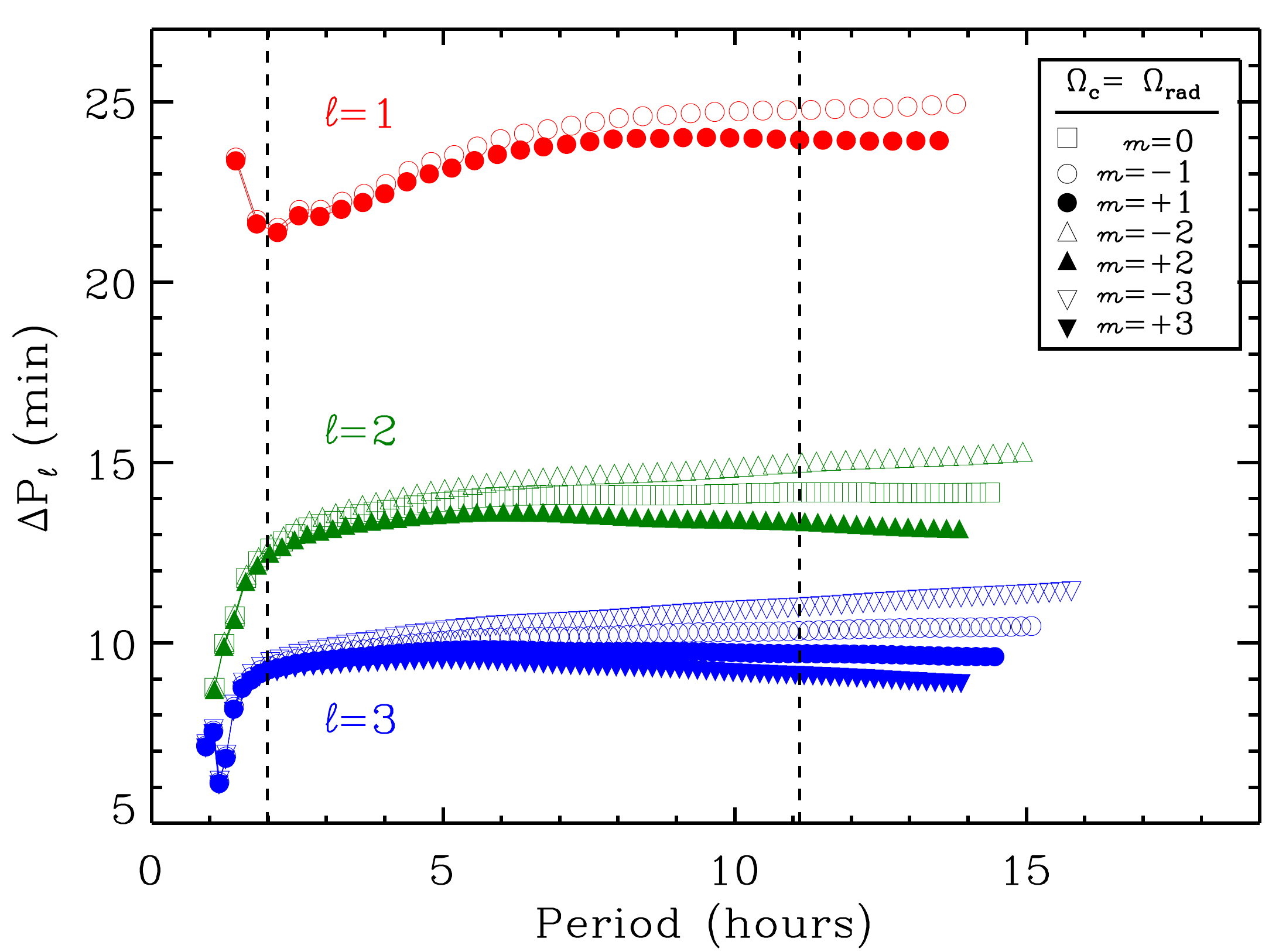}
\end{center}
\caption[g-mode period separations for modes with $\ell$=1, 2 and 3]{\label{Fig_DPrig} g-mode period separations for modes with $\ell$=1, 2 and 3 expressed in minutes versus the period of each gravity mode expressed in hours and assuming a rigid rotation rate in the radiative interior. See the text for details.}
\end{figure}

Figure~\ref{Fig_DPrig} shows the separations in Period, $\Delta P_\ell$,	between consecutive radial orders ($n$, $n$ + 1) gravity modes of the Sun for $\ell$= 1, 2, and 3 (red, green, and blue, respectively), using the theoretical frequencies from the seismic model \citep{2007ApJ...668..594M}. The constant periodicity is achieved at 6, 4, and 2 hours for the modes $\ell$ = 1, 2, and 3, respectively. The g modes are split in frequency due to the rotation. They are very sensitive to the dynamics inside the radiative core \citep{2008A&A...484..517M}. To make the plot we have assumed a rigid rotation in the Sun in which $\Omega_c$ is the angular velocity of the solar core, and $\Omega_{rad} \simeq 433$ nHz, is the angular frequency of the remaining radiative zone, and $m$ is the azimuthal order of the modes. Inside the zone limited by the two vertical dashed lines (from $\sim$~2 to $\sim$~11 hours, corresponding to 25 to 140 mHz), we expect periodicities between 21 and 24 min for the dipole modes, between 12 and 14 min for the quadrupole modes, and between 9 and 11 min for the octupole modes.

\subsubsection{Mixed modes}
The distinction between pure acoustic modes and pure gravity modes is not always clear and sometimes they can be coupled together as it has already been described theoretically in previous works \citep[e.g.][and references there in]{1975PASJ...27..237O,1991A&A...248L..11D,2004SoPh..220..137C,2009A&A...506...57D}. In such cases, we name these waves as mixed modes. 
Mixed modes are very interesting because  they propagate as pressure waves in the convective envelope, and as gravity waves in the radiative interior. Therefore, they can probe the very inner core, while having enough amplitude in the surface to be detectable. The first observations of mixed modes in solar-like stars were reported from ground-based observations of $\eta$ Boo, by \citet{1995AJ....109.1313K} and confirmed later by \citet{2003AJ....126.1483K}, and \citet{2005A&A...434.1085C}. They were in very good agreement with theoretical predictions by e.g. \citet{1995ApJ...443L..29C}. 

Latter many observations of mixed modes have been done from ground-based observations but also from space thanks to CoRoT  \citep[e.g.][]{2010A&A...515A..87D} and \emph{Kepler} observations \citep{2010ApJ...713L.169C,2011A&A...534A...6C,2011ApJ...733...95M}. \citet{2011Sci...332..205B} first reported the existence of mixed modes in red-giant stars. Latter, \citet{2011Natur.471..608B} and \citet{2011A&A...532A..86M} showed the power of these modes to measure the evolutionary status of red giants, with a clear difference between stars ascending the red-giant branch (RGB) and those in the clump.

\section{Spectral analyses}
The natural domain to analyze the information embedded in the rich spectrum of the Sun and the stars is through the Fourier spectrum. Let's start with some general definitions.

\begin{figure}[!htb*]
\begin{center}
\includegraphics[width=.8\textwidth]{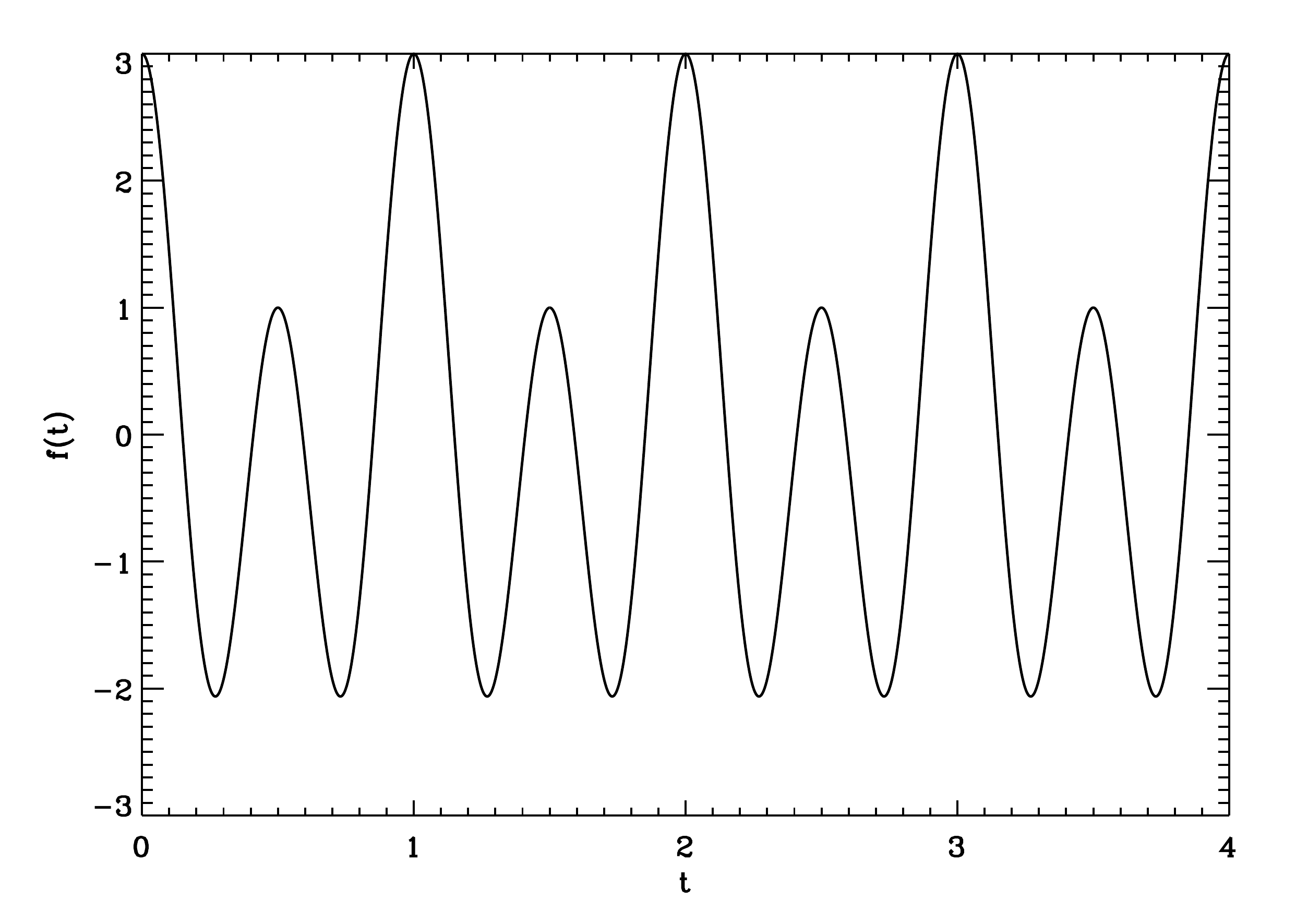}
\end{center}
\caption{\label{periodic1} Example of a periodic function $f(t)=\cos(2\pi t) + 2 \cos(4\pi t)$.}
\end{figure}

A signal, $f(t)$, is periodic of period $T_0$, if there is a period $T_0 > 0$ such as $f(t+T_0)=f(t)$ for all t. In this case, every integer number $n$ of the fundamental period $T_0$, is also a period: $f(t+nT_0)=f(t)$, with $n=0,\pm1, \pm2,...4$. 
In the Fourier domain, the frequency associated to $T_0$, $1/T_0$, is called the fundamental frequency or the first harmonic of the signal.  Any integer multiple of the  fundamental frequency, $n/(T_0)$, with $n > 1$, is called and overtone or the 2nd harmonic (if $n=2$), 3rd harmonic (if $n=3$), etc. Examples of periodic functions are Sin(t) and cos(t). An example of a periodic function is given in Fig,~\ref{periodic1}.

In 1822 Joseph Fourier established the basis of the spectral decomposition by demonstrating that any arbitrary function can be decomposed in a combination of sine and cosine functions (i.e. harmonic functions). Therefore, it was established that for a given function $f(t)$, fulfilling the necessary conditions of continuity and finiteness, its continuous Fourier transform of an infinite time series $f(t)$ could be defined as:
\begin{equation}
\overline{f(t)}=F(\nu)=\int\limits_{-\infty}^{\infty} {f(t) \, e^{-2\pi i \nu t}dt} \,\,\, ,
\label{intFourier}
\end{equation} 
where $i^2=-1$. In general, the Fourier transform $F(\nu)$ is a complex function.

Apart from the sign of the exponential, the Fourier transform is its own reverse:
\begin{equation}
f(t)=\int\limits_{-\infty}^{\infty} {F(\nu) \, e^{2\pi i \nu t}d\nu}  \,\,\, ,
\end{equation} 
which is usually called the reversed or inverse transform. Using the compact notation we have:
\begin{equation}
\overline{\overline{f(t)}}=\overline{F(\nu)}=f(t) \,\,\, .
\end{equation}

 \subsection{Examples of common Fourier transform pairs}
 Each combination of a function $f(t)$ and its Fourier transform $F(\nu)$ is usually called a Fourier transform pair. In this section we are going to review some Fourier transform pairs usually found in the analysis of real seismic data. For a complete review on Fourier transforms and Fourier transform pairs please refer to  \citet{2000fta..book.....B}.
 
\begin{figure}[!htbp*]
\begin{center}
\includegraphics[width=.95\textwidth]{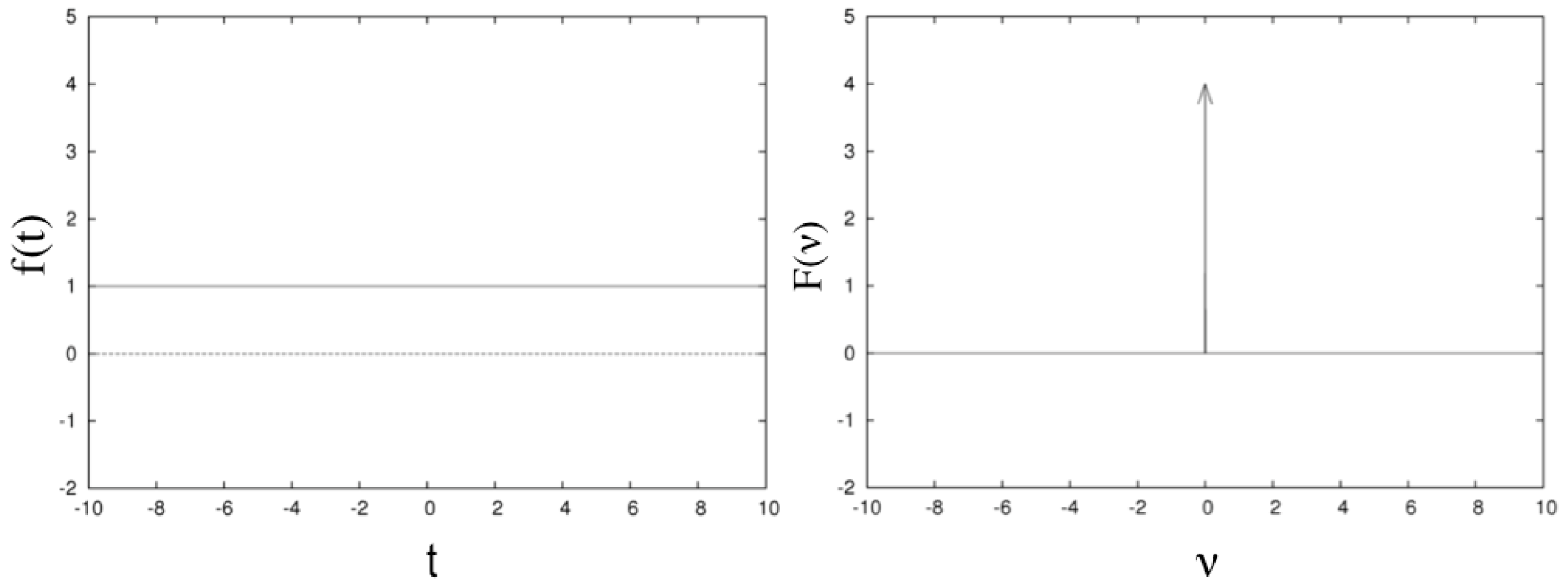}
\end{center}
\caption{\label{constant} Fourier Transform pair of a constant and a Dirac function.}
\end{figure}

\begin{figure}[!htbp*]
\begin{center}
\includegraphics[width=.95\textwidth]{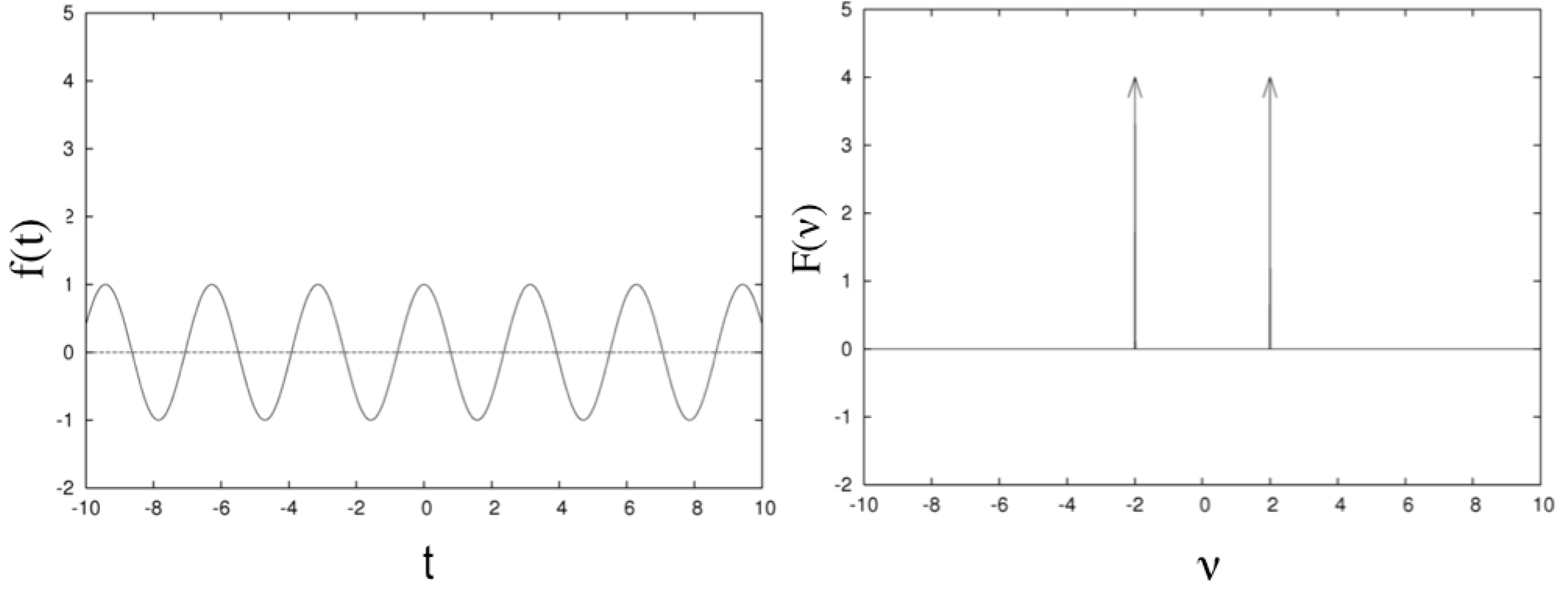}
\end{center}
\caption{\label{cos} Fourier Transform pair of a cosinus function and a pair of Dirac deltas.}
\end{figure}

\begin{figure}[!htbp*]
\begin{center}
\includegraphics[width=.95\textwidth]{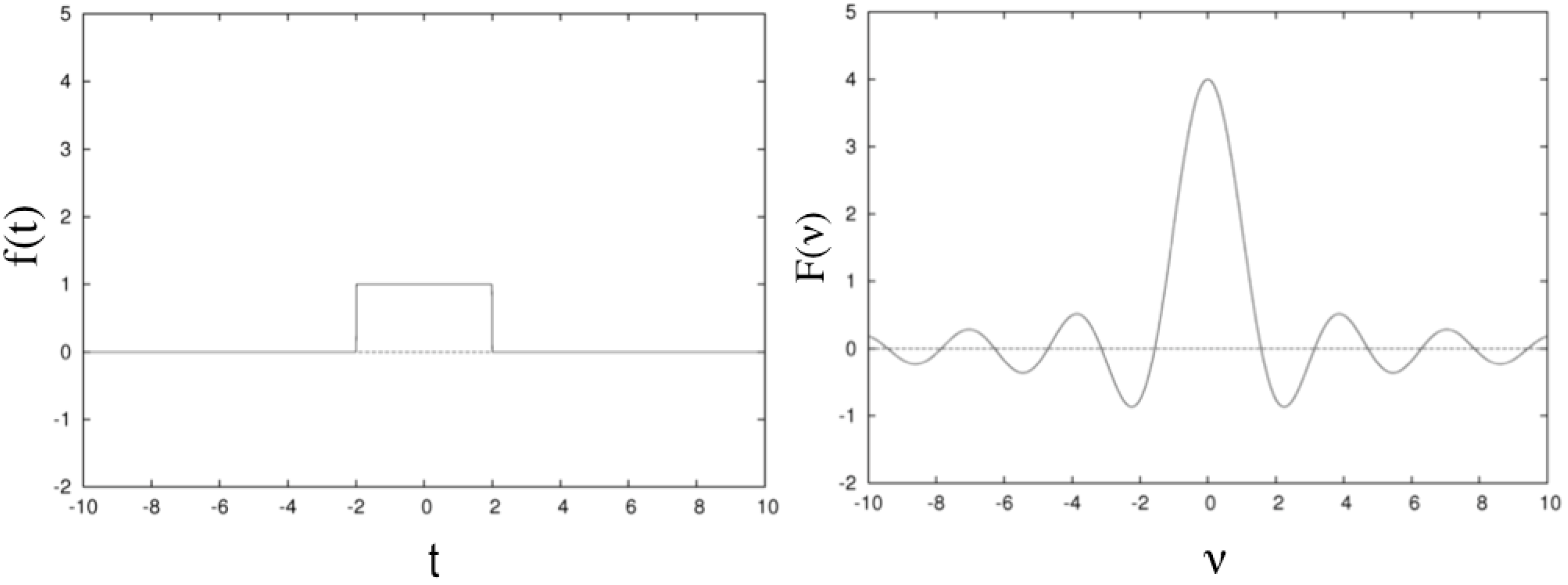}
\end{center}
\caption{\label{box} Fourier Transform pair of a boxcar function and a sinc.}
\end{figure}

\begin{figure}[!htbp*]
\begin{center}
\includegraphics[width=.95\textwidth]{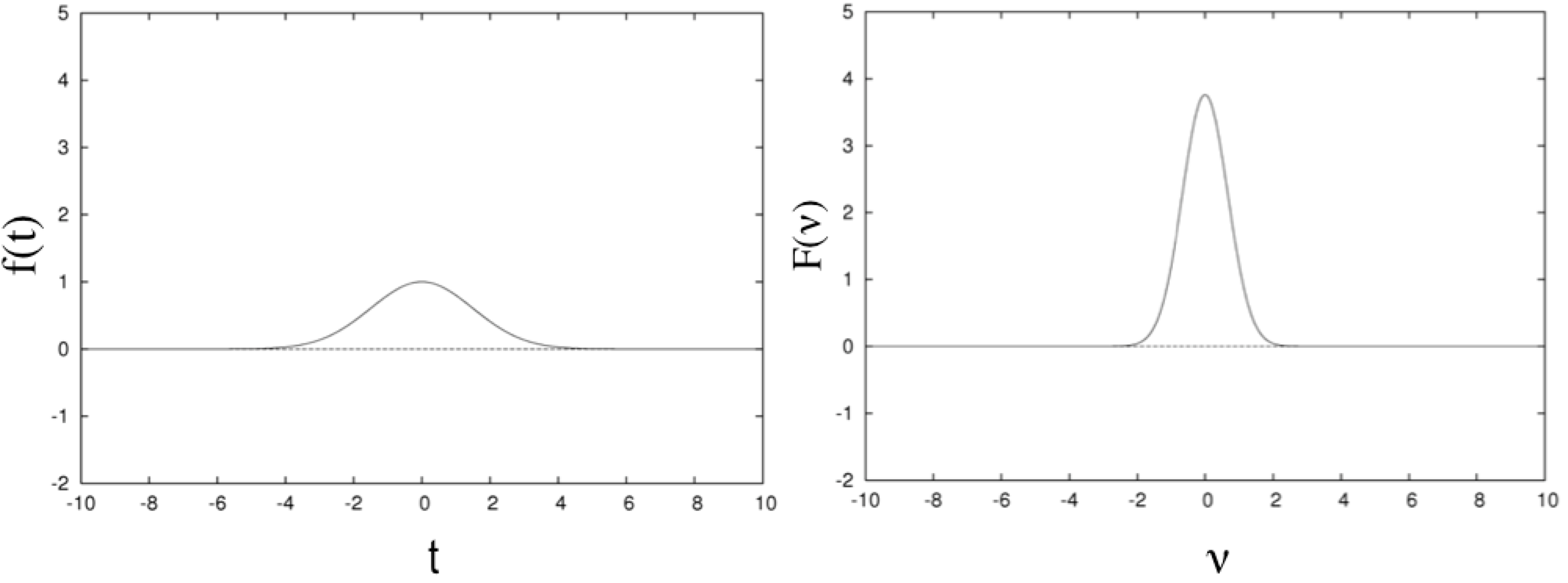}
\end{center}
\caption{\label{Gauss} Fourier Transform pair of a Gaussian function.}
\end{figure}

\begin{figure}[!htbp*]
\begin{center}
\includegraphics[width=.8\textwidth]{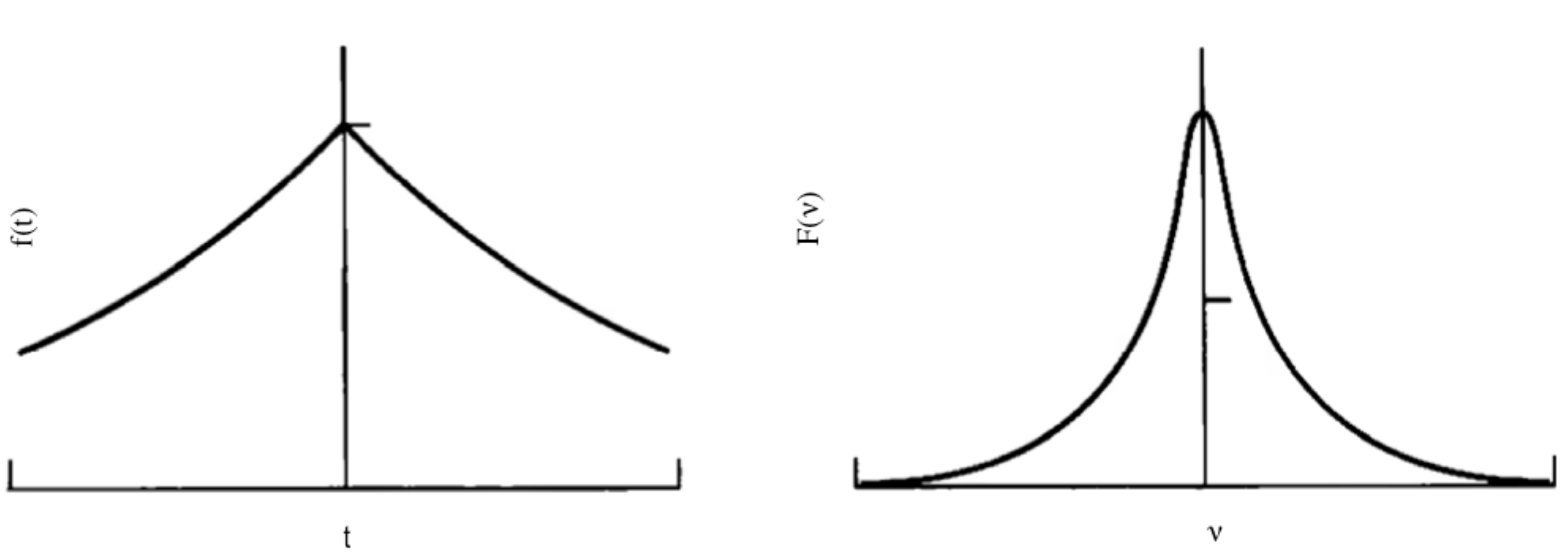}
\end{center}
\caption{\label{exp} Fourier Transform pair of a exponential and a Lorentzian function.}
\end{figure}

\begin{figure}[!htbp*]
\begin{center}
\includegraphics[width=.95\textwidth]{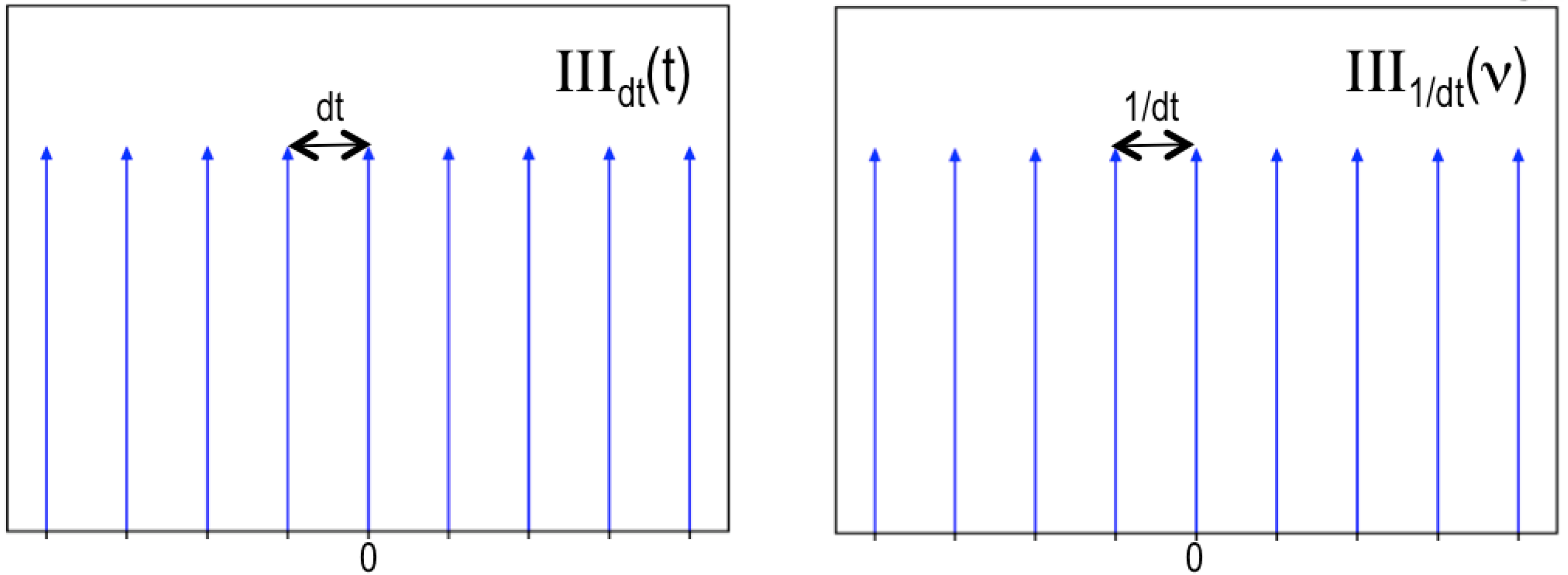}
\end{center}
\caption{\label{Comb} Fourier Transform pair of a Comb function.}
\end{figure}

 \begin{itemize}
\item $f(t)$= constant ;  $F(\nu)$ is a Dirac delta (see Fig.~\ref{constant}):\\ $F(\nu)=\delta(\nu)$.

\item $f(t)= cos(at)$ ;  $F(\nu)$ are two Dirac deltas centered at $\pm a$ (see Fig.~\ref{cos}):\\ $F(\nu)=\delta(\nu-a) + \delta(\nu+a)$.

That implies that for sinusoidal signals, $F(\nu)$ is only different from zero at $\nu = \pm \nu_a$. Therefore, in the case of multi periodic signals (defined as the superposition of several sinusoidal functions) of frequencies $\nu_1$,..., $\nu_N$, and amplitudes $A_1,..., A_N$, its Fourier transform, $F(\nu)$, can be written as a sum of harmonic functions with:
\begin{equation}
\overline{f(t)}=F(\nu)=\sum_{k=1}^{N} A_k \delta(\nu-\nu_k) \;\;\; .
\end{equation}

\item $f(t)$ = Boxcar function of width a ; $F(\nu)$ is a Sinc function of width at half maximum 1/a (see Fig.~\ref{box}): \\$F(\nu)= 2a \; \mathrm{sinc}(a \nu \pi)=2a \sin(a \nu \pi)/a \nu \pi$.

\item $f(t)$ = Gaussian of  width a: $f(t) = e^{-t^2/2a^2}$  ; $F(\nu)$ is another Gaussian (see Fig.~\ref{Gauss}): \\$F(\nu) = (2\pi)^{1/2} a e^{-(a\nu)^2/2}$.

\item $f(t)$= Exponential, $f(t)=e^{-\mid{t}\mid}$ ; $F(\nu)$ is a Lorentzian function (see Fig.~\ref{exp}):\\ $F(\nu)=2/(1+(2\pi\nu)^2)$.

\item $f(t)$= Dirac Comb function defined as a series of Dirac deltas separated by $dt$, $f(t)=\rm{III}_{dt}(t)= \sum_{k=-\infty}^{\infty} {\delta_{kdt}}$;  $F(\nu)$ is another Dirac Comb function but spaced $1/dt$ (see Fig.~\ref{Comb}):\\ $F(\nu)=\rm{III}_{1/dt}(\nu)$.
 \end{itemize}


\subsection{Some properties of the Fourier transform}

\subsubsection{The addition theorem}
The Fourier Transform is a linear transformation. Hence, given two functions $f(t)$ and $g(t)$, whose Fourier Transforms are $F(\nu)$ and $G(\nu)$, respectively, the Fourier Transform of any linear combination of $f(t)$ and $g(t)$ is:
\begin{equation}
\overline{h(t)}=\overline{af(t)+bg(t)}= aF(\nu)+bG(\nu)
\end{equation}
where $a$ and $b$ are any real or imaginary constants. 
\begin{figure}[!htb*]
\begin{center}
\includegraphics[width=.8\textwidth]{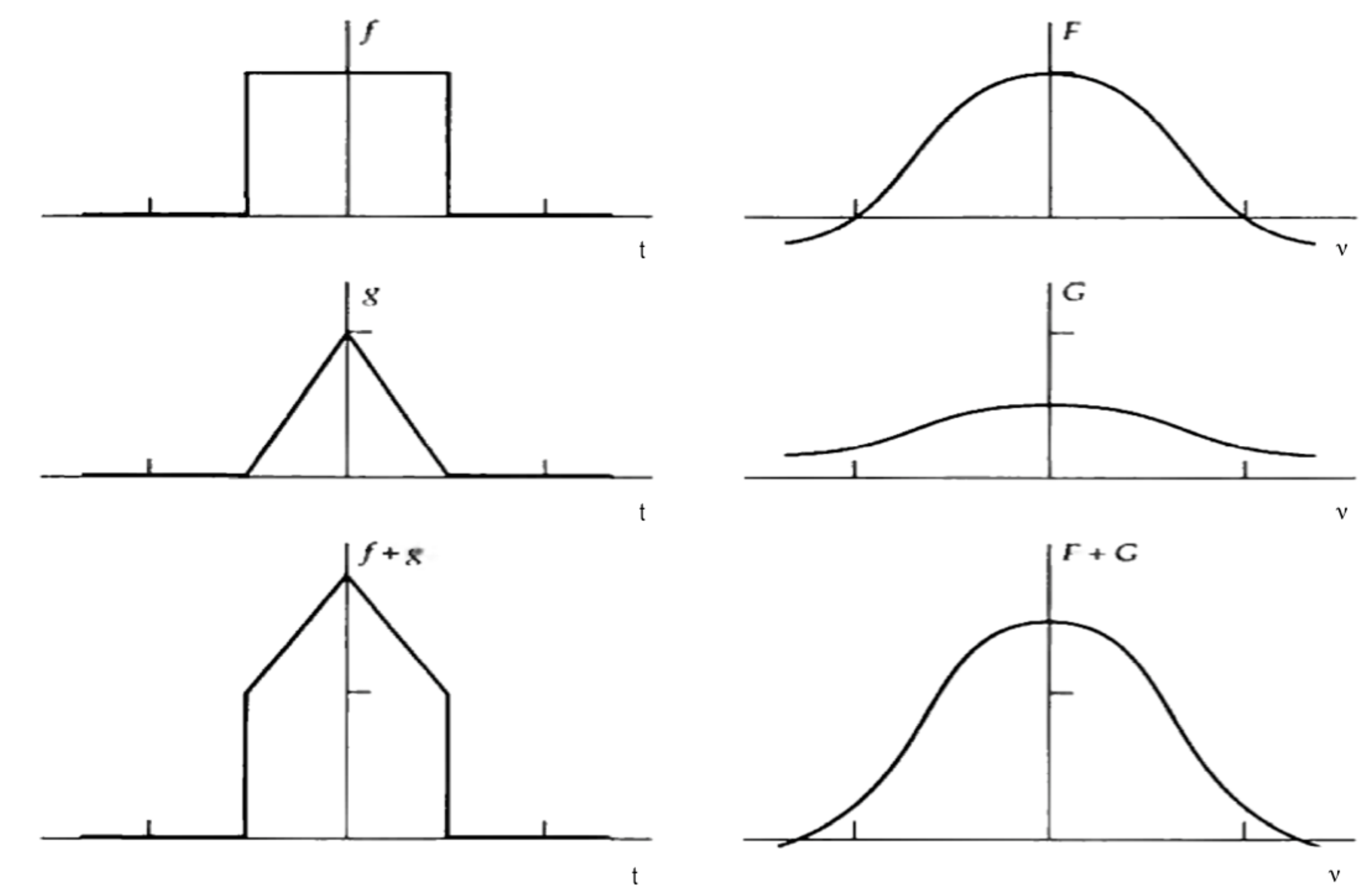}
\end{center}
\caption{\label{addition} Graphical illustration of the addition theorem \citep[from][]{2000fta..book.....B}.}
\end{figure}

\subsubsection{The shift theorem}
\begin{figure}[!htb*]
\begin{center}
\includegraphics[width=.8\textwidth]{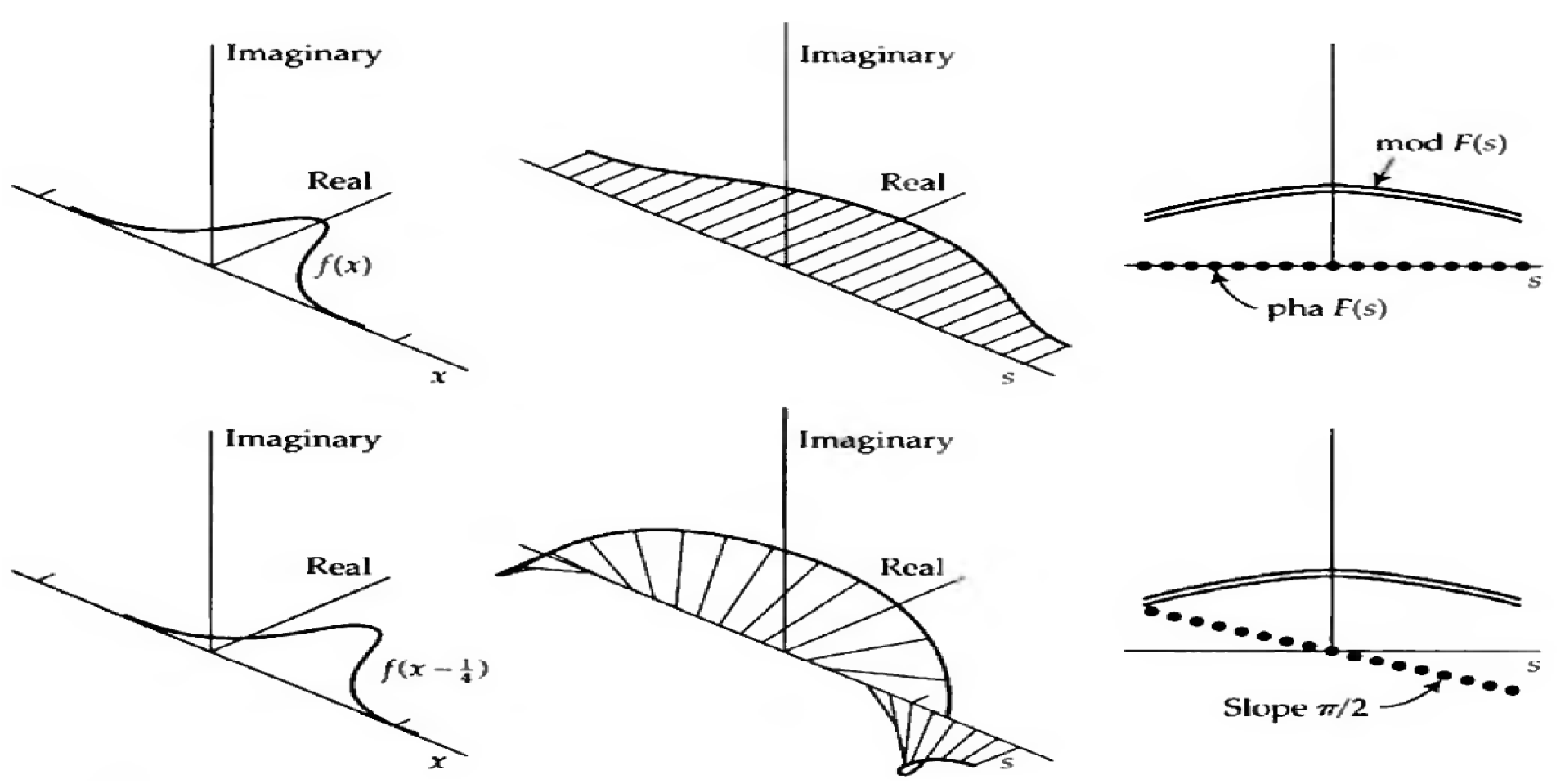}
\end{center}
\caption{\label{shift} Graphical illustration of the shift theorem \citep[from][]{2000fta..book.....B}.}
\end{figure}

The Fourier transform of a function shifted in time by a real number {\it a}, is a function that does not change in amplitude but in phase:
\begin{equation}
\overline{f(t-a)}=e^{-2\pi i a \nu} F(\nu)  \,\,\, .
\end{equation}
In other words, each Fourier component will be delayed by a factor which is proportional to the frequency $\nu$, the higher the frequency, the greater the change in the phase angle. See a graphical illustration of the shift theorem in Fig.~\ref{shift}.

\subsubsection{The scaling (or similarity) property}
A compression of the time scale corresponds to the expansion of the frequency scale:
\begin{equation}
\overline{f(at)}=\frac{1}{a} F \left( \frac{\nu}{a} \right) \,\,\, .
\end{equation}
However, as one member of the transform pair expands horizontally, the other not only contracts horizontally but also grows vertically. In such way, the area beneath the functions is preserved. For periodic signals, an expansion of a function in time corresponds to a stretching of the frequencies in the Fourier domain.  In other words, a ``wide" function in the time-domain is a``narrow" function in the frequency-domain.  This is the basis of the uncertainty principle in quantum mechanics and the diffraction limits of radio telescopes.

\begin{figure}[!htb*]
\begin{center}
\includegraphics[width=.8\textwidth]{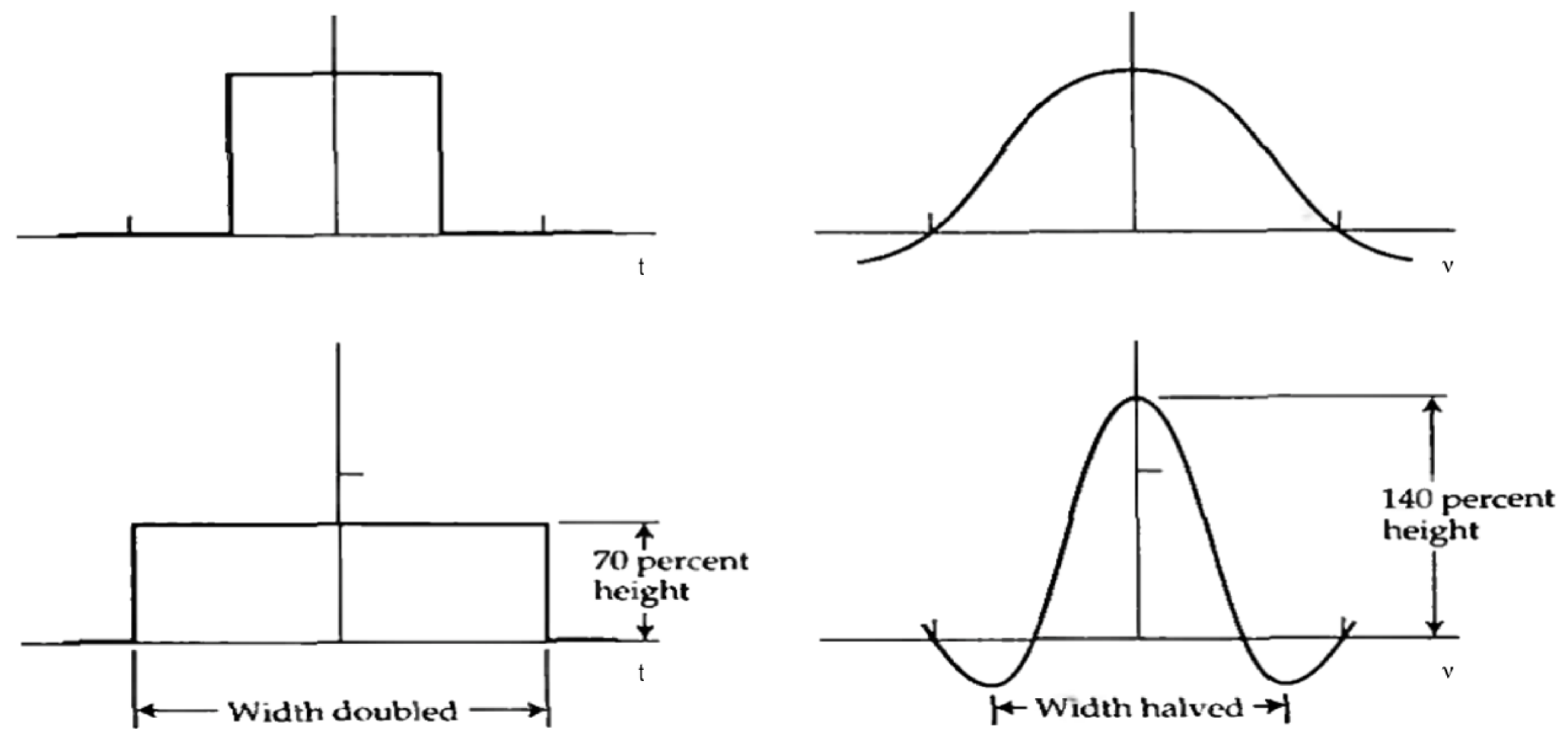}
\includegraphics[width=.8\textwidth]{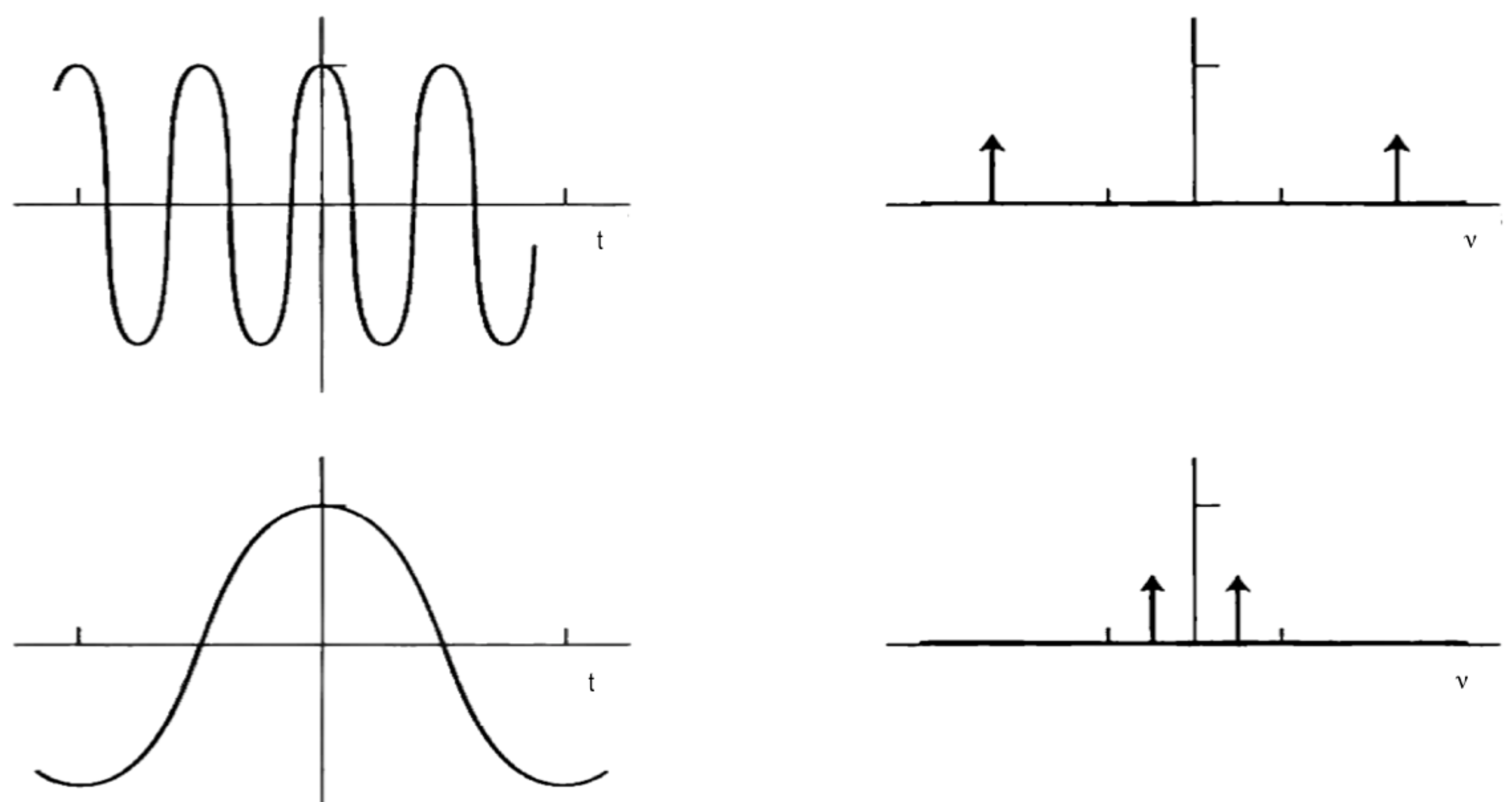}
\end{center}
\caption{\label{scaling} Graphical illustration of the scaling theorem for two boxcar functions (upper two panels) and two cosinus functions (lower two bottom panels). From \citet{2000fta..book.....B}.}
\end{figure}

\subsubsection{The convolution function}
The convolution function gives the area overlapped between the two considered functions, $f(t)$ and $g(t)$, as a function of the amount that one is translated in respect to the other one. In this case, $u$ is the time lag between the two:
\begin{equation}
h(u) = f \ast g = \int\limits_{-\infty}^{\infty} {f(t) g(u-t) dt}  \,\,\, .
\end{equation}

\begin{figure}[!htb*]
\begin{center}
\includegraphics[width=.8\textwidth]{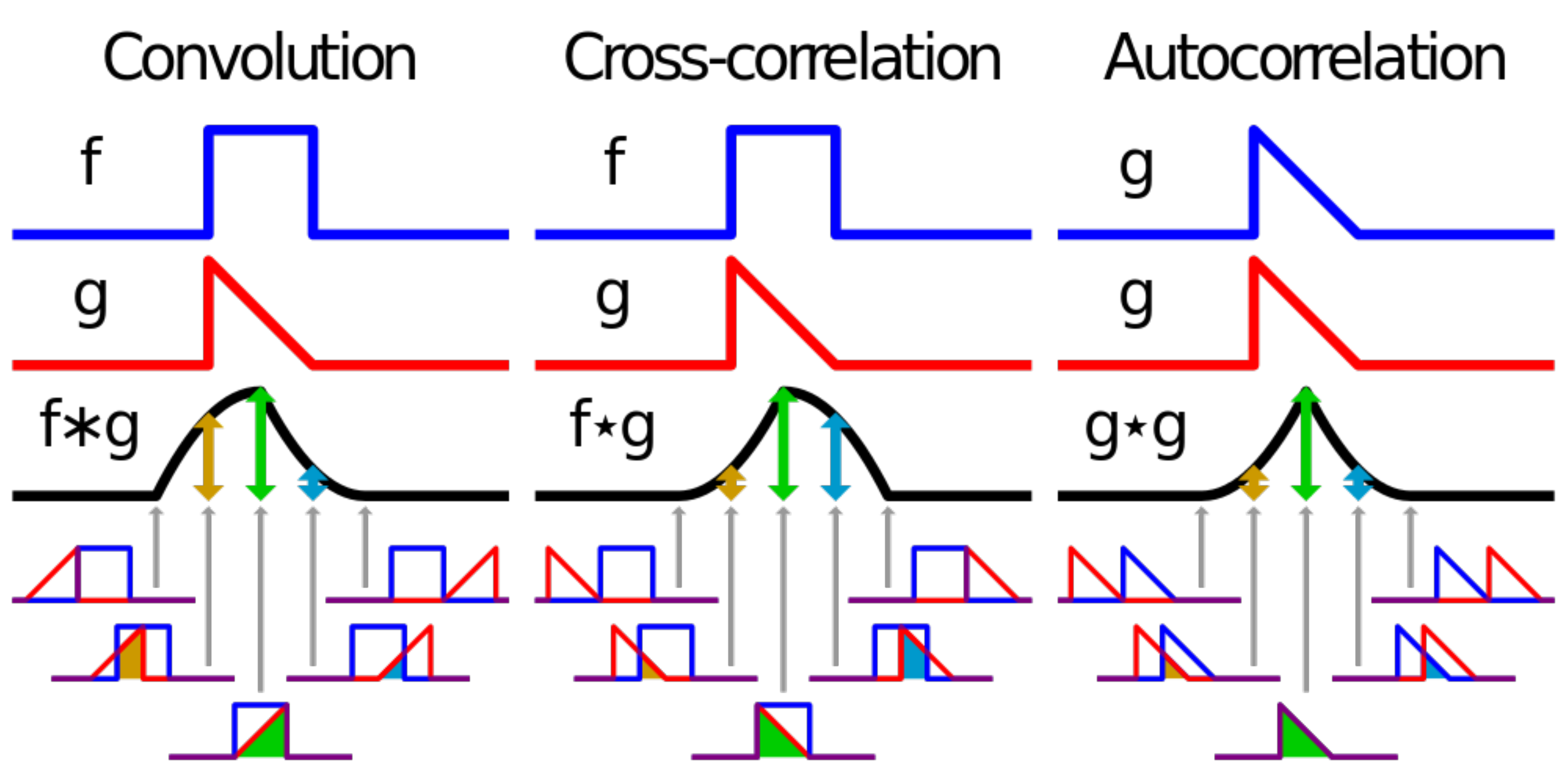}
\end{center}
\caption{\label{crosscorr} Graphical comparison between convolution, cross-correlation and autocorrelation (credits Cmglee).}
\end{figure}

To ensure commutability, the two functions move in opposite directions (see Fig.~\ref{crosscorr} for a graphical representation). It can be seen that the Fourier transform of the product of two functions is the convolution of the Fourier transform of each one:
\begin{equation}
\overline{fg}=\overline{f} \ast \overline{g}
\end{equation}

\subsubsection{Cross-correlation and autocorrelation}
It is a measure of similarity of two waveforms as a function of a time lag applied to one of them (see Fig.~\ref{crosscorr} for a graphical representation):
\begin{equation}
h(u) = f \star g = \int\limits_{-\infty}^{\infty} {f^*(t) g(t+u) dt}  \,\,\, .
\end{equation}

In the special case in which the cross-correlation is applied with the same function, it is called autocorrelation. It provides the linear dependence of a variable with itself at two points in time:
\begin{equation}
h(u) = g \star g= \int\limits_{-\infty}^{\infty} {g^*(t) g(t+u) dt}\,\,\, .
\end{equation}
In this case, if a function g(t) has its Fourier transform $G(\nu)$ then the Fourier transform of its autocorrelation is $\mid G(\nu) \mid^2$. 

The autocorrelation has always a maximum in zero. For stationary processes, the autocorrelation between any two functions only depends on the time lag between them ($u$).

\subsubsection{The Rayleigh's (Parseval) theorem}
The Rayleigh's theorem (applied for the first time by Lord Rayleigh in 1879 while investigating blackbody radiation), is sometimes also called the Plancherel's theorem (because he established in 1910 the conditions under which it holds) and is related to the Parseval's theorem (1799) for Fourier series. It shows that the integral of the squared modulus of a function is equal to the integral of the modulus of its spectrum. In other words, it establishes that the energies in the time and frequency domains are equal: 
\begin{equation}
\int\limits_{-\infty}^{\infty} {\mid f(t) \mid^2 dt} =  \int\limits_{-\infty}^{\infty} {\mid F(\nu) \mid^2 d\nu} \,\,\, .
\label{eq:Pars}
\end{equation}

\subsubsection{Shannon's or sampling theorem}
In the case of a function with a limited spectral response (between $\nu_{\rm{min}}$ and $\nu_{\rm{max}}$), its sampling frequency should be $\Delta\nu > 2 \nu_{\rm{max}}$. If the signal is periodic, to properly sample such function in the time domain, it is necessary that $\Delta t < T_{o} / 2$. In other words, it is necessary to have more than two distinct points per fundamental period to properly sample the function. This theorem established the basis of the discretization.

\subsection{Real observations}
When dealing with observations of real physical phenomena, it is natural to start the observations at a given moment and perform the measurements during a given time at a given rate. Therefore, we need to set up the basis to move from a continuous mathematical description to a discrete  framework which physicist usually deals with. 

\subsubsection{Sampling rate and Nyquist frequency}
\begin{figure}[!htb*]
\begin{center}
\includegraphics[width=.8\textwidth]{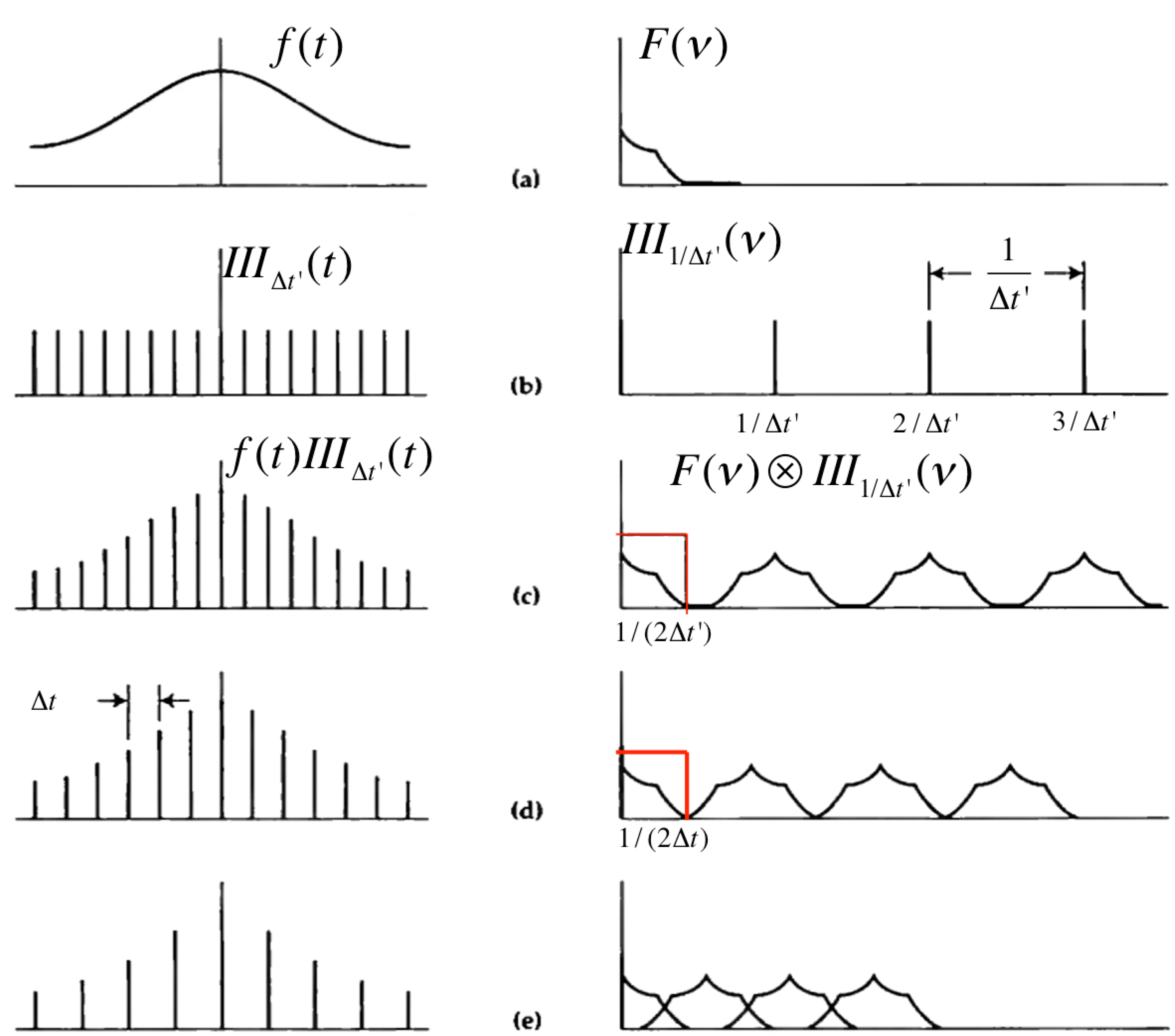}
\end{center}
\caption{\label{discrete} Graphical representation of the discretization of a signal and the aliasing. (a) $f(t)$ is a band-limited signal whose Fourier transform $F(\nu)$ is zero outside a given interval. (b)  $\rm{III}_{\Delta t'}(t)$ is a Comb function of separation $\Delta t'$ whose Fourier transform is another Comb function of separation $1/\Delta t'$. It is used to discretize $f(t)$ by multiplying both functions in the temporal domain (c). The Nyquist frequency, $\nu_{\rm{Nyquist}}=1/(2\Delta t)$, determines the longest sampling rate $\Delta t$ that a band-limited signal can be measured (d). Panel (e) illustrates the case when the signal $f(t)$ is undersampled and the signal is aliased in the spectrum around the Nyquist frequency. Based on \citet{2000fta..book.....B}. }
\end{figure}

Let's assume that $f(t)$ is a band-limited signal, i.e., a signal whose Fourier transform is identically zero outside a finite interval (see panel (a) in Fig.\ref{discrete}) . 

In this case, applying the Shannon's theorem, the signal $f(t)$ can be completely recovered if it is sampled with a rate $\Delta\nu > 2 \nu_{\rm{max}}$. 

Let's define a Comb function of separation $\Delta t' $ whose Fourier transform is another Comb function of separation $1/\Delta t' $ in such way that $1/\Delta t'  > 2 \nu_{\rm{max}}$ (see panel (b) in Fig.\ref{discrete}).
The process of measuring the signal $f(t)$ implies the discretization of the signal. If the measurement is done at a rate of $\Delta t'$ (the temporal resolution), then the measurement consists of multiplying the function $f(t)$ by de Comb function $\rm{III}_{\Delta t'}(t)$. In the Fourier domain, the signal is represented by the convolution $F(\nu) \ast \rm{III}_{1/\Delta t'}(\nu)$. Hence, $F(\nu)$ is repeated at every multiple of $1/\Delta t'$ as illustrated in the panel (c) of Fig.~\ref{discrete}. Because $f(t)$ is band-limited, $F(\nu)$ is a perfect representation of $f(t)$ if the sampling rate $\Delta t$ is such that  $\nu_{\rm{max}}=1/(2\Delta t)$ (see Fig.~\ref{discrete} (d)). This high frequency cut-off of the system for a given sampling rate $\Delta t$ is known as the Nyquist frequency, $\nu_{\rm{Nyquist}}=1/(2\Delta t)$. For any sampling rate longer than the $\Delta t$  mentioned above, the signal would be undersampled. Any frequencies present in the original signal above $\nu_{\rm{Nyquist}}$ (also called super-Nyquist regime) will be reflected or {\it aliased} back into the frequency region below $\nu_{\rm{Nyquist}}$ mixing with the real signal lying in this region as illustrated in panel (e) of Fig.~\ref{discrete} \citep[or further details on super-Nyquist seismology with \emph{Kepler} see][]{2013MNRAS.430.2986M,2014MNRAS.445..946C}. As a consequence, if the signal $F(\nu)$ was band limited and it is sampled at a higher cadence than $\nu_{\rm{Nyquist}}$, no aliasing will appear in the spectrum in accordance to the Sampling Theorem.

\begin{figure}[!htb*]
\begin{center}
\includegraphics[width=.75\textwidth]{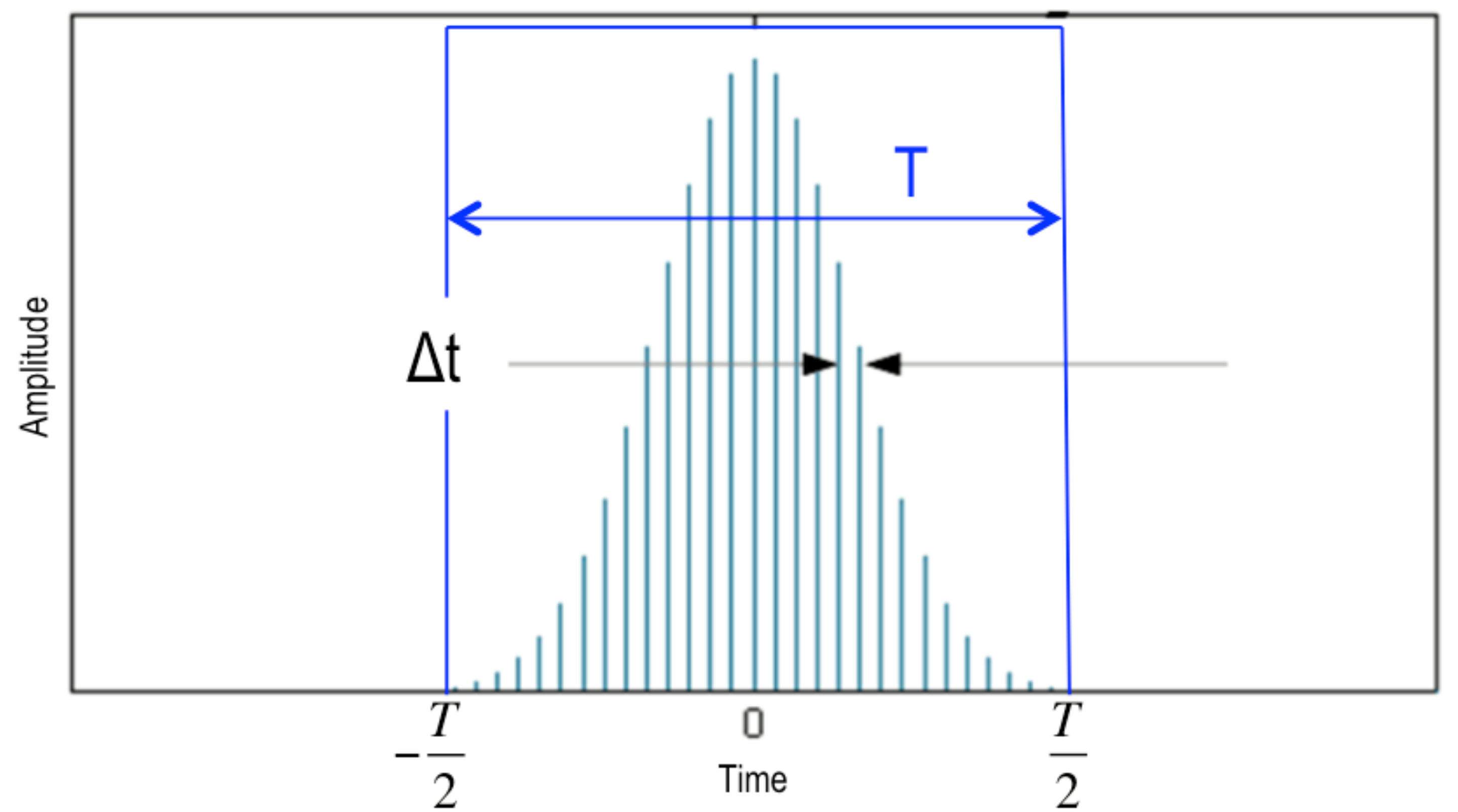}
\includegraphics[width=.75\textwidth]{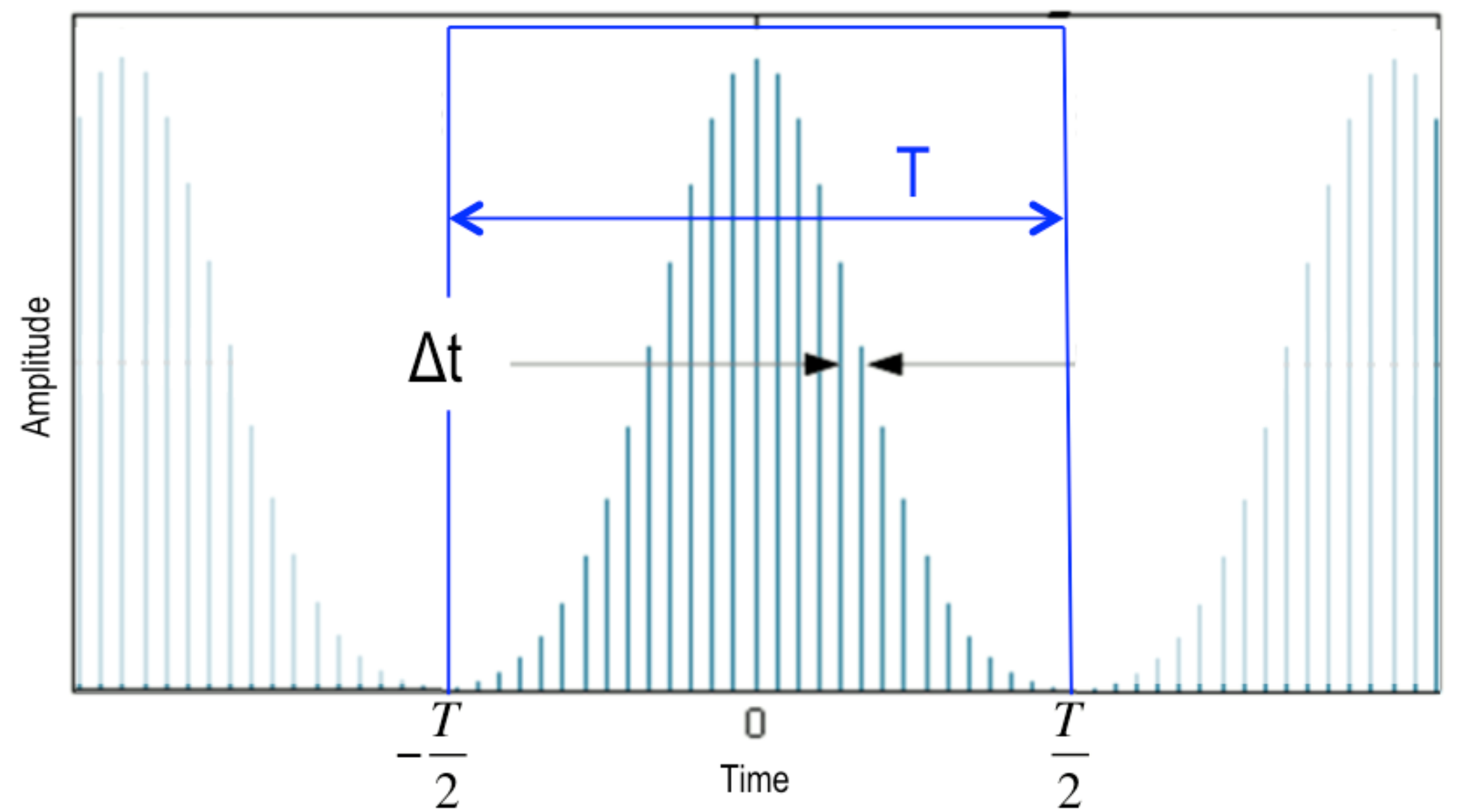}
\includegraphics[width=.75\textwidth]{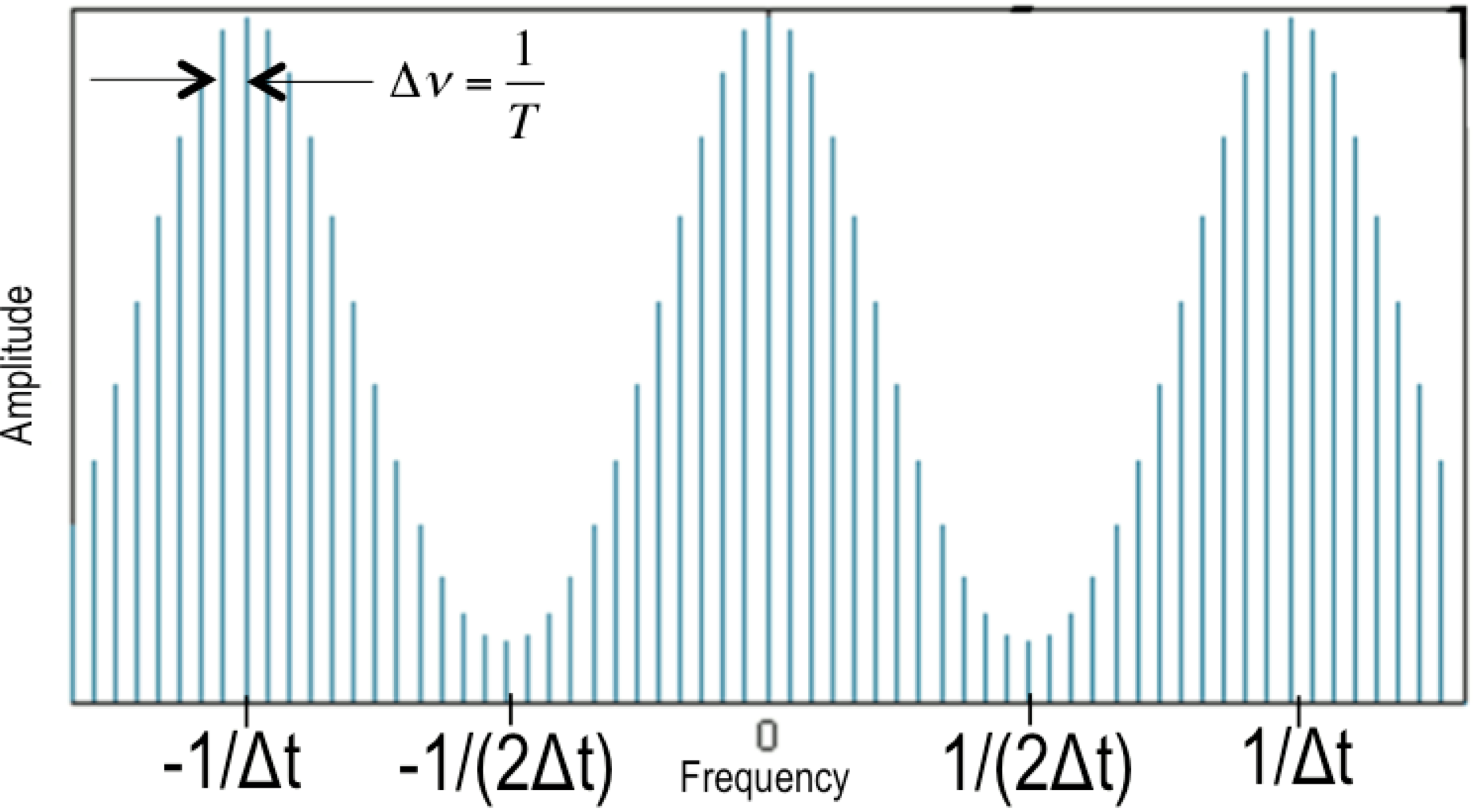}
\end{center}
\caption{\label{fig:finite} Gaussian function, $f(t)$, observed at a sampling rate of $\Delta t$ multiplied by a window function of width $T$ (top panel). Because $f(t)$ is implicitly periodic (middle panel), $F(\nu)$ is sampled at a rate $1/T$. When $f(t)$ is not band limited, there is an aliasing around the Nyquist frequency in $F(\nu)$ (bottom panel).}
\end{figure}

\subsubsection{Finite observations}
Most of the physical processes in which seismology is interested in, correspond to signals that are not band limited. During the observing time, $T$, only part of the signal is sampled $N$ times at a rate given by $\Delta t$: $f_N(t)$. Mathematically this is equivalent to multiplying a given infinite (or longer than the observations) signal sampled $N$ times $f_N(t)$ by a {\it window function} represented by a square function, $\sqcap_T(t)$ defined as being 1 between the range $[-T/2,T/2]$ and zero elsewhere. Therefore, our observations can be represented as the product: 
\begin{equation}
f(t)\, \rm{III}_{\Delta t}(t)\,  \sqcap_T(t) = f_N(t) \, \sqcap_T(t) \;\;\; .
\end{equation}
An illustration can be seen in the top panel of Fig.~\ref{fig:finite} for an infinite Gaussian function sampled at $\Delta t$ and observed during a time span of $T$.

Because the Fourier transform assumes that $f_N(t)$ is implicitly periodic, it is defined as follows $f(t)\rm{III}_{\Delta t}(t) \sqcap_T(t) \ast \rm{III}_T(t)$ (see middle panel in Fig.~\ref{fig:finite}).

The Fourier transform $F_N(\nu)=\overline{f_N(t)}$ can then be defined as:
\begin{equation}
F_N(\nu)=\overline{f_N(t)}=F(\nu) \ast \rm{III}_{1/\Delta t}(\nu) \ast \rm{sinc}(\nu) \,   III_{1/T}(\nu)    \;\;\; . 
\label{finite_obs}
\end{equation}
An illustration of $F_N(\nu)$ can be seen in the bottom panel of Fig.~\ref{fig:finite}. Each vertical line is in reality a Sinc function. For the sake of clarity, these Sinc functions have not been drawn.
The function $F_N(\nu)$, has a frequency resolution $\Delta\nu=1/T$ and has a Nyquist frequency of $1/(2\Delta t)$ as explained before. Because $f_N(t)$ is assumed to be periodic, $F_N(\nu)$ is also periodic at multiples of $1/\Delta t$. Moreover, there is an aliasing around the cut-off frequency because the function $f_N(t)$ is not band limited.

\begin{figure}[!htb*]
\begin{center}
\includegraphics[width=.95\textwidth]{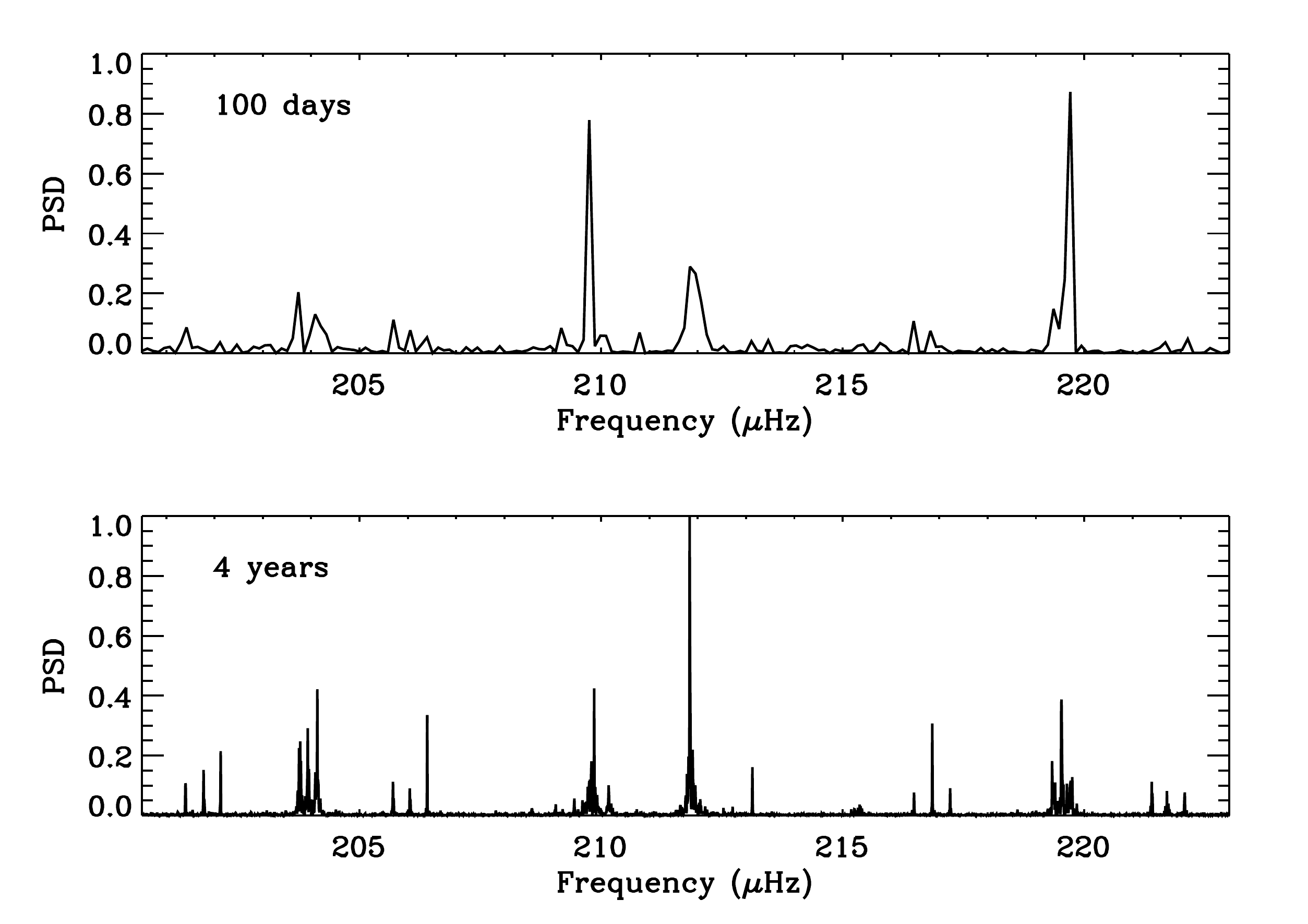}
\end{center}
\caption{\label{resolution} Power density spectrum of the red giant KIC~5356201 observed by \emph{Kepler}. On top, during 100 days, on the bottom panel, during the full length of the mission, 4 years.}
\end{figure}

An example of the effect of the window size is shown in Fig.~\ref{resolution} for the \emph{Kepler} red giant KIC~5356201 with $\nu_{\rm{max}}=212$ $\mu$Hz. When this star is observed during 100 days (typical length of observations of K2 and some CoRoT fields), only the rough characteristics of the modes can be observed and, for example, the splittings of the mixed modes cannot be measured. However, when the observations stands for 4 years, the fine structure of the multiplets in the mixed modes are clearly distinguishable, and a precise measurement of the core internal rotation can be obtained \citep{2012Natur.481...55B}.

To properly resolve the rotational split components of a multiplet, as in Fig.~\ref{resolution}, it is necessary to have enough frequency bins between the two maxima. This will depend on the rotation rate of the cavity traversed by the modes, but also on their line widths \citep[e.g.][]{BalGar2006}. When the modes are stochastically excited (Lorentzian profile) it is commonly   assumed that the length of the time series $T$ needs to be 10 times longer than the mode lifetime $\tau$ (defined as the time for the amplitude to decay by the factor e, $\tau=1/(\pi \Gamma)$) in order to resolve the mode profile  \citep{ChapTongGar,AppTongGar}. For example, for lifetimes of 3.7 days ($\Gamma=1$ $\mu$Hz, typical for main-sequence solar-like stars with $T_{\rm{eff}}=5770$ K), the length of the observations required to properly characterize the modes is 37 days. It is also important to notice that when the ratio between the mode lifetime and the length of the observations is small ($T/\tau \leq 2$), the observed profile tends to a Sinc-squared function.

\subsection{The discrete Fourier Transform}
Following the concepts defined in the previous sections, it is possible to define the Discrete Fourier Transform (D.F.T.) as the finite sum of complex sinusoids, ordered by their frequencies, representing a regularly sampled temporal function:

\begin{equation}
D.F.T.=\overline{f(t_n)}=F_N(\nu_k)=\Delta t \sum_{n=1}^{N} f(t_n)\, e^{(-i2\pi \nu_k t_n)} \;\;\; ,
\end{equation}
where $\nu_k = k/T = k\Delta\nu$, $t_n = n \Delta t$, and $T = N\Delta t$. This expression is equivalent to the truncated original Fourier series and it is implicitly periodic at $\Delta \nu = 1/\Delta t$, i.e., at frequencies double of the Nyquist frequency.

\begin{figure}[!htb*]
\begin{center}
\includegraphics[width=.9\textwidth]{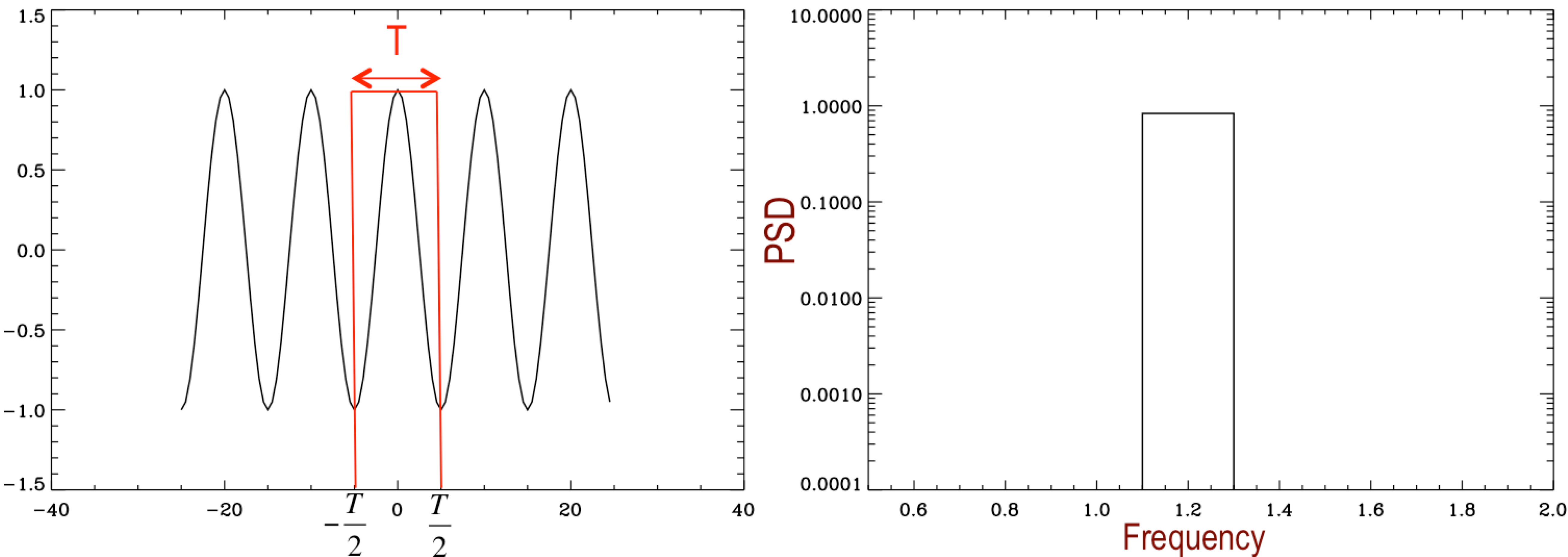}
\includegraphics[width=.9\textwidth]{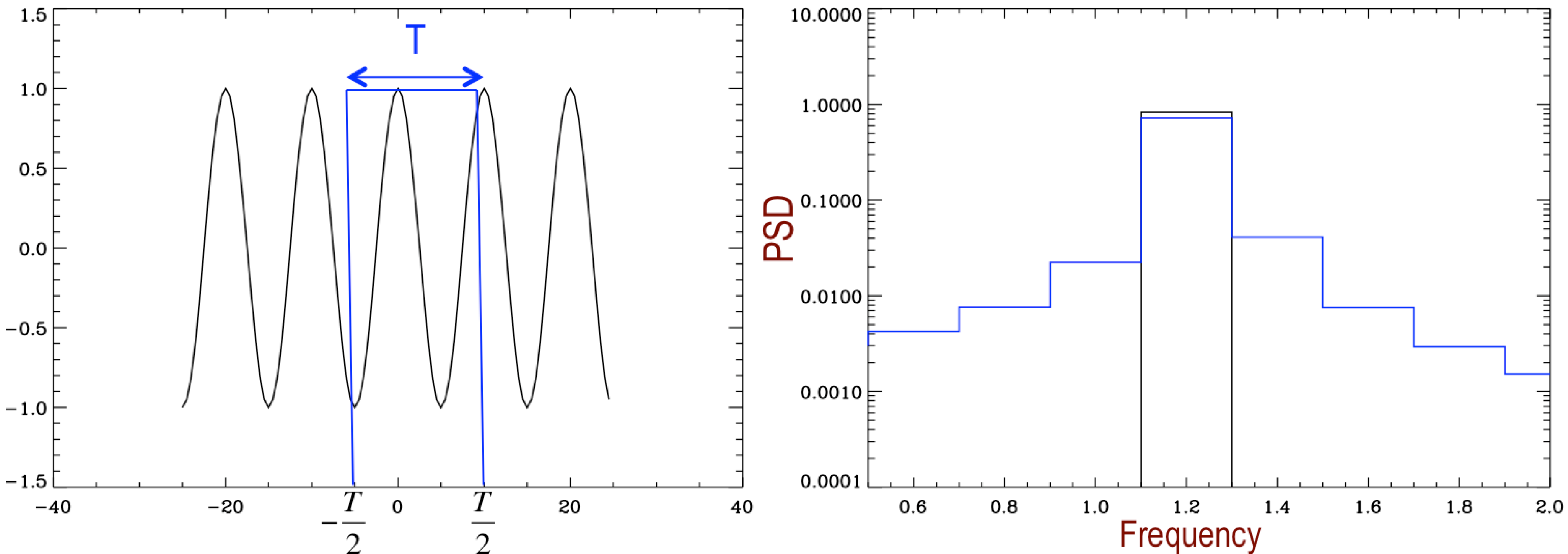}
\end{center}
\caption{\label{periodic} Periodic function in the time domain (left-hand panels) observed during an integer number of periods (top) or with another observational length (bottom). The right hand-side panels correspond to the Fourier transform. In the first case, the Fourier transform is a sinc function in which the zeroes coincide with the points used to compute the discrete Fourier transform. It can be approximated as a Dirac. In the second case, it is a broad Sinc function (blue line in the bottom right-hand panel). To facilitate the comparison we have also overplotted the Sinc function of the previous case (in black). }
\end{figure}

A particular interesting case is when the function $f(t)$ is periodic and it is observed during a period $T$ which is an integer number of periods of the original signal. In such case, the zeroes of the sinc function in the Fourier transform coincide with the points used to compute the discrete Fourier transform and it can be approximated by a Dirac (in non oversampled power spectra, see Fig.~\ref{periodic}).

An interesting consequence of the above is that the amplitude of a periodic signal with period $P$ changes in the spectrum depending on the remainder of $T/P$. In the best case, when the reminder of this ratio is zero, $T$ is proportional to the period of the signal and all the power will be concentrated in the central bin. However, when the remainder is bigger, the Fourier transform is a Sinc function and some power leaks into the adjacent bins while the central one is reduced. This should be taken into account when low-amplitude peaks are searched in a noisy spectrum. In such case, it can be interesting to oversample the signal or to slightly change $T$ by removing some points of the temporal signal \citep[e.g.][]{GabBau2002,STCGar2004}. 

\begin{figure}[!htb*]
\begin{center}
\includegraphics[width=.9\textwidth]{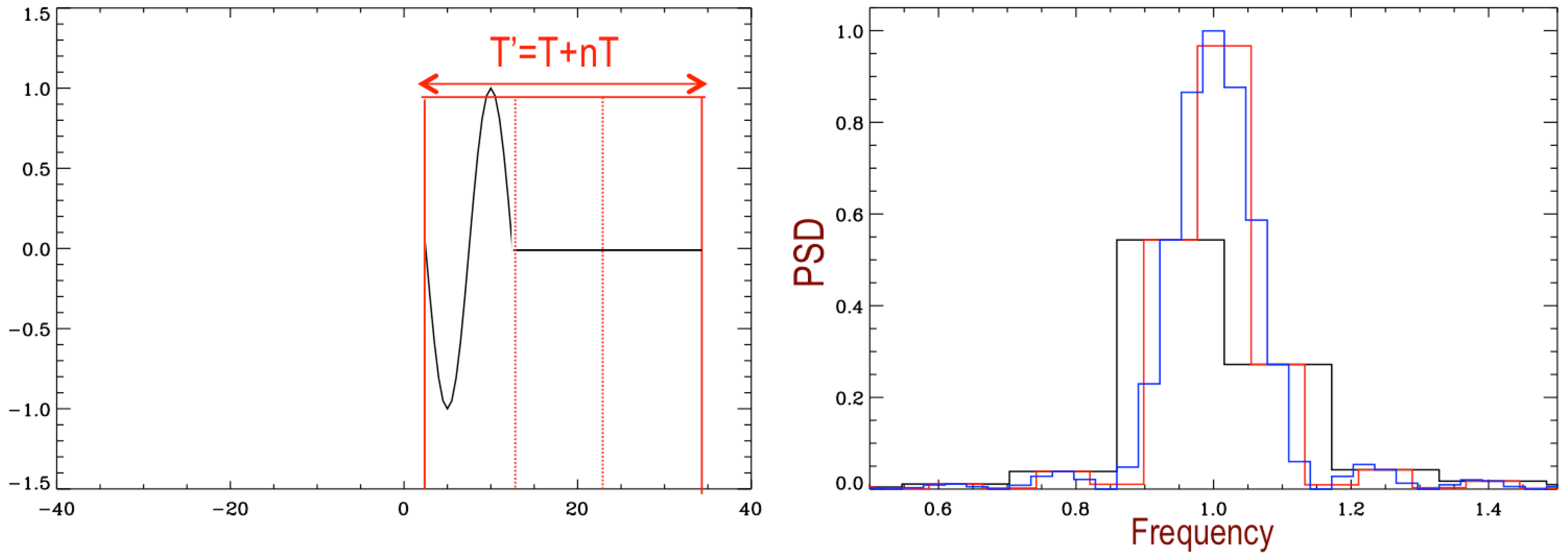}
\end{center}
\caption{\label{zpad} Time series of a periodic signal (left panel) and corresponding Fourier transform (Sinc function in the right panel) for different zero padded time series $T+nT$, with n=0, 3, and 6; black, red, and blue courbes respectively. }
\end{figure}

It has been demonstrated that an oversampling of a factor of 6 (adding zeroes at the end of the time series by 5 times the length of the series) the central bin of the Sinc function recovers $\sim$~97$\%$ of the total signal \citep[for more details see][]{GabBau2002}.

\subsection{Power Spectrum: Fast Fourier Transform}
It is common to work with the power spectrum instead of the amplitude spectrum. It is defined as the modulus squared of the Fourier transform:
\begin{equation}
\mid F_N(\nu_k) \mid^2 = 
P_N(\nu_k)=\frac{1}{N} \left\{ \left( \sum_{n=1}^{N} f(t_n)\cos (2\pi\nu_k t_n)  \right)^2 +\left( \sum_{n=1}^{N} f(t_n)\sin (2\pi\nu_k t_n)  \right)^2 \right\}
\;\;\; .
\label{eq:ps}
\end{equation}

The power spectrum has no information on the phase of the original function. It is generally defined between [0,$\nu_{\rm{Nyq}}$]. When this is the case it is called ``single-sided'' power spectrum, in opposition to the ``double-sided'' power spectrum defined in the full range of frequencies between [$-\nu_{\rm{Nyq}},\nu_{\rm{Nyq}}$] \citep[see also the extended discussion in][]{1992nrfa.book.....P}. This has an impact on the calibration of the power spectrum. The proper way to calibrate the power spectrum is by applying the Parseval's Theorem (see Eq.~\ref{eq:Pars}), i.e., by ensuring that the integral of the squared modulus of the function in the time domain is equal to the integral of the squared of its spectrum. In the case of the single-sided power spectrum, the power of each bin is multiplied by two to take into account the power in the negative frequency region. In seismology it is very common to work with the power spectrum density (or PSD) which is the power spectrum divided by the resolution in the Fourier domain. This representation has the advantage that it takes into account the variable size of the frequency bin for different lengths of observations $T$, allowing simple and direct inter comparations between spectra computed from datasets of different lengths.

Assuming that $f(t)$ is normally distributed for $N>>1$, then the real and imaginary parts will also be normally distributed. Therefore, the power spectrum will follow a $\chi^2$ with 2 degrees of freedom statistics. 

When the signal $f(t)$ is evenly sampled the sums involved in Eq.~\ref{eq:ps} requires $N^2$ operations. This can take quite some amount of time and several algorithms of {\it Fast Fourier Transform} have been developed to increase the speed. They usually require a number of operations that are of the order of $n \log N$ \citep[see some examples in][and references therein]{1992nrfa.book.....P}.

\subsubsection{Case of unevenly distributed points: Lomb-Scargle Periodogram}

When the sampling rate is not regular and the data are unevenly distributed, the Fourier spectrum does not follow in the general case a $\chi^2$ with 2 degrees of freedom statistics. To overcome this issue, \citet{1982ApJ...263..835S} developed the so-called Lomb-scargle (LS) periodogram:
\begin{equation}
F_{LS}(\nu_k)=\frac{1}{\omega(\tau)} \sum_{n=1}^{N} f(t_n)\cos (2\pi\nu_k (t_n-\tau))  + i \frac{1}{\nu(\tau)} \sum_{n=1}^{N} f(t_n) \sin (2\pi\nu_k (t_n-\tau))  \;\;\; ,
\label{eq:LS}
\end{equation}
where:
\begin{equation}
\omega(\tau) = \sum_{n=1}^{N} cos^2 (2\pi\nu_k (t_n-\tau)) \;\;\; ,
\end{equation}
\begin{equation}
\nu(\tau) = \sum_{n=1}^{N} sin^2 (2\pi\nu_k (t_n-\tau))
\end{equation}
and $\tau$ is selected to keep the invariant with time of equation Eq.~\ref{eq:LS} as follows:
\begin{equation}
\tan(2\pi \nu \tau) = \frac{\sum_{n=1}^{N} \sin(2\pi\nu t_n)}{\sum_{n=1}^{N} \cos(2\pi\nu t_n)} \;\;\; .
\end{equation}

With this formulation, it can be demonstrated that the power spectrum $\mid F_{LS}(\nu_k)\mid^2$ follows a $\chi^2$ with 2 degrees of freedom statistics \citep{Press1989}. The Lomb-Scargle periodogram implicitly adds zeroes at the positions of the missing points, which means that the bins are correlated as in any time series with gaps. It is also important to notice that to speed up the calculations the LS periodogram is sometimes approximated by an interpolation into a regular mesh of points and the use of a FFT algorithm properly normalized \citep{1982ApJ...263..835S}.

\begin{figure}[!htb*]
\begin{center}
\includegraphics[width=.9\textwidth]{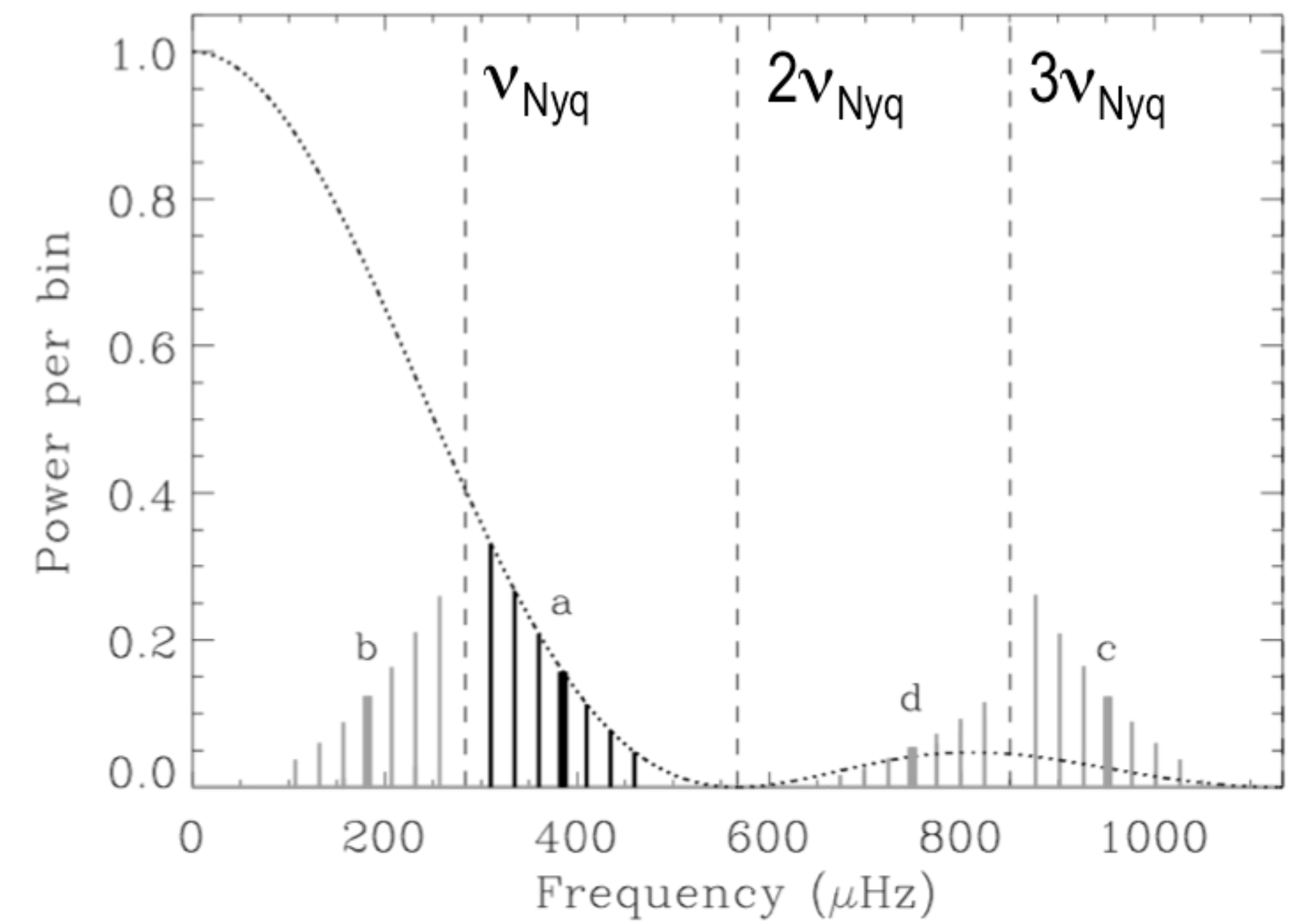}
\end{center}
\caption{\label{fig:chalias} Power Spectrum of a series of pure sinusoids of unit amplitude having frequencies between 310 and 460 $\mu$Hz irregularly sampled in time. The black vertical lines are the true frequencies while the grey ones are the aliases. The dotted line marks the Sinc-squared attenuation envelope. See the discussion in the text about the general shape of the spectrum. The vertical dashed lines are the multiples of $\nu_{\rm{Nyq}}$. Adapted from  \citet{2014MNRAS.445..946C}. }
\end{figure}

Unevenly sampled data, as the ones obtained by the \emph{Kepler} satellite, can be useful to disentangle real peaks from aliases at frequencies close to the Nyquist cut-off frequency \citep{2013MNRAS.430.2986M}. In Fig.~\ref{fig:chalias} we show a series of pure sinusoids of unit amplitude with frequencies between 310 and 460 $\mu$Hz above the Nyquist frequency and  irregularly sampled following the \emph{Kepler} timing \citep[see more details on the \emph{Kepler} timing in][]{2014A&A...568A..10G}. 

\begin{figure}[!htb*]
\begin{center}
\includegraphics[width=.95\textwidth]{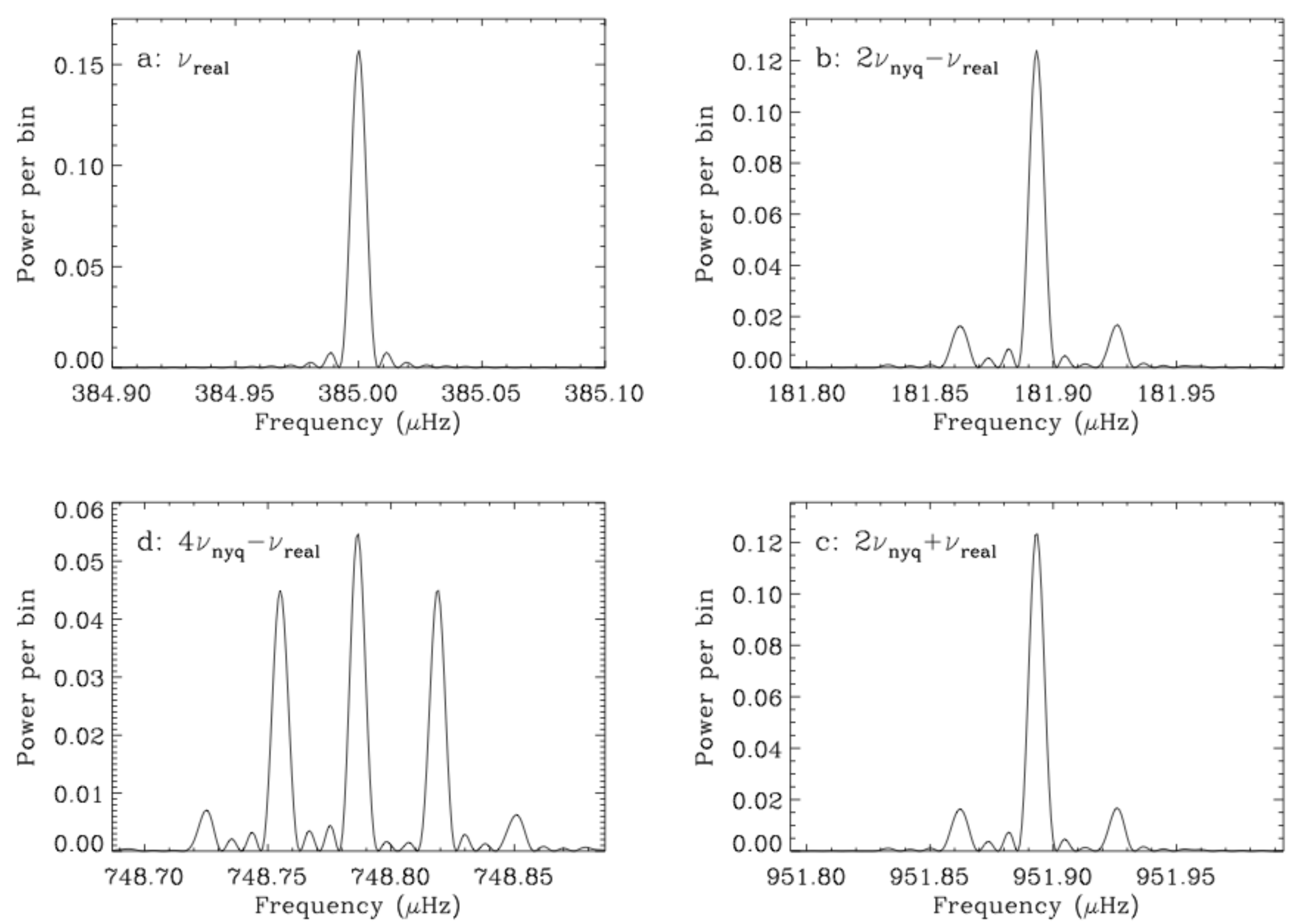}
\end{center}
\caption{\label{fig:chalias2} Zoom of the Power Spectrum of the peaks marked ``a, b, c, and d'' in Fig.~\ref{fig:chalias}  \citep[from][]{2014MNRAS.445..946C}. }
\end{figure}

Because the sampling is not regular, the aliased peaks into the frequency band below $\nu_{\rm{Nyq}}$ are not perfect (see a zoom of the peaks marked ``a, b, c, and d'' in Fig.~\ref{fig:chalias2}). As explained by \citet{2014MNRAS.445..946C} the exact structure of the alias peaks is different and depends on the relation of their frequencies with the sampling frequency. For coherent pulsators (with narrow peaks), the detailed study of the sidebands in the peaks allows to discriminate real peaks to the aliased ones \citep{2013MNRAS.430.2986M}. However, in the case of solar-like pulsators with modes of short lifetimes, the situation is more complicated because the structure of the aliases is mixed with the natural stochastic excitation of the modes \citep[][]{2014MNRAS.445..946C}.

The general appearance of the full power spectrum is affected by the effective integration time per sampling unit used to collect the data (usually called cadence). Assuming an integration time per cadence of $\Delta t'$, the amplitude of any signal at a given frequency $\nu$ is given by $\eta=\rm{sinc}(\pi\nu\Delta t')$ \citep{2012PhDT.......389C}. When the integration time is close to the sampling time (i.e., the dead time per measurement is close to zero), $\Delta t' \sim \Delta t$, the amplitudes follows the attenuation factor defined as $\eta=\rm{sinc}(\pi\nu/(2\nu_{\rm{Nyq}}))$ \citep[e.g.][and references therein]{2014MNRAS.445..946C}.

 \subsection{Regular gaps in the time series}
 \label{Sec:gaps}
 Continuous observations of real phenomena are usually difficult. In ground-based astronomy, the observation of the Sun and stars is conditioned by the rotation of the Earth. Thus, unless observing at high latitudes, it is generally impossible to observer longer than 10-15 hours continuously the same object. In seismology, the existence of regular gaps in the data has ominous effects in the Fourier domain.
 
 Short regular gaps in the time series can be represented mathematically as the product of the signal we are interested in, $f(t)$, with a Comb function of period $T_1$:
 \begin{equation}
 \overline{\rm{III}_{T_1}(t)} = \rm{III}_{1/T_1}(\nu)
 \end{equation}
 
In the case of longer gaps, the mathematical representation is a product of a rectangular window convolved with a Comb function:
  \begin{equation}
 \overline{\sqcap_{T_2} (t) \ast \rm{III}_{T_1}(t)} = \rm{sinc}(\nu) \, \rm{III}_{1/T_1}(\nu)
 \end{equation}

\begin{figure}[!htb*]
\begin{center}
\includegraphics[width=.95\textwidth]{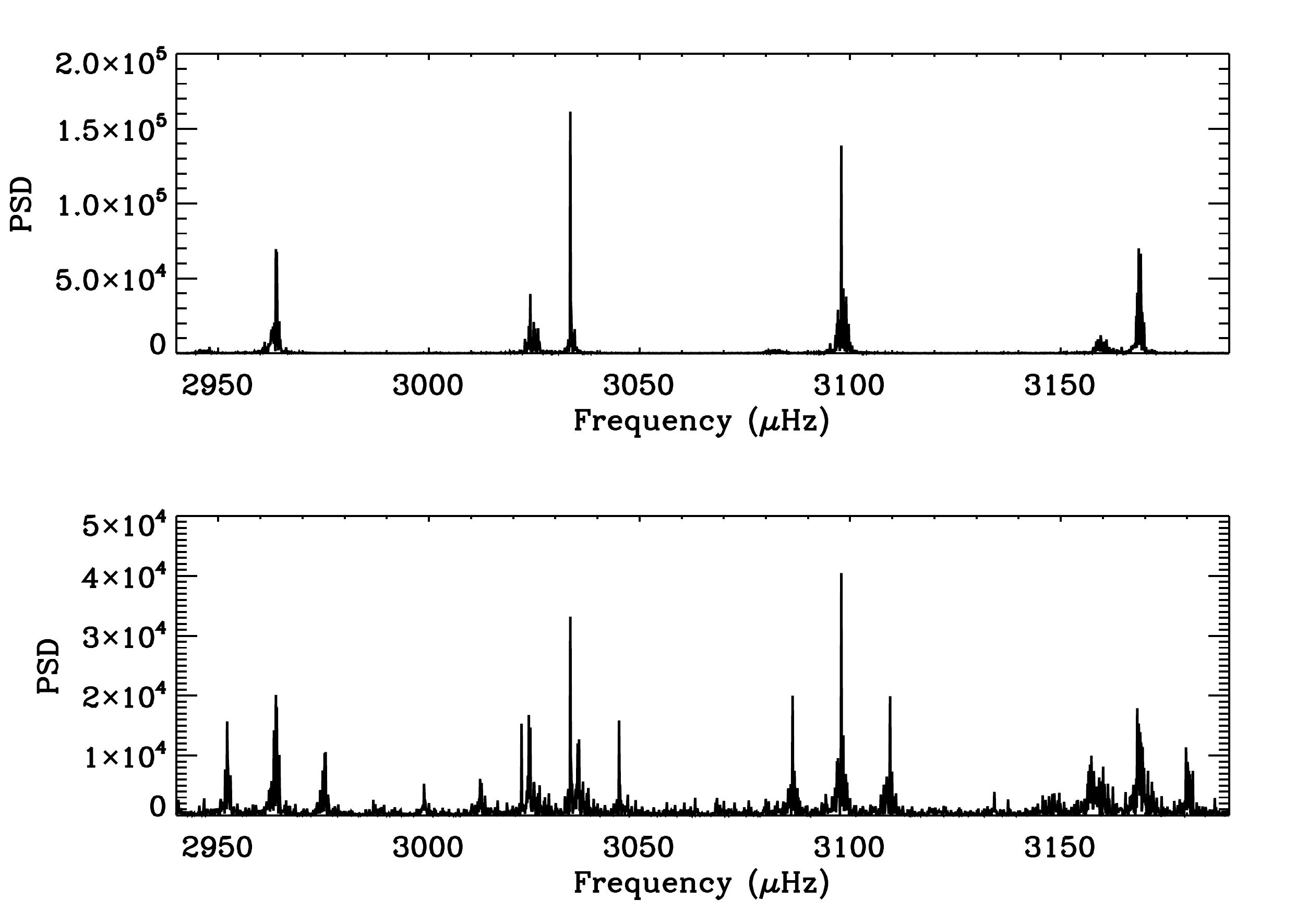}
\end{center}
\caption{\label{fig:mki} Power spectrum density of 100 days helioseismic GOLF data with nearly 100\% coverage (Top panel). The bottom panel represents the PSD of the same data but after multiplying by a realistic window function of 23\% duty cycle obtained from the Observatorio del Teide.  }
\end{figure}

Due to the day-night alternance, ground-based seismic observations of a single site of the Sun or stars imply that each mode in the power spectrum will appear at frequencies multiples of 1/24 h, i.e., 11.57 $\mu$Hz, usually called daily sidebands. In Fig.~\ref{fig:mki} the GOLF power spectrum density is shown. In the top panel the one obtained from 100-day time series with a duty cycle close to 100$\%$. In the bottom panel, the PSD obtained after multiplying the same time series by a window mask corresponding to the Mark-I instrument --one of the BiSON helioseismic network \citep{1996SoPh..168....1C} located at Observatorio del Teide--  with a duty cycle of 23$\%$.  As expected, in the simulated ground-based data, every single solar mode is replicated at frequencies multiple of 11.57 $\mu$Hz, complicating the analysis.
 
\begin{figure}[!htb*]
\begin{center}
\includegraphics[width=.95\textwidth]{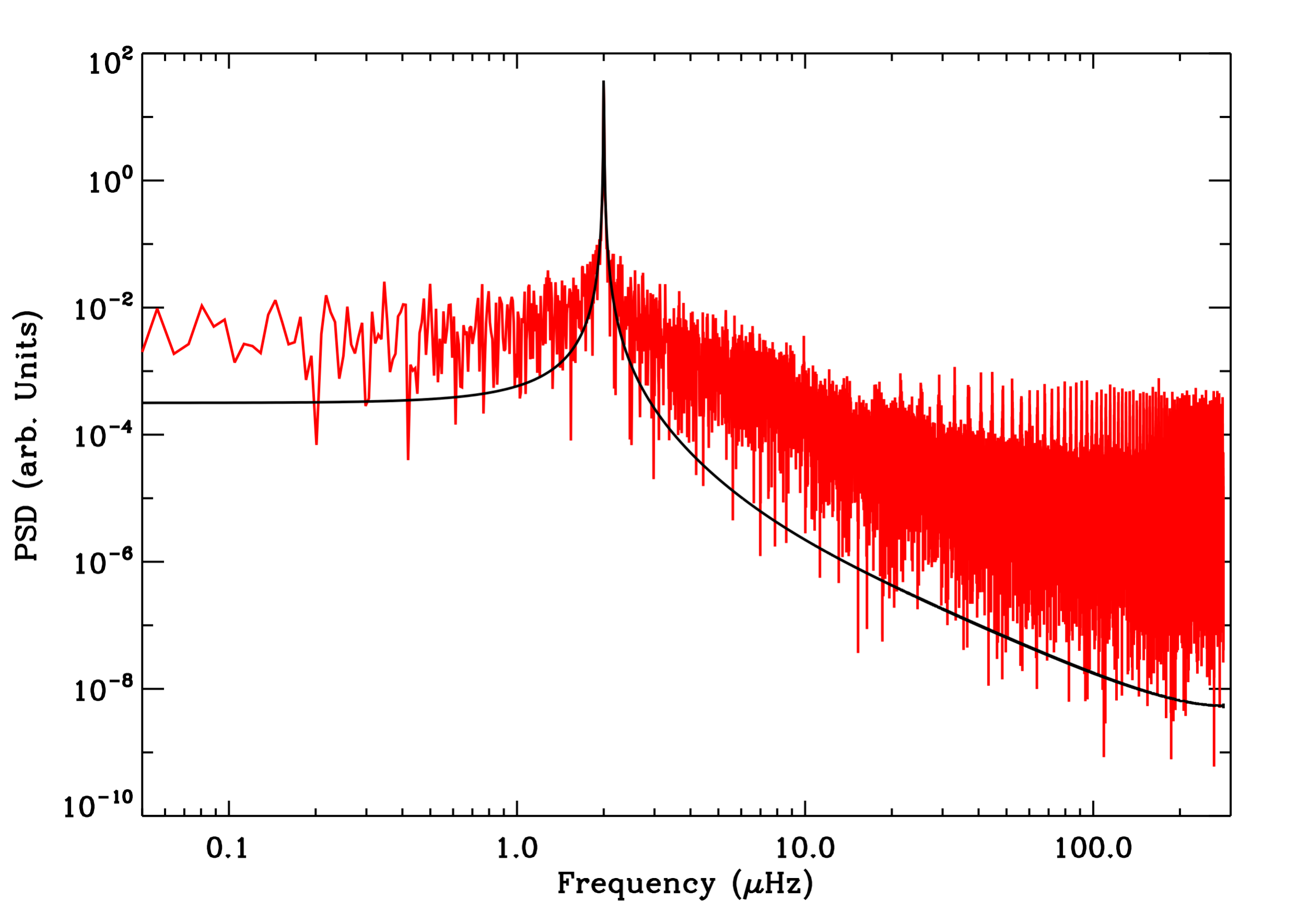}
\end{center}
\caption{\label{fig:gaps_kep} Power spectrum density on logarithmic scale of a 2 $\mu$Hz simulated sine wave (black) and after multiplying by the \emph{Kepler} window function (red).}
\end{figure}

Regular short (typically one or two points) or longer gaps (up to a day or so) can be found in space missions such as CoRoT or \emph{Kepler}. For example, the normal operations of this latter spacecraft involves the angular momentum dump of the reaction wheels every 3 days producing typical regular gaps of one long-cadence (29.42 min) or several short-cadence (58.85 s) measurements \citep[][]{Christiansen2013_data_handbook}. Moreover, every month the satellite stops the scientific observations program to point towards the Earth and downlink all the data recorded on board. This interruption has a typical size of about a day \citep[see for a detailed explanation of the \emph{Kepler} gaps][]{2014A&A...568A..10G}. In both cases the effects, if they are not corrected, are the addition of harmonics of the stellar signals at all frequencies multiple of $\sim$~1/3 and $\sim$~1/30 days respectively. An example of the \emph{kepler} window function over a perfect 2 $\mu$Hz sinusoid is shown in Fig.~\ref{fig:gaps_kep}. The power of the wave leaks at higher frequencies at multiples of the inverse of the gap's frequencies \citep{2014A&A...568A..10G}.

To solve the problem imposed by the regular gaps, several techniques have been used in asteroseismology. One widely used algorithm to study classical pulsators --where the modes are highly coherent-- is CLEAN \citep{1987AJ.....93..968R,1995AJ....109.1889F}. It is based on an iterative procedure that searches for consecutive maxima in the power spectrum. CLEAN starts by finding the highest peak in the periodogram, removing it in the time domain, recomputing the amplitude spectrum, and iterating for the next highest peak until a given amplitude threshold is reached. Some of the main caveats of the algorithm are that sometimes false peaks can be removed as being part of the signal, and  any error on the properties of the retrieved peaks  will introduce significant errors into the resulting ÒcleanedÓ periodogram. Finally, the use of this algorithm with stochastically excited modes is more complicated.
 
Another approach consists of interpolate the data in the gaps. Since the pioneer's work on helioseismology, several methods have been proposed to interpolate these datasets \citep[e.g.][]{1982MNRAS.199...53F,1990ApJ...349..667B}. Although they work quite well, they require in general some a-priori knowledge of the signal to be treated. Although this worked well for the Sun because we know pretty well its main seismic properties, those algorithms are not well suited to treat thousands of unknown asteroseismic targets as it is the case on present and future space missions. Hence a simpler approach was first adopted by the CoRoT project consisting to perform linear interpolation  \citep{2009A&A...506..411A,2007astro.ph..3354S} on the main solar-like targets observed in the asteroseismic field \citep[e.g.][]{2008A&A...488..705A,2009arXiv0907.0608G}. However, in some cases a more refined interpolation algorithm was used in the analysis of these CoRoT data due to the limitations of the linear approach \citep[e.g.][]{2009A&A...506...33M,2011A&A...530A..97B}. 

\begin{figure}[!htb*]
\begin{center}
\includegraphics[width=.95\textwidth]{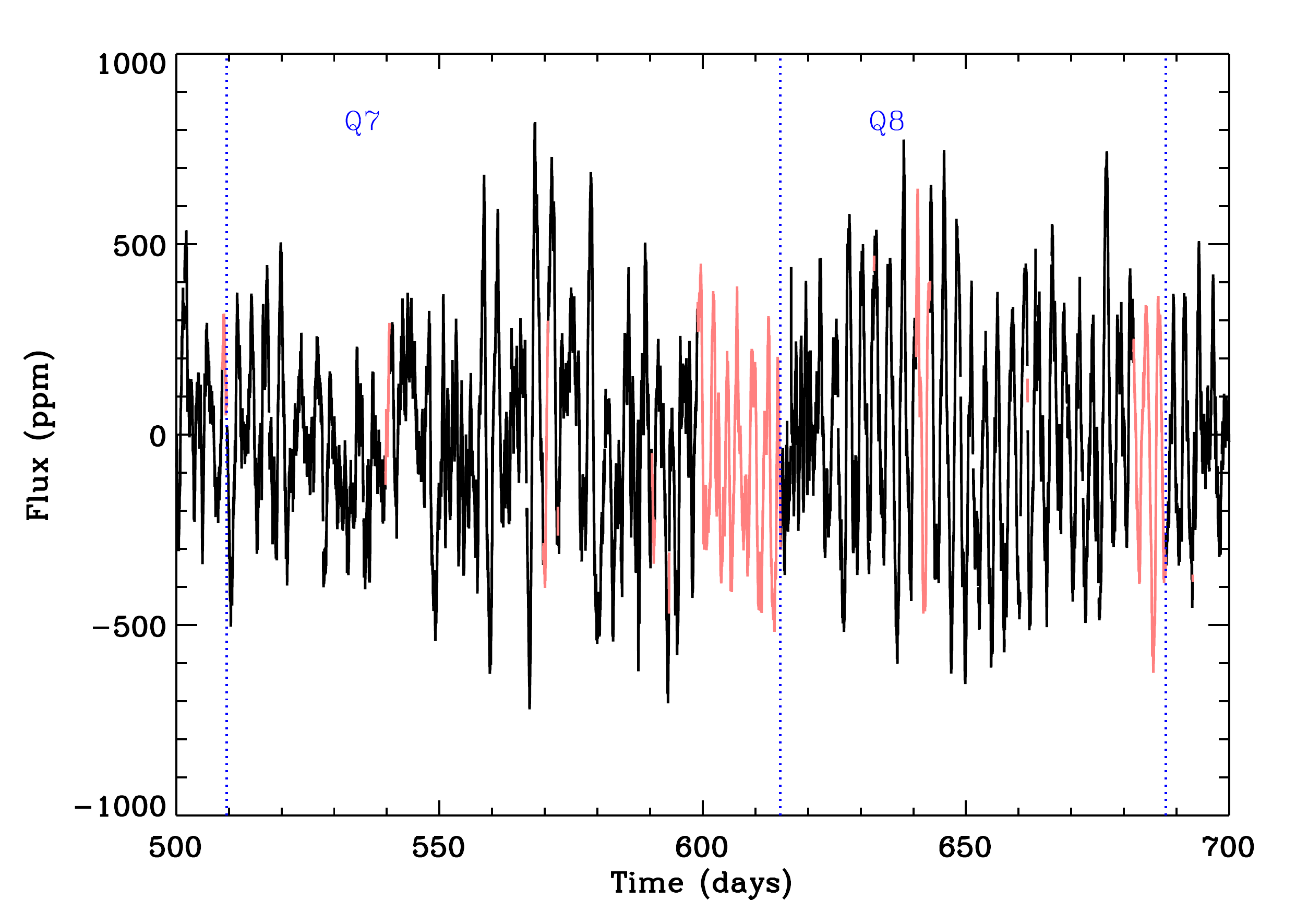}
\end{center}
\caption{\label{fig:gaps_SK} Light curve of the \emph{Kepler} F star KIC~3733735 taken during quarters 7 and 8 (black). The vertical blue dotted lines mark the separation between quarters. The pink segments of the light curves has been obtained by the inpainting interpolation.}
\end{figure}

Recently, an interpolation algorithm based on in-painted techniques \citep{2015A&A...574A..18P} has been applied to asteroseismic data from CoRoT \citep[e.g.][]{2010A&A...518A..53M,2015A&A...574A..18P} and  \emph{Kepler}, providing very good results to minimize the impact of the multiples of the orbital frequency at 161.7 $\mu$Hz and the gaps due to the perturbed data collected during the crossing of the South Atlantic Anomaly. An example of the application of these algorithm to the \emph{Kepler} active F star KIC~3733735 \citep{2014A&A...562A.124M} is shown in Fig.~\ref{fig:gaps_SK}. Indeed, the final procedure to correct the CoRoT data will propose this interpolation. 

Inpaint methods \citep{Elad05} are based on a prior of sparsity. In other words, the sparsity concept assumes that there is a representation of the signal in which most of the coefficients are close to zero. In the case of a single sine wave for example, the sparsest representation would be the Fourier transform because most of the Fourier coefficients would be zero except one (hence sparse), which is sufficient to represent the sine wave in the frequency space. Therefore in asteroseismology, and to deal with the large variation of gap sizes (from 1 short-cadence data point to $\sim$16 days), the best representation is the Discrete Cosine Transform  \citep[see for more details][and references therein]{2015A&A...574A..18P}. 

\subsection{Precision and detectability of modes}
To detect a mode in a periodogram it is required that the peaks rise above the general noise by a given factor, which can be translated into a minimum detection probability of 90$\%$, or a more conservative value of 99$\%$ confidence level. In principle, the stellar background noise, which is the main source of noise in the frequency region where stochastically excited modes lie, cannot be reduced with a single set of observations. The situation is slightly better in helioseismology where stereoscopic observations of the Sun through two different viewing angles  \citep[e.g. STEREO mission,][]{2008SSRv..136....5K} allows to simultaneously observe two non-coherent backgrounds. The other important sources of noise are the instrumental and statistical noise. Let's see other methods to compute the periodogram that can reduce the noise and enhance the signal-to-noise ratio of the observations.

Before that, it is important to say a word on the frequency precision that can be obtained. \citet{1992ApJ...387..712L} deduced an expression for the frequency precision that can be obtained in a typical seismic observation:
\begin{equation}
\sigma_\nu = \sqrt{f(\beta)\frac{\Gamma}{4\pi T}} \;\;\; ,
\end{equation}
where $1/T$ is the frequency resolution, $\Gamma$ is the line width of the modes, and $\beta$ is the inverse of the SNR. $f(\beta)$ is given by the expression:
\begin{equation}
f(\beta)= \sqrt{1+\beta}\left[\sqrt{1+\beta} + \sqrt{\beta}\right]^3 \;\;\; .
\end{equation}
This expression, which is a generalization of the case without background \citep[$\beta$=0,][]{1990ASSL..159..253D}, is only accurate when the observation time is much longer than the mode lifetime ($T>>\Gamma^{-1}$). It says that the mode precision is proportional to the square root of the mode linewidth and is proportional to the square root of the frequency resolution \citep[for more details see,][]{2014aste.book..123A}.

The amplitude of stochastically excited modes in solar-like stars follows \citep{1995A&A...293...87K}:
\begin{equation}
\frac{A}{A_\odot}\approx \frac{L}{M}\left(\frac{T_\odot}{T_{\rm{eff}}} \right)^s \;\;\; ,
\end{equation} 
where the exponent $s$ is 0 for Doppler velocity measurements and $s=2$ for photometric observations.

Assuming a Lorentzian profile for the modes, the maximum mode height in the power spectrum at $\nu_{\rm{max}}$ is directly related to the maximum mode amplitude and the mode lifetime as follows  \citep{2009A&A...500L..21C}:
\begin{equation}
H=\frac{2A^2}{\pi \Gamma} \;\;\; .
\end{equation}
Therefore, we can derive a relation between the height, $H$ and the frequency of maximum power, $\nu_{\rm{max}}$:
\begin{equation}
H=\frac{2A_\odot^2}{\pi \Gamma_{\rm{max}}} \left(\frac{T_{\rm{eff}}}{T_\odot} \right)^{7-2s}\left(\frac{\nu_\odot}{\nu_{\rm{max}}} \right)^2 \;\;\; .
\end{equation}
It is important to note that the CoRoT and \emph{Kepler} observations seem to show that there is a relation between the line width of the modes and the effective temperature of the star \citep{2011A&A...529A..84B,2012A&A...537A.134A,2015A&A...579A..83C}.

\begin{figure}
\begin{center}
\includegraphics[width=.9\textwidth]{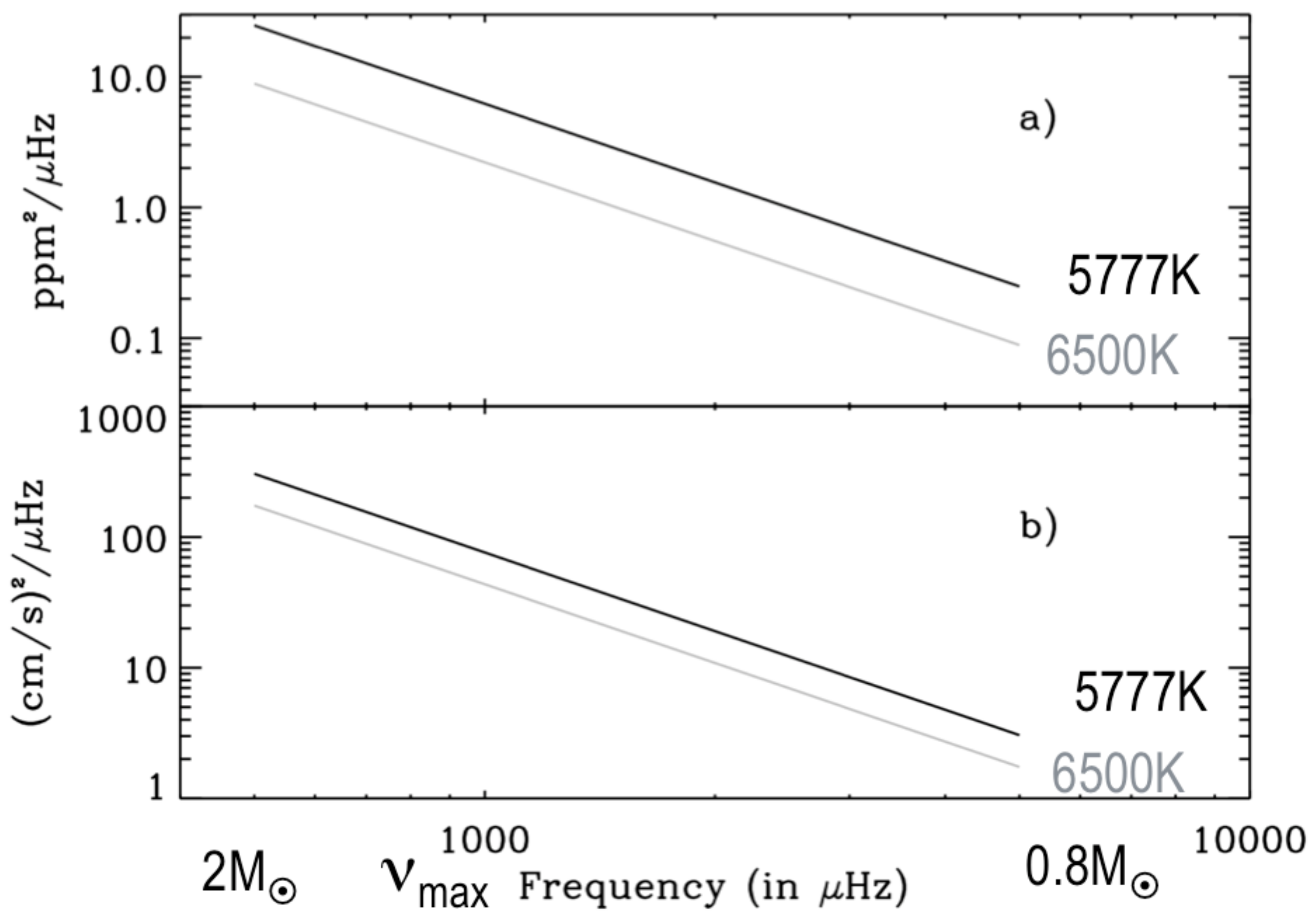}
\end{center}
\caption{\label{fig:width_teff}Background noise as a function of $\nu_{\rm{max}}$ required to detect acoustic modes with a SNR of 10 for Doppler intensity (top) and velocity (bottom) observations and for two main-sequence stars, a cooler one (black) with $T_{\rm{eff}}=5777$K (mode line widths of 1 $\mu$Hz and a hotter F star (grey) with $T_{\rm{eff}}=6500$K and a mode line width of 4 $\mu$Hz. Adapted from \citep{AppTongGar}.}
\end{figure}

It is represented in Fig.~\ref{fig:width_teff} the background noise as a function of $\nu_{\rm{max}}$ required to detect acoustic modes with a SNR of 10 for Doppler velocity and intensity observations and for two main-sequence stars, a cooler one with $T_{\rm{eff}}=5777$K (mode line widths of 1 $\mu$Hz) and a hot F star with $T_{\rm{eff}}=6500$K and a mode line width of 4 $\mu$Hz \citep[see for more details the discussion in][]{AppTongGar}.

\subsection{Other Periodogram estimators}
In the previous sections we have described the two most common ways to compute a periodogram in asteroseismology: the FFT when we have to deal with evenly distributed time series, and the Lomb-Scargle periodogram for irregularly sampled time series. In the rest of the section other ways to compute the periodogram will be described. These methods (or variations of the previous ones) have the advantage that they can increase the SNR of the signals in some particular circumstances. Indeed, the Fourier spectrum is very well adapted to periodic functions.

\subsubsection{Average Power Spectrum}
Sometimes, it can be useful to split the observations into several smaller chunks of data and average the independent spectra, instead of doing one single periodogram corresponding to the full time series. Thus, the average Power Spectrum (AvPS) of N independent time series can be calculated as:
\begin{equation}
AvPS = \sum_{i=1}^{N} \mid F_i(\nu)\mid^2 \;\;\; .
\end{equation}
The AvPS has the advantage of reducing the variance of the incoherent noise and improving the statistics.  This periodogram can also be applied when two observations are separated by a long gap. This was the case of the CoRoT observations of HD~49933 in which the two first observing runs were separated by $\sim$~1 year \citep{2009A&A...507L..13B}. In such case, it is more convenient to compute the average of the two independent observations than computing the full periodogram with a 1 year gap in between. It is important to note that the AvPS has a reduced frequency resolution (corresponding to the size of the individual chunks of data) and thus, it can only be used when the resolution is enough for the problem we want to study. 

It can be demonstrated \citep{2003A&A...412..903A} that the statistics of the AvPS of independent time series follows a $\chi^2$ with $2N$ degrees of freedom. As a practical recipe, the AvPS can be fitted with a standard maximum likelihood estimator and the error bars can be normalized by $\sqrt{N}$ \citep{2003A&A...412..903A}.

\subsubsection{Multitaper spectral analysis}
Multitaper Spectral Analysis (MTSA) methods consist of multiplying a single light curve by a series of functions (windows) called tapers. MTSA methods are an extension of single-taper spectral analysis where the time series are multiplied or apodized by a single window function such as a Hanning window. This taper gives less weight to the ends of the time series than to the center, reducing the effect of the squared window (the Sinc function in Eq.~\ref{finite_obs}) in the periodogram \citep{1982IEEEP..70.1055T}. The multitaper approach uses a variety of orthogonal tapers, some of which give more weight to the ends of the time series and others to the center, with a good compromise between any biases and the reduction of the global variance \citep{1993sapa.book.....P}. 

In practice MTSA analysis involves the calculation of the windowed functions $f_k(t)=f(t)h_k(t)$, where $f(t)$ is the time series and $h_k(t)$ represents the $k$ taper. Thus the Multi Taper (MT) spectrum is the average of the power spectrum of the individual windowed functions:
\begin{equation}
MT\; {\rm{Spectrum}}=\sum_{k=1}^N{\overline{\mid f_k(t)}\mid^2}	\;\;\; .
\end{equation}
The statistics of the MT spectrum built in such ways are a $\chi^2$ with $2N$ degrees of freedom  \citep{1982IEEEP..70.1055T}. It can also be demonstrated that the variance of the MT spectrum is reduced by a factor $\sim 1/N^3$ \citep{1999ApJ...519..407K}.

\begin{figure}[!htb]
\begin{center}
\includegraphics[width=1.03\textwidth]{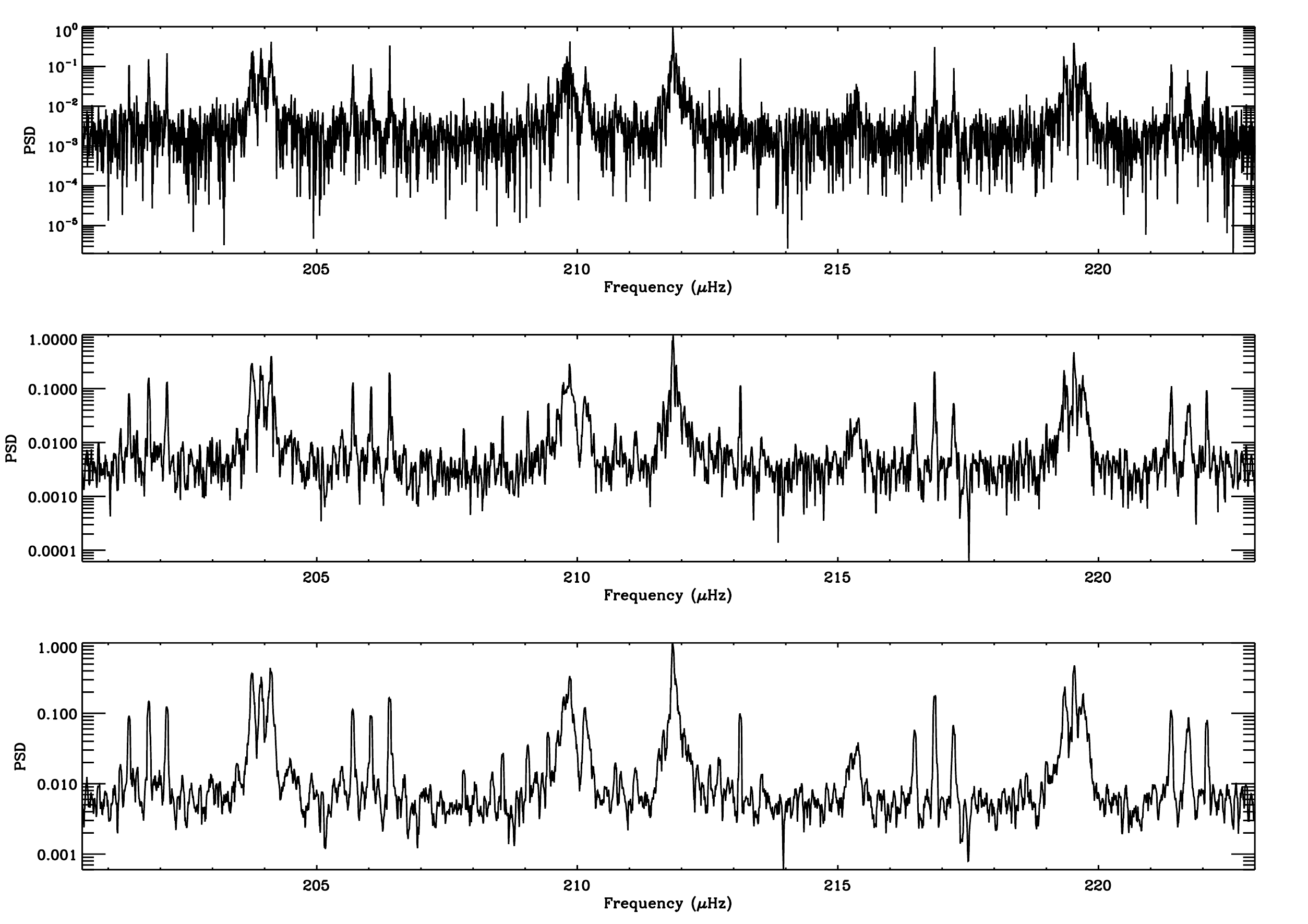}
\end{center}
\caption{\label{fig:MT} Comparison between the Lomb-Sargle periodogram (top) represented in log scale and a MT spectrum computed with 3 (middle) and 6 (bottom) sinusoidal tapers for the \emph{Kepler} target KIC~5356201.}
\end{figure}

Several functions can be used as tapers \citep[e.g. Slepian tapers,][]{1978ATTTJ..57.1371S}, but they are difficult to calculate. Indeed a simpler approach is to use as tapers sinusoidal functions in which the first taper is similar to a Hanning window \citep{1999ApJ...519..407K}. In Fig.~\ref{fig:MT} we compare a Lomb-Scargle spectrum with a MT spectrum computed with 3 (middle) and 6 (bottom) sinusoidal tapers. The higher the order of the taper, the smaller the variance of the spectrum. However, the MT spectrum tends to enlarge the width of the modes. Therefore the number of tapers that can be used would be limited by the lifetimes of the modes we want to measure.

\subsubsection{Average Cross Spectrum: Temporal and Spatial}
The observation of the same physical phenomena with two independent measurements provide a reduction of noise by a factor $\sqrt{2}$ in amplitude (2 in power). Using cross-correlation techniques we can define the averaged cross-spectrum (AvCS) as the average of the complex product of the Fourier transform of one data set, $A(\nu)$, by the complex conjugate of the Fourier transform of another, $B(\nu)$ \citep{1994MNRAS.269..529E,GarPal1998}:
\begin{equation}
AvCS =\frac{1}{N} \sum_{k=1}^N{A_k(\nu)B_k^*(\nu)}
\end{equation}
The significance of the AvCS can be computed from the coherency:
\begin{equation}
\rm{Coherency}(\nu) = \frac{\sum_{k=1}^N{A_k(\nu)B_k^*(\nu)}}{\sqrt{\sum_{k=1}^N{A_k(\nu)A_k^*(\nu)}\sum_{k=1}^N{B_k(\nu)B_k^*(\nu)}}}
\end{equation}

\begin{figure}[!htb]
\begin{center}
\includegraphics[width=.95\textwidth]{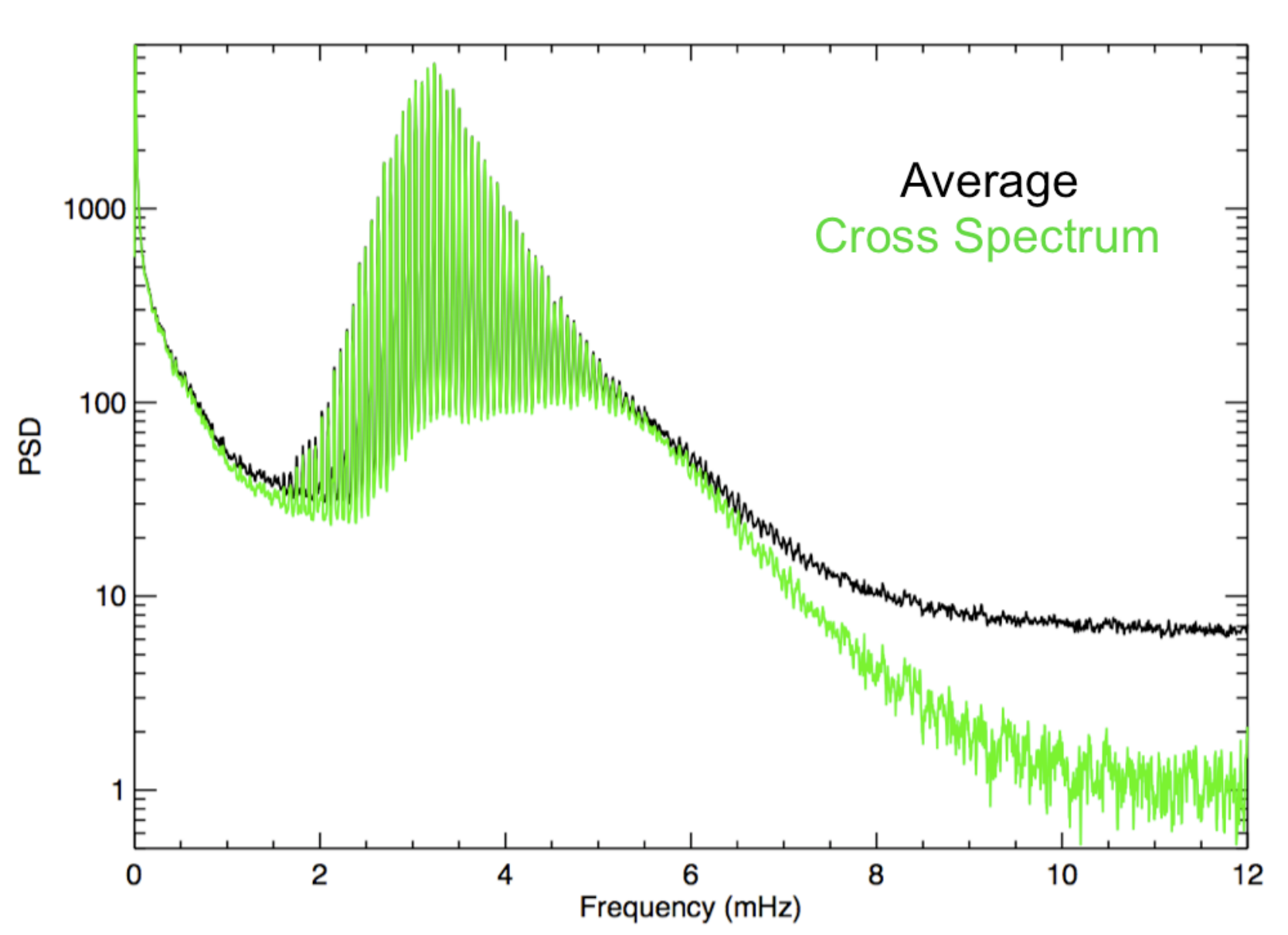}
\end{center}
\caption{\label{fig:AvCS} Comparison between the AvCS (in green) and the standard average (black) of 157 subseries of 4 days smoothed by a 5 points boxcar function.}
\end{figure}

\citet{2007A&A...463.1211A} demonstrated that the mean of the AvCS tends to zero for independent series (instead of to a value of $2\sigma^2$ for standard averaged power spectra) while the sigma remains the same compared to the standard averaging of the power spectrum of independent series. Therefore, the average level in the AvCS is reduced compared to the mean spectrum while the dispersion stays the same. This implies that the SNR of resolved peaks will increase. This methodology was successfully applied to 157 four-day time series of the two independent channels of the GOLF instrument \citep[properly calibrated in velocity following][]{GarSTC2005} in order to reduce the photon noise at high frequency (see Fig.~\ref{fig:AvCS}). This allowed to uncover the existence of high-frequency peaks in Sun-as-a-star observations \citep{GarPal1998}.

Unfortunately, we do not always have two simultaneous and independent measurements of the same physical phenomenon. Alternatively, when we are not interested in the high-frequency part of the spectrum (above $\nu_{\rm{Nyq}}/2$), \citet{GarJef1999}
demonstrated that we can built two independent time series by splitting the original time series in two, one containing the odd measurements and the other, the even points. The Interleave-shift Cross Spectrum (ISCS) is then the AvCS where the two independent series are the two that were just built. Because there is a small time delay between the two time series, it was proposed to shift the phases of the second channel by the sampling time. By construction, the new time series have a sampling rate doubled, which implies that $\nu_{\rm{Nyq}}$ of the ISCS is reduced to half. 

The same procedure could be applied in the space instead of the temporal domain to imaged helioseismic instruments such as SoHO/MDI or SDO/HMI in order to reduce the convective background \citep[for more details on this procedure see][]{2009ASPC..416..297G}. In this case, the granulation noise has a small correlation from one pixel to the next and the overall background level is reduced.

\section{Preparing the time series}

Any seismic analysis starts by collecting time series of a given phenomenon, e.g., the photometric variation of the flux of a star or the Doppler velocity displacement of the spectral lines formed in the photosphere of the Sun or other stars. Unfortunatelly, in many cases, the time series obtained from the observations are not directly exploitable and some preparation is needed. This involves correction from any known instrumental drifts and perturbations occurred during the observations, the inter-calibration of the observations taken by different instruments (for example when a network of ground-based telescopes is used), etc. Due to the nature of the seismic analyses, special care is always required in the handling of the timing of every measurement. All the subsequent analyses done from the time series rely on an accurate timing of the data points. 

Although the calibration of any time series is completely dependent on the instrument(s) that collected the data and the scientific objective of the analysis, there are several common steps that we are going to summarize in the following section based on the calibration procedure of the \emph{Kepler} data for seismic analysis including the surface dynamics, which means keeping the stellar signal at low frequencies. A more detailed description can be found for example in  \citet{2011MNRAS.414L...6G}, \citet{ThompsonRel21}, and \citet{2014MNRAS.445.2698H}.

{\it Kepler} is located in a 372.5-day, Earth-trailing, heliocentric orbit. This requires to perform 90$^{\circ}$ rolls about its axis every 93 days to maintain the solar panels illuminated and the radiator, which cools the focal-plane arrays, pointed away from the Sun \citep{2010ApJ...713L.115H}. Data are consequently subdivided into quarters (denoted Qn or Qn.m, where n is the quarter number and m, the month), starting with the initial 10 days commissioning run (Q0), followed by a 34 days long first quarter (Q1) and subsequent three months quarters (Q2, Q3,...), up to the last observations during Q17. 

For each star, two types of observations are available: `The pixel-data files'' and the integrated light curves. In both cases they are corrected for some instrumental effects, although not all. The pixel-data files are CCD stamps (also known as ``imagettes in the CoRoT  community) centered at each star and covering all the pixels that contains signal from the star. They allow individual scientist to perform their own aperture photometry following their own requirements. The integrated light curves are part of the products provided by NASA from the Pre-Data-Conditioning (PDC)  allowing to search for exoplanet transits  \citep[][]{2010ApJ...713L..87J}.

While these PDC datasets, either the PDC-SAP (Simple Aperture Photometry) or the PDC-msMAP (multi scale Maximum A Posteriori methods) are in constant evolution and new and more refined procedures are established, part of the low-frequency stellar signal (such as the one produced by starspots or low-frequency modes) could be filtered. Therefore, for solar-like oscillating stars as well as some classical pulsators ($\delta$-Scuti and $\Gamma$-Doradus stars), it is better to take the pixel-data files and develop a specific set of corrections that takes into account the particularities of the oscillating signal that we are interested in.

\begin{figure}
\includegraphics[trim =5mm 90mm 1mm 16mm, clip, height=0.3\textheight, width = 1\textwidth]{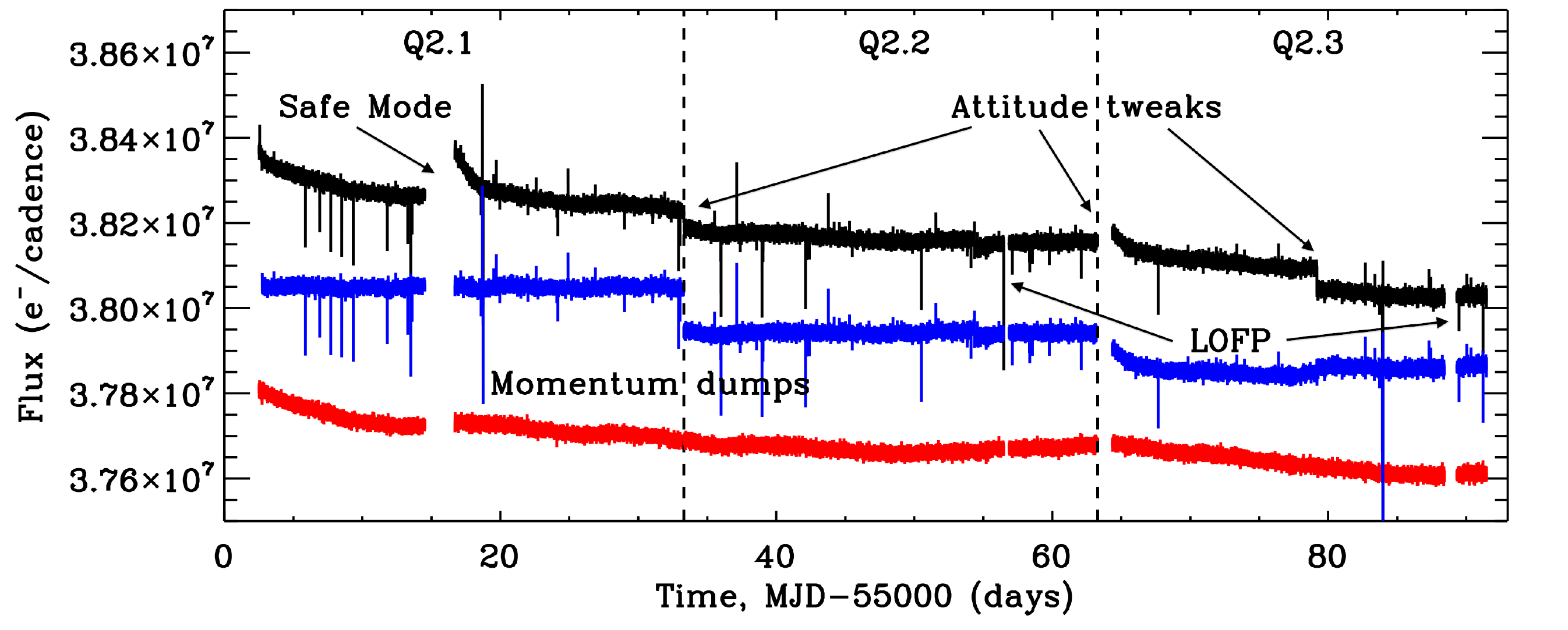}
\caption{\label{Q2} {Raw (black), PDC corrected (blue) and corrected --using the procedure described in this paper (red)-- light curves of the solar-like target: KIC~11395018 (Mathur et al. in preparation). The corrected light curve has been shifted down, by $ 4\; 10^5 \;\rm{e^-/cadence}$, for the clarity of the comparison. The origin of the time axis is in Modified Julian dates (MJD) - 55000. The points in which the flux fall down are most of them due to momentum-dump operations. LOFP stands for: Loss Of Fine Pointing.} Image from \citet{2011MNRAS.414L...6G}.}
\end{figure}

In general, light curves should be corrected for three types of instrumental effects (see Fig~\ref{Q2}): outliers, jumps, and drifts. 
Outliers are those measurements showing a too large point-to-point deviation. Following the procedures described by \citet{2011MNRAS.414L...6G} that are based on the ones developed for GOLF/SoHO \citep{GarSTC2005}. The deviation  is computed on the backward difference function of the light curve \citep{2008A&A...477..611G} with a threshold greater than 3$\sigma$, where $\sigma$ is defined as the standard deviation of the backward difference of the time series. This correction removes $\sim$ 1\% of the data points. 

Jumps are defined as sudden changes in the mean value of the light curve due, for example, to attitude adjustments or because of a sudden pixel sensitivity drop. Finally, drifts are small low-frequency perturbations, which are in general due to temperature changes (after, for example, a long safe mode event) that last for a few days and are corrected using polynomial fits.

Once these corrections are applied, we build a single time series after equalizing the average counting-rate level between all the quarters (red curve in Fig.~\ref{Q2}). A change of the average counting rate can also happen inside a roll when the aperture mask is changed. To do this equalization and to convert into parts per million (ppm) units, we use a low-pass filter of the data. The details on the filter and its cutoff frequency depends on every calibration procedure. 

The light curves from CoRoT or \emph{Kepler} suffer from some discontinuities. As seen in Sect.~\ref{Sec:gaps} those gaps are usually interpolated.

\section{The observed stellar power spectrum}
\label{sec:freqSep}
The natural way to study stellar oscillations is by analyzing the Fourier components of the signal. In the analysis of solar like stars, it is common to ignore the phase information and work with the power spectrum, or the power spectrum density (PSD). 

\subsection{Generalities}
\begin{figure}[!htb]
\begin{center}
\includegraphics[width=.95\textwidth]{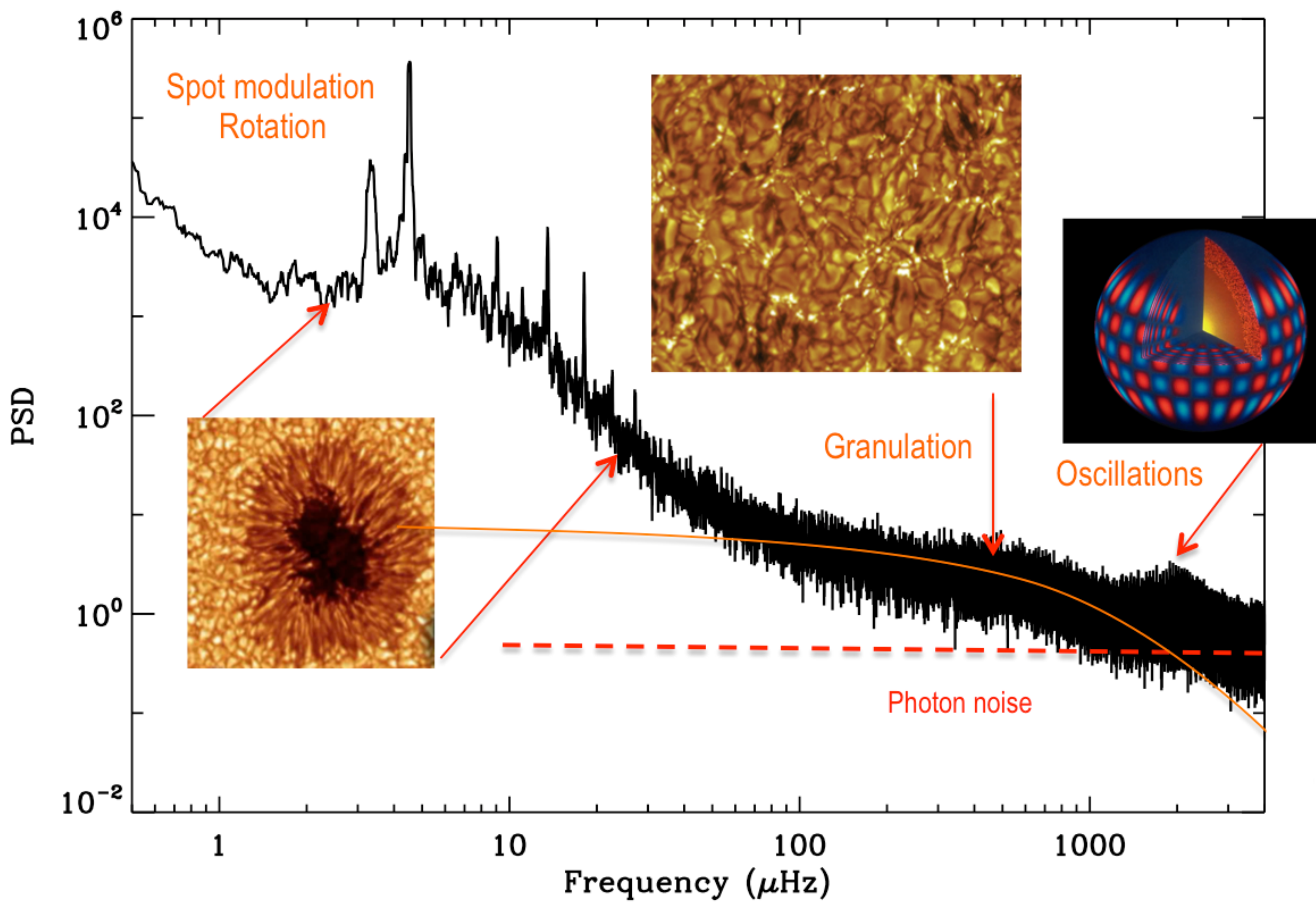}
\end{center}
\caption{\label{fig:PSD} PSD of the \emph{Kepler} target KIC~3733735 observed in short cadence during three years. Physical phenomena associated with each region of the PSD are indicated: photon noise, oscillations, convection (granulation), and rotation through the spot modulation of the emited stellar flux. }
\end{figure}

The PSD of the \emph{Kepler} target KIC~3733735 is shown in Fig.~\ref{fig:PSD}. This star is a typical hot F main-sequence solar-like star. Depending on the frequency range, the spectrum is dominated by the features related with a different physical phenomena.

Starting by the low-frequencies (between 1 and 10 $\mu$Hz), the spectrum is dominated by a series of high-amplitude peaks and their harmonics. This is the signature of the surface differential rotation of the star through the modulation induced by the stellar spots crossing the visible stellar disk. The surface differential rotation of this star exhibits two active bands, one spinning at $\sim$~3 days and another one at $\sim$~2.5 days \citep{2014A&A...562A.124M,2014A&A...572A..34G}. At higher frequencies, between 50 and 1000 $\mu$Hz, the spectrum is dominated by  convection (granulation). At even higher frequencies, it is possible to distinguish the bump of the acoustic modes centered at around 2000 $\mu$Hz. Finally, close to the Nyquist frequency, the spectrum is flat and it is dominated by the photon noise of the instrument.

\subsection{Global oscillation parameters: $\Delta\nu$ and $\nu_{\rm{max}}$}
\begin{figure}[!htb]
\begin{center}
\includegraphics[width=.95\textwidth]{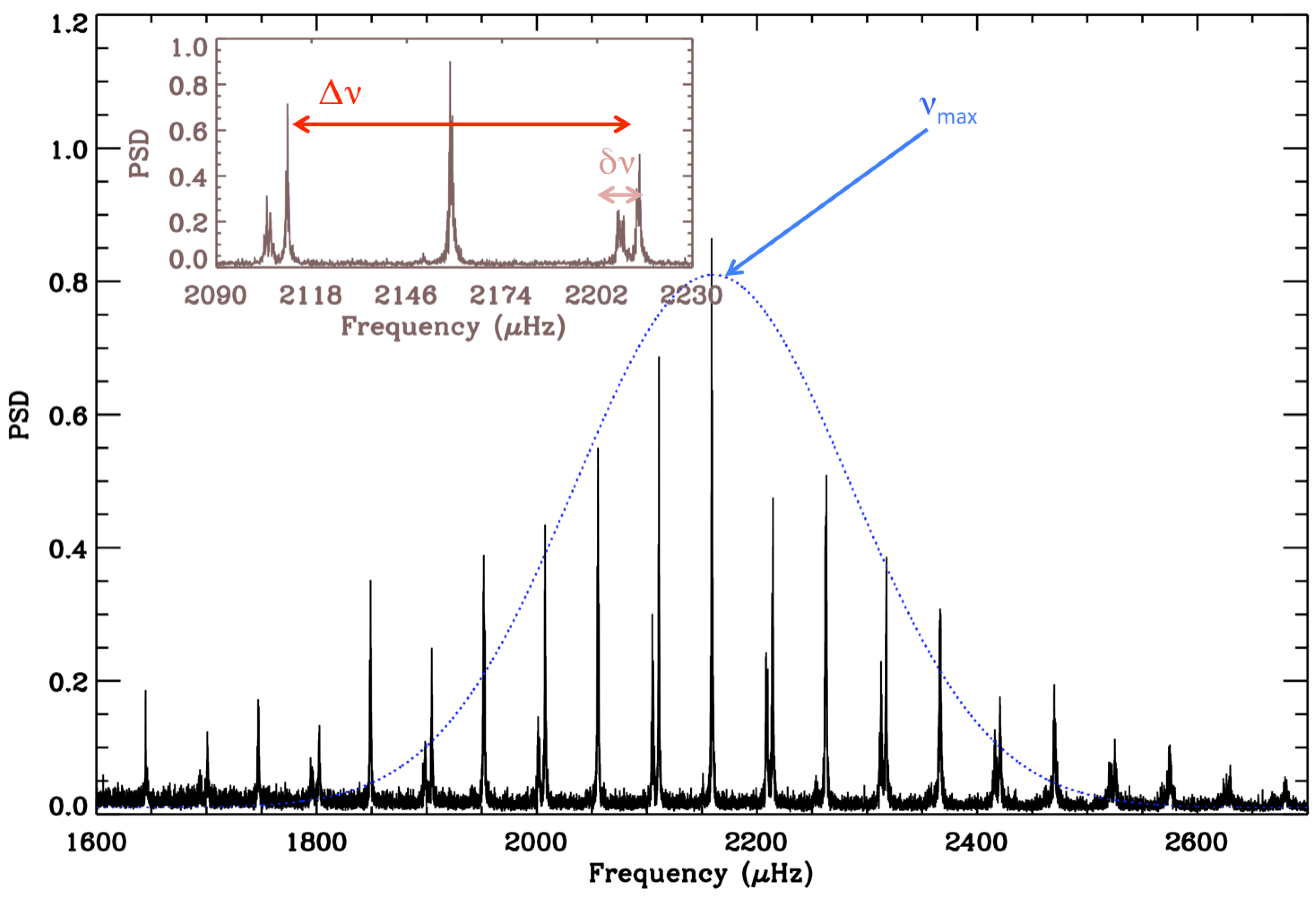}
\end{center}
\caption{\label{Separation} PSD (in arbitrary units) of the \emph{Kepler} target 16~Cyg A. The blue doted line represents the Gaussian fit to obtain the frequency at maximum power, $\nu_{\rm{max}}$. The inset is a zoom showing the large frequency separation, $\Delta\nu$, between two modes $\ell$=0 and the small separation $\delta \nu_{0,2}$ }
\end{figure}

As seen in Fig.~\ref{fig:PSD}, when the SNR is enough \citep[e.g.][]{2011ApJ...732...54C}, the power spectrum of any solar-like star shows a power bump in which a repetitive structure or pattern of modes is visible (see Fig.~\ref{Separation} for the \emph{Kepler} star 16~Cyg A). The power bump allows the definition of the frequency of maximum power, $\nu_{\rm{max}}$, which is related with the acoustic cutoff frequency \citep{1991ApJ...368..599B,1995A&A...293...87K,2011A&A...530A.142B,2013ASPC..479...61B}.  To measure $\nu_{\rm{max}}$, it is common practice to fit a Gaussian function within the convective background fit (see for more details Sect.~\ref{sub_back}).

Looking in more details to the p-mode bump, it is composed by a repetitive sequence of odd and even degree modes. Thus, two important seismic variables can be defined: the large- and the small-frequency separations or simply large and small separations.

The large separation of low-degree p modes is given by (see also Fig.~\ref{Separation}):
\begin{equation}
\Delta \nu_{\ell} (n)= \nu_{n,\ell}-\nu_{n-1, \ell} \;\;\; .
\end{equation}

This large separation depends inversely on the sound-travel time between the center and the surface of the star \citep[see e.g.][]{JCD2002} which means, it is proportional to the square root of the mean density in the cavity in which the modes propagate:
\begin{equation}
\Delta \nu_{\ell} (n)=\left[2 \int_0^R \frac{dr}{c_s}\right]^{-1}  \;\;\;,
\end{equation}
where R is the stellar radius, and $c_s$ is the sound speed. In the case of the Sun, the mean large separation has a value of $\sim$135 $\mu$Hz.

The small separation of  low-degree p modes is given by (see also Fig.~\ref{Separation}):

\begin{equation}
\delta \nu_{\ell, \ell+2} (n)= \nu_{n,\ell}-\nu_{n-1, \ell+2} \;\;\; .
\end{equation}

This difference is mainly dominated by the sound-speed gradient near the core and, therefore, it is sensitive to the chemical composition in the central regions.  Indeed, the small separation is the difference of two modes with nearly identical eigenfunctions in the surface (similar outer turning points) and being only different in the deeper layers, with different inner turning points (see right panel in Fig.~\ref{Fig_prop_modes}). It is important to notice that we can also define another small separation between the radial and the dipole modes. In this case we define $\delta_{0,1}$ to be the amount by which the modes $\ell$=1 are offset from the midpoint of the modes $\ell$=0 on either side:

\begin{equation}
\delta \nu_{0,1} (n)=\frac{1}{2} ( \nu_{n,0}+\nu_{n+1,0})-\nu_{n,1} \;\;\; .
\end{equation}

Using the asymptotic theory it can be shown that \citep{1991soia.book..401C}:

\begin{equation}
\delta \nu_{\ell, \ell+2} (n) \simeq -(4\ell + 6) \frac{\Delta \nu_\ell (n)}{4 \pi^2 \nu_{n,\ell}}\int_0^R \frac{dc_s}{dr} \frac{dr}{r} \;\;\;.
\end{equation}

As the frequencies of both modes are very close, they have similar near-surface effects and the small separation is less  affected by such effects. However, some residuals can still remain and therefore, it has been demonstrated that the ratio of the small separation to the large separation, defined as $r_{0,2}  \equiv r_{0,2} (n) = \delta \nu_{0,2} (n) / \Delta \nu_\ell (n)$ , can exclude such effects to a great extent \citep[see for more details][]{2003A&A...411..215R}. 

Although the oscillation spectrum is not perfectly regular and they slightly vary with frequency, it is possible to use this regularity to look for the large period spacing. Either by computing the power spectrum of the power spectrum \citep[e.g.][]{2010MNRAS.402.2049H,2009CoAst.160...74H,2010A&A...511A..46M} or by computing the autocorrelation of the signal \citep[e.g.][]{2007MNRAS.379..801R,2009A&A...508..877M}, it is possible to determine the large (and the small) frequency separations in a global way.
Another way consists on fitting a few modes around $\nu_{\rm{max}}$ and extract in such way the large frequency spacings as well as other global parameters as the phase shift term $\epsilon$ \citep{2012A&A...541A..51K}. We will describe in more details the fitting techniques in the next section of this chapter.

It is important to notice that the global techniques does not normally extract the large spacing but half of it because the main periodicity retrieved is the distance between the odd and the even modes. However, there is a family of evolved stars called: ``depressed dipole mode stars'' \citep{2014A&A...563A..84G,2012A&A...537A..30M}, in which the $\ell=1$ modes have lower-than-expected amplitudes and thus the value retrieved by the automatic pipelines is directly the large separation.

\subsection{The \'echelle diagram}
\begin{figure}[!htb]
\begin{center}
\includegraphics[width=0.98\textwidth]{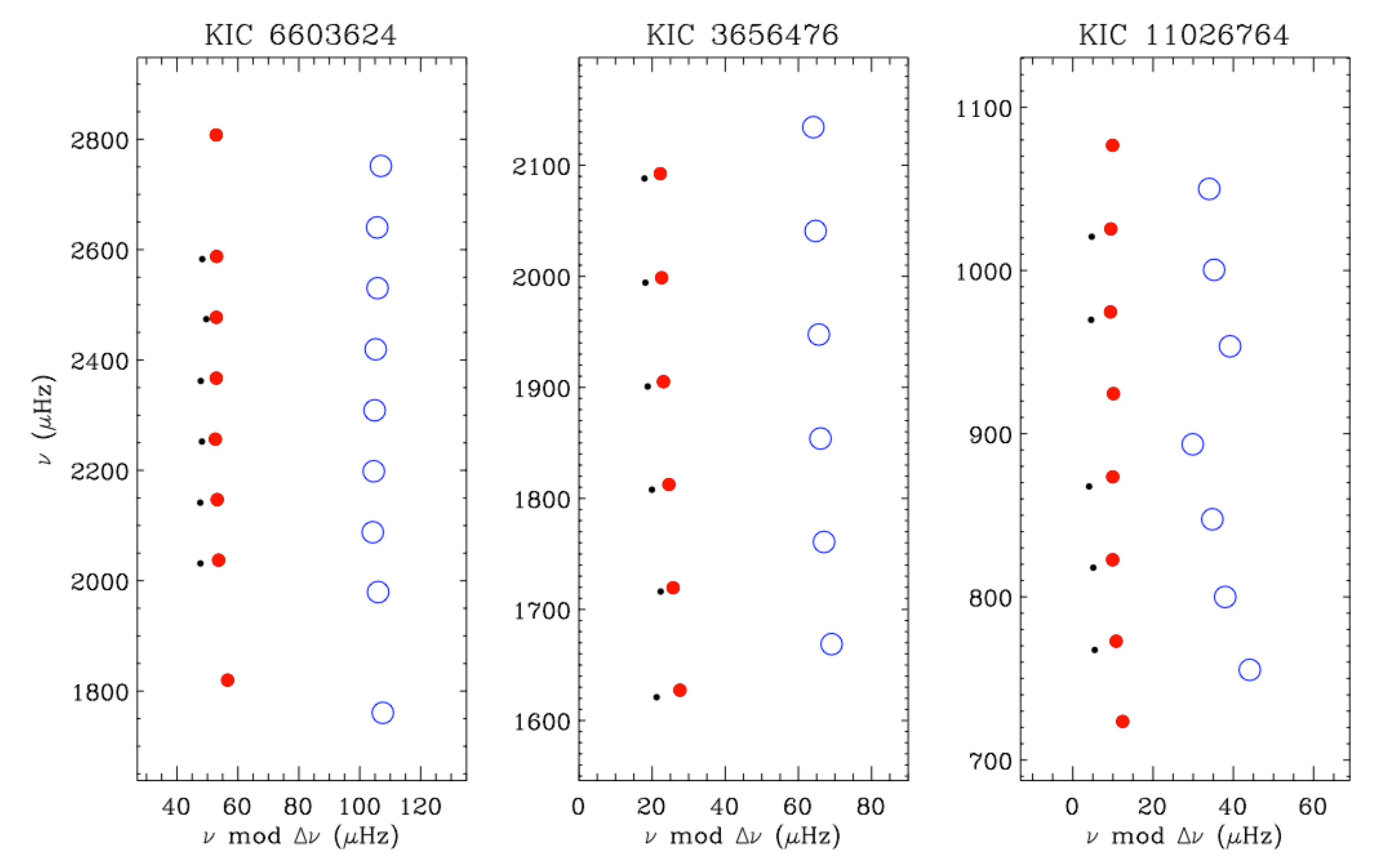}
\end{center}
\caption[\'Echelle diagram of 3 solar-like stars measured by {\it Kepler}]{\label{ED169392} \'Echelle diagrams of 3 solar-like stars observed by {\it Kepler}, showing the $\ell$ = 0 (filled red symbols), $\ell$ = 1 (open blue symbols), and $\ell$ = 2 (small black symbols) ridges. Extracted from \citet{2010ApJ...713L.169C}.
}
\end{figure}

The \'echelle diagram is a 2-D representation of the power spectrum in which the frequency is plotted as a function of the frequency modulo the large frequency separation. In other words, it is build by cutting the spectrum in segments of multiples of the large-frequency spacing and stacking them one in top of the next. In such way, modes that are equally spaced by this quantity are aligned forming vertical ridges. It was originally introduced in helioseismology by \citet{1980Natur.288..541G} to correctly identify the modes of the power spectrum of the Sun measured from the South Pole. Nowadays, this type of diagram is commonly used in asteroseismology to properly tag the degree and the order of the modes. Moreover, any departures from regularity is clearly visible as a curvature in the individual ridges of each mode degree. For example, variations in the small separations appear as a convergence or divergence of the corresponding ridges. It is very useful to identify bumped modes (those displaced from its original position due to the presence of a mixed mode in the vicinity) as well as the presence of mixed modes. In Fig.~\ref{ED169392} the \'echelle diagram of 3 solar-like stars observed by {\it Kepler} is shown. The ridges corresponding to the even modes are clearly shown on the left-hand side of the diagrams, while the ridge of the $\ell$=1 is visible onto the right. Because the amplitude of the $\ell$=3 modes is very small the odd ridge is composed of only one set of modes in most of the stellar observations. From left to right, stars are more and more evolved. Indeed the \'Echelle digram of KIC~11026764 (right-hand side panel) shows a bumped $\ell$=1 mode (a mode that have been moved from its original position by the presence of a mixed mode or by the mode immediately below) at $\sim$ 900 $\mu$Hz. This is a clear signature that this star is a sub-giant star. This kind of diagrams clearly help in the identification of the modes. 


\section{Characterizing the p-mode spectrum}

In this section we describe the methodology to characterize the acoustic modes over the entire p-mode range in both, Doppler velocity and luminosity observations. We will explain the main difference between the solar and the stellar case, as well as the difference when dealing with sub giants and giants in which the apparition of mixed modes complicates the analysis.  

To extract the properties of the modes it is necessary to fit a given model to the data, while providing statistical tools for {\it hypothesis testing} and take a decision on whether or not the model and/or the hypothesis is accepter or rejected. To reach this goal, there are two different approaches, the {\it frequentist} and the {\it Bayesian}. In the frequentists approach, the laws of physics are deterministic and future realizations of an event are conditioned by  past realizations. If the result of an experiment or event has occurred 20$\%$ of the times, a new realization of the same event will have this probability that the solution would be the same. In the Bayesian approach \citep{Bayes1763}, each realization will be conditioned by other considerations that could change the probability attributed to this particular realization, i.e., there is an a-posteriori evaluation of the chances of each possible result. 

In this course we will not go further in the discussion between this two statistical  approaches and we will only provide the general framework of how to model the stellar spectra and the general steps that are required to fit the spectrum. An example will be given following the frequentist approach. I recommend the reader the excellent review by \citet{2014aste.book..123A} to uncover the details of both statistical approaches.

 \subsection{Modeling the acoustic spectrum}

A useful analogy for the stellar acoustic modes is that of an ensemble of
harmonic oscillators, excited stochastically, and damped intrinsically
by turbulence in the outer layers of the convection zone \cite[e.g.][]{1977ApJ...212..243G,1988ApJ...326..462G}. Each resonant component can be represented in
the form:
 \begin{equation}
 \frac{d^2x(t)}{dt}+2\eta \frac{dx(t)}{dt}+(2\pi\nu_o)^2x(t)=f(t) 
 \end{equation}
where $x(t)$ is the displacement of the oscillator, $\eta$ its damping
rate, $\nu_o$ the frequency of the undamped oscillator and $f(t)$ the
random forcing function. The Fourier transform of the oscillator
equation gives a Lorentzian shape for the expected power spectrum of
the signal in the vicinity of the resonance, under the assumption that
$F(\nu)$ -- the power spectrum of $f(t)$ -- is a slowly-varying
function of $\nu$, and $\eta \ll \nu_{o}$.  The maximum height (power
density) of the resonant peak in the frequency domain is then given
by:
 \begin{equation}
 H= \frac{F(\nu)}{16\pi\nu{_o}^2\eta^2},
 \end{equation}
and its width at half-height by:
 \begin{equation}
 \Gamma=\frac{\eta}{\pi}.
 \end{equation}
Even though the solar low-$\ell$ mode peaks exhibit small amounts of
asymmetry---typically of the order of a few percent \cite[e.g.][]{1998ApJ...505L..51N,1998ApJ...506L.147T,1999MNRAS.308..424C,ThiBou2000}---the magnitude of
this is so small that for the moment we will assume that  a Lorentzian description is sufficiently
accurate for our purposes here. Indeed it is usually neglected in asteroseismic analyses. The integrated power\footnote{For
Doppler velocity observations, this will correspond to the integrated
velocity power; while for photometric observations this will provide a measure of the total
power in the intensity fluctuations associated with the mode.}, $P$,
of a single mode is then given by:
 \begin{equation}
 P = \frac{\pi}{2} H\Gamma.
 \end{equation}
The total energy in the mode, $E$, is taken to be the sum of 
the kinetic and potential energy and can be written as:
 \begin{equation}
 E ={\cal I} P,
 \end{equation}
where ${\cal I}$ is the corresponding mode inertia. Since we will ignore
possible changes in ${\cal I}$, variations in $P$ and $E$ are
identical. The rate at which energy is dissipated in the modes \cite[also
sometimes referred to as the acoustic noise generation rate, e.g.,][]{1999A&A...351..582H} can be derived by using the analogy of the
harmonic damped oscillator \cite[e.g.][]{2000MNRAS.313...32C}:
 \begin{equation}
 \frac{dE}{dt}=\dot{E}=2\pi H \Gamma^2.
 \end{equation}

From the above equations, we should expect: the linewidth
to provide a direct measure of the damping rate; the mode power (or
mode energy) to provide a measure of the balance between the
excitation and damping of the modes; and the energy supply rate to
provide a diagnostic of the forcing or excitation.

 \subsection{Extraction of mode parameters}
 \label{extr_pmodes}
 \subsubsection{Helioseismology}
 The close proximity of modes in the power spectrum of each time series demanded that the low-p modes be fitted in pairs (i.e., monopole modes with quadrupole modes, and dipole modes with octupole modes, see an example in Fig.~\ref{fit_mp_lf}) to avoid any bias in the extraction of the mode parameters due to their proximity in frequency. We modeled the power in each modal component of radial order $n$ and angular degree $\ell$ with azimuthal order $m$ with an asymmetric Lorentzian profile \citep{1998ApJ...505L..51N}, as:
 
\begin{figure}[!htb]
\center
 \includegraphics[width=1\textwidth]{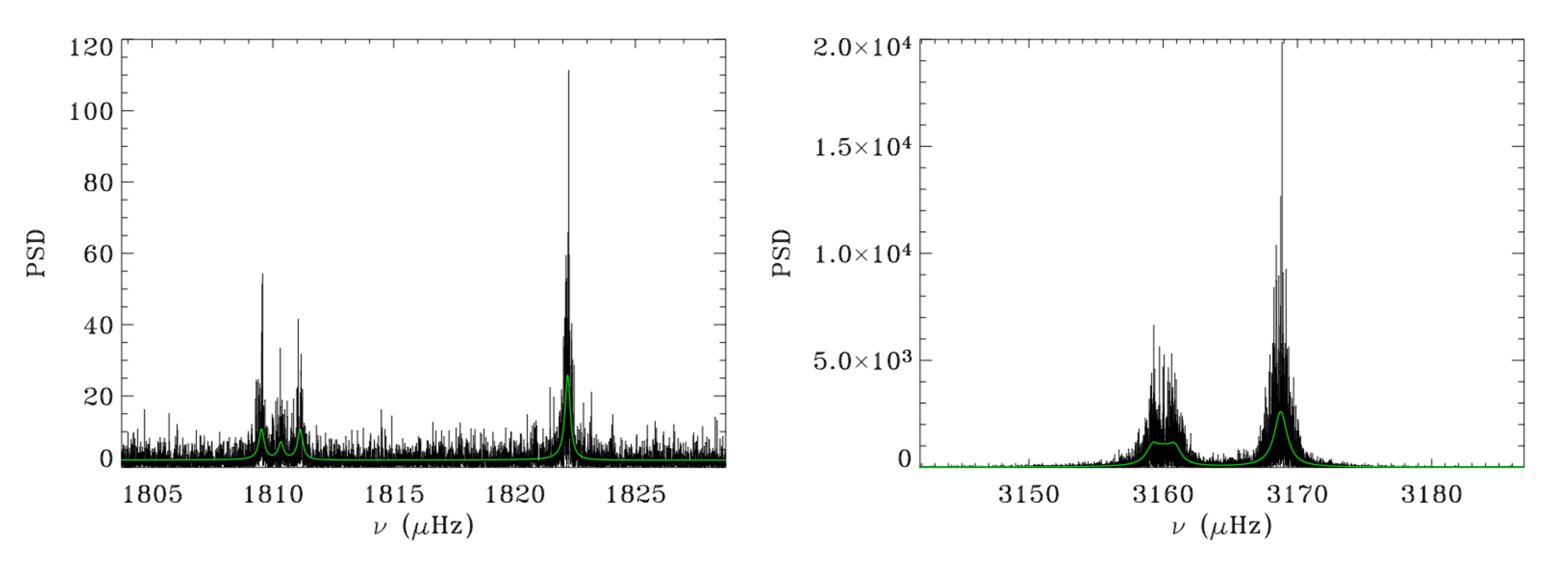}\\
 \includegraphics[width=1\textwidth]{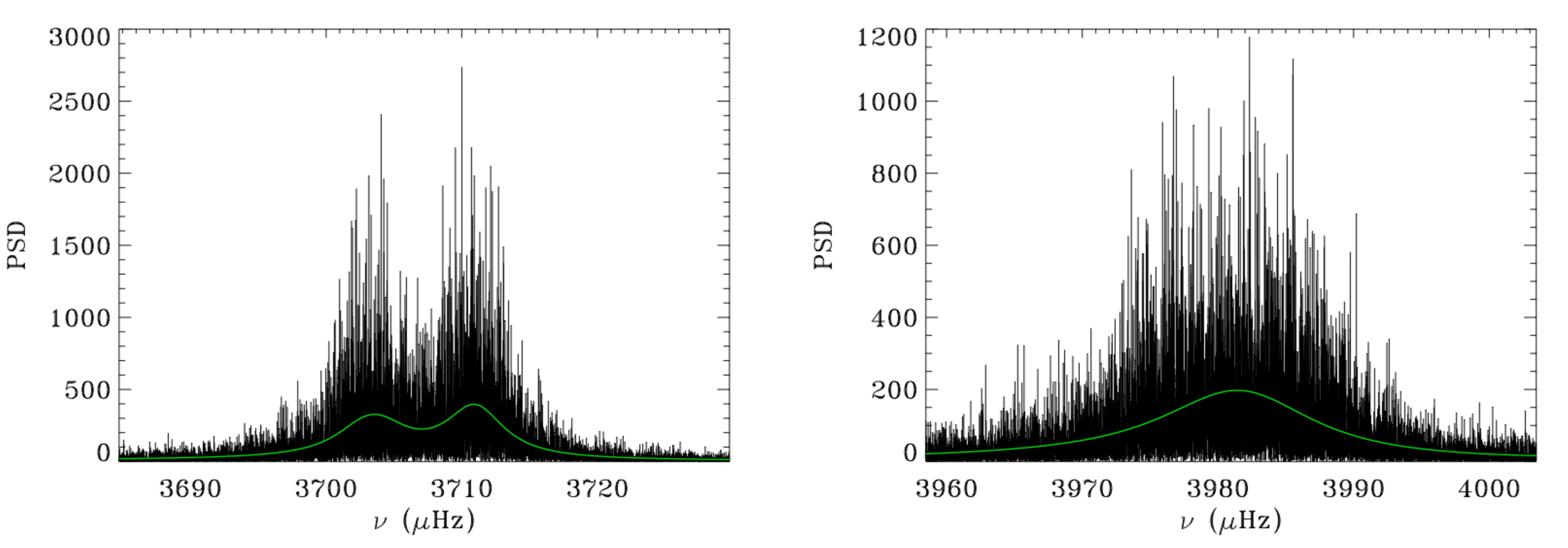}
  \caption[Examples of fits of low-order $\ell$=2, 0 and $\ell$=3, 1 modes at low and high frequency]{\label{fit_mp_lf} Example of fits (green line) $\ell$=2, 0 and $\ell$=3, 1 modes, left- and right-hand side panels respectively at low (top) and high (bottom) frequencies of a typical GOLF spectrum. In the first case, the lifetimes of the modes are longer and therefore the linewidths are smaller than for the modes at high frequency.}
\end{figure}

\begin{equation}
P(x)=H\frac{(1+bx)^2+b^2}{1+x^2}
 \label{eq:peak}
 \end{equation}
where $x=2(\nu-\nu_o)/\Gamma$. $H$ is the maximum power spectral density (often called the mode ``height''), and the parameter $b$ provides a measure of the fractional asymmetry of the peak. 
The mode-pair model, $M$, to be fitted is then described in full by:
 \begin{eqnarray}
 M(x,\vec{a})= \sum_{m=-\ell,\ell} \beta^{\ell}_{|m|} P(x)_{n,\ell} + 
 \nonumber \\
 \sum_{m=-\ell,\ell} \beta^{\ell+2}_{|m|} P(x)_{n-1,\ell+2} + N,
 \end{eqnarray}\\
where: $\vec{a}$ is the vector of parameters to be fitted;
$\beta^{\ell}_{|m|}$ are the $m$-component height ratios in each
multiplet; and $N$ is the uncorrelated background (assumed to be
constant across the frequency range defining the window of the fit).

Since the observed power is distributed about the limit spectrum with $\chi^2$ 2-d.o.f. statistics \cite[e.g.][]{AppGiz1998}, the probability (likelihood) function that must be maximized in the frequenctist approach (to give the model that makes
the data most likely) takes the form:
 \begin{equation} 
 \label{dpf} f(X,\vec{a})=
 \prod_{i=1}^n \frac{1}{M(x_{i},\vec{a})}
 \exp\left[-\frac{X(x_{i})}{M(x_{i},\vec{a})}\right]. 
 \end{equation}
One seeks to find the vector of model parameters $\vec{a}$ that
maximizes $f(X,\vec{a})$ across the $n$ frequency bins in the fitting
interval. In practice we used a modified Newton method (Press et
al. 1992) to minimize the negative logarithm of the likelihood
function.  The covariance matrix of the vector $\vec{a}$ is well
approximated by the inverse of the Hessian matrix. The uncertainties
on each fitted parameter are therefore taken as the square roots of
the diagonal elements of the inverted matrix.

We imposed the following constraints when fitting each mode pair in order to reduce the number of free parameters and stabilize the peak-fitting procedure:

 \begin{enumerate}

 \item All components within a given multiplet (i.e., for a given
 $\ell$) were assumed to have the same linewidth

 \item A single height -- that of the outer, sectoral components --
 was fitted for each mode. The relative $m$-component height ratios,
 $\beta^{\ell}_{|m|}$, were assumed to take fixed theoretical values as
 calculated, a priori, for each instrument \cite[see a complete discussion on this point in][]{2011A&A...528A..25S}

 \item The components of both multiplets in a pair were assumed to
 possess the same peak asymmetry.

 \item The natural logarithm of the height, width and background terms
 were varied---not the straightforward parameters themselves---in
 order to give a quasi-normal fitting distribution.

\item Prior to fit each pair of modes, we first compute the background parameters and we either divide the PSD by the fitted background model or we fix it in the fitting of the modes. See Section~\ref{sub_back} for further details.

 \end{enumerate}

It is important to notice, that sometimes, we let all the parameters free for each mode or just a combination of them.

We also found that the size of the fitting window had important implications for the extracted asymmetry parameter. This was largely a result of the influence of neighboring $\ell$ = 4 and 5 modes \cite[a long discussion on this bias can be found in][]{ChaApp2006}. While these higher $\ell$ values are much less prominent in the Sun- as-a-star data than their fitted $\ell \le$ 3 counterparts, they nevertheless appear at sufficient amplitude to (subtly) affect parameter extraction. This is because the fitting models do not usually account for the presence of ``leakage'' from $\ell$ = 4 and 5 into the fitting window (as we also omitted to do it here), although they can be measured in, for example, Sun-as-a-star observations using 12 years of VIRGO/SPM data \citep[e.g.][]{2014ApJ...782....2L}.

 \subsubsection{Asteroseismology}

In opposition to the ``local fitting scheme'' traditionally used in helioseismology in which narrow frequency bands are fitted, it has been proved that a global approach is better suited in the case of asteroseismology \citep[e.g.][]{2008A&A...488..705A,2012A&A...543A..54A,2009A&A...506...51B,2009A&A...507L..13B,2010A&A...515A..87D,2011ApJ...733...95M,2012ApJ...749..152M}. 

Although fitting low-degree p-mode profiles in helioseismology and asteroseismology might appear very similar, the unknown stellar inclination angle makes the fitting of asteroseismic data much more difficult (see for example \cite{2008A&A...488..705A} for the case of the star HD~49933, \cite{2014MNRAS.439.2025D} for the case of the Sun, and \cite{2003ApJ...589.1009G} for Monte-Carlo simulations with artificial p-mode profiles).
It is not only the lower signal-to-noise ratio of the p-mode asteroseismic signal that makes the fitting difficult,  
but also the high correlation between the inclination and the rotational splitting \citep{BalGar2006,2008A&A...486..867B}.
Because of that, the determination of these two parameters can be rather poor and will consequently
affect the determination of the other parameters (frequencies, widths, heights, etc). Therefore, instead of fitting each
multiplet or pair of modes individually -- as commonly done in helioseismology -- we chose to perform a global fitting of all the multiplets above a given amplitude threshold around the maximum of the p-mode hump, assuming that the rotational splitting is independent of the frequency \cite[see][for all the details]{2008A&A...488..705A}. This type of global method was pioneered by \cite{1999ESASP.448..135R} using solar data. By doing so, the splitting and the inclination angle are better constrained, even though the stars are then modeled as a rigidly rotating star. This condition can then be relaxed to allow fitting individual splittings for each mode while fixing the inclination angle \cite[e.g.][]{2012Natur.481...55B,2012ApJ...756...19D,2014A&A...564A..27D}.

\begin{figure}[!htb]
  \centering
   \includegraphics[width=0.9\textwidth]{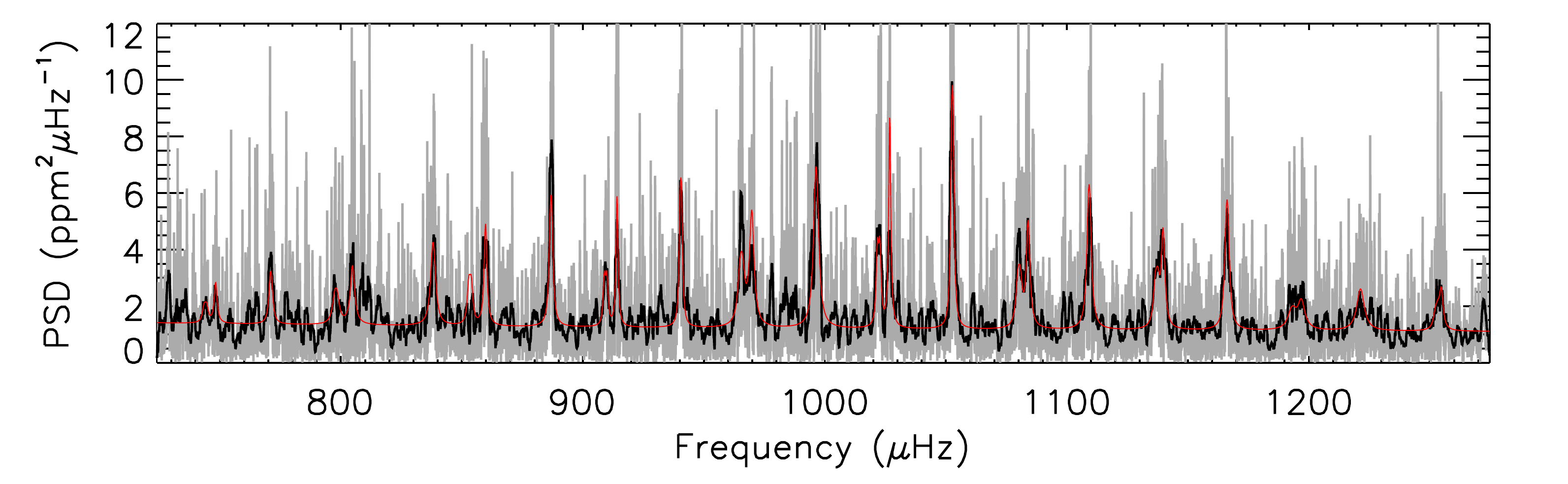}
      \caption[Example of a global fitting done over 11 orders]{Power spectral density (PSD) of HD 169392 in the p-mode region at full resolution (grey) and smoothed over 15 bins wide boxcar (black).  The red line corresponds to the global fitting performed over 11 orders.             }
         \label{fpsd}
\end{figure}

 Each multiplet is described by five parameters: the central frequencies of the modes $\ell=0,1,2$, one line width (the same for all modes within a large separation), and one mode height. In general, the same visibility ratio between angular degrees is assumed \citep{2011A&A...528A..25S}, unless there are indications that this relation does not hold \citep[e.g. in the case of depressed dipolar mode stars,][]{2014A&A...563A..84G}. An example of such global fit can be seen in Fig.~\ref{fpsd} corresponding to the analysis of 11 orders of the CoRoT star HD~169392 \citep{2013A&A...549A..12M}.

It is important to note that in the solar case the variation induced in the p-mode parameters by magnetic activity  (e.g, frequency shifts of the modes, amplitude modulation, etc) can be measured at different time scales \citep[e.g.][]{1985Natur.318..449W,1992A&A...255..363A,2001MNRAS.322...22C,2009A&A...504L...1S,2010ApJ...718L..19F,2015A&A...578A.137S,2015arXiv150906894H}. In ateroseismology, these magnetic effects are not taken into account yet, although they can be measured in some targets \citep[e.g.][]{2010Sci...329.1032G,2011A&A...530A.127S}. The good news is that magnetic activity seems to inhibit stellar pulsations and thus, the solar-like pulsating stars are those with the weakest magnetic effects \citep{2010Sci...329.1032G,2011ApJ...732L...5C}.

\subsection{Background fitting}
\label{sub_back}

As said before, the low-frequency part of the PSD can be explained by a model in which each source of convective motions is described by an empirical law --initially proposed by  \cite{harvey85} for the Sun-- corresponding to an exponentially decaying time function. To properly fit this convective background, one or two Lorentzian functions are also fitted to take into account for the extra power coming from the $p$ modes \citep[e.g.][]{2002ESASP.506..897V,2008A&A...490.1143L} and a constant for the photon noise. Therefore, in the solar case, the model of the global spectrum, including both non-periodic and periodic components, are  expressed by:

\begin{equation}
P(\nu) = N_{ph} + \sum^{N}_{i=1}\frac{4\sigma_i^2\tau_i}{1+(2\pi\nu\tau_i)^{b_i}} + \sum^{M}_{j=1}A_j\left[\frac{\Gamma_j^2}{(\nu-\nu_{0_j})^2+\Gamma_j^2}\right]^{c_j} 
\end{equation}
where
\begin{itemize}
	\item $P(\nu)$ is the power spectral density; 
	\item $N_{ph}$ is the photon noise;
	\item $i$ corresponds to the non-periodic motions;
	\item $j$ corresponds to the periodic component;
	\item $\sigma_i$ and $\tau_i$ are respectively the rms-variations and the characteristic time of the $i$-th background component (the limit of the first sum, $N$, varies depending on the number of non-periodic background components of the spectrum to fit);
	\item $A_j$ and $\nu_{0_j}$ are the power and the central frequency of the Lorentzian profiles to fit to the periodic components at the higher frequency region of the spectrum, while $\Gamma_j$ sets its width. These $M$ possible peaks to fit can be identified as the so-called photospheric or/and the chromospheric component;
	\item finally, $c_j$ (as well as $b_i$) are decay rates.
\end{itemize}

In the solar case, and because of the large length of the time series (19 years in the case of SoHO and even longer form GONG and BiSON ground-based networks) it is possible to perform local averages to reduce the number of points while taking into account that the fitting is done in a logarithmic space and thus the points should be equally spaced after doing this transformation. 

If signatures of rotation are visible in the PSD at low frequency, it is possible to add a power law and a sequence of Lorentzian peaks at the frequencies of rotation and its harmonics. 

\begin{figure*}[!htb]
\centering
\begin{tabular}{cc}
	\includegraphics[width=0.5\textwidth]{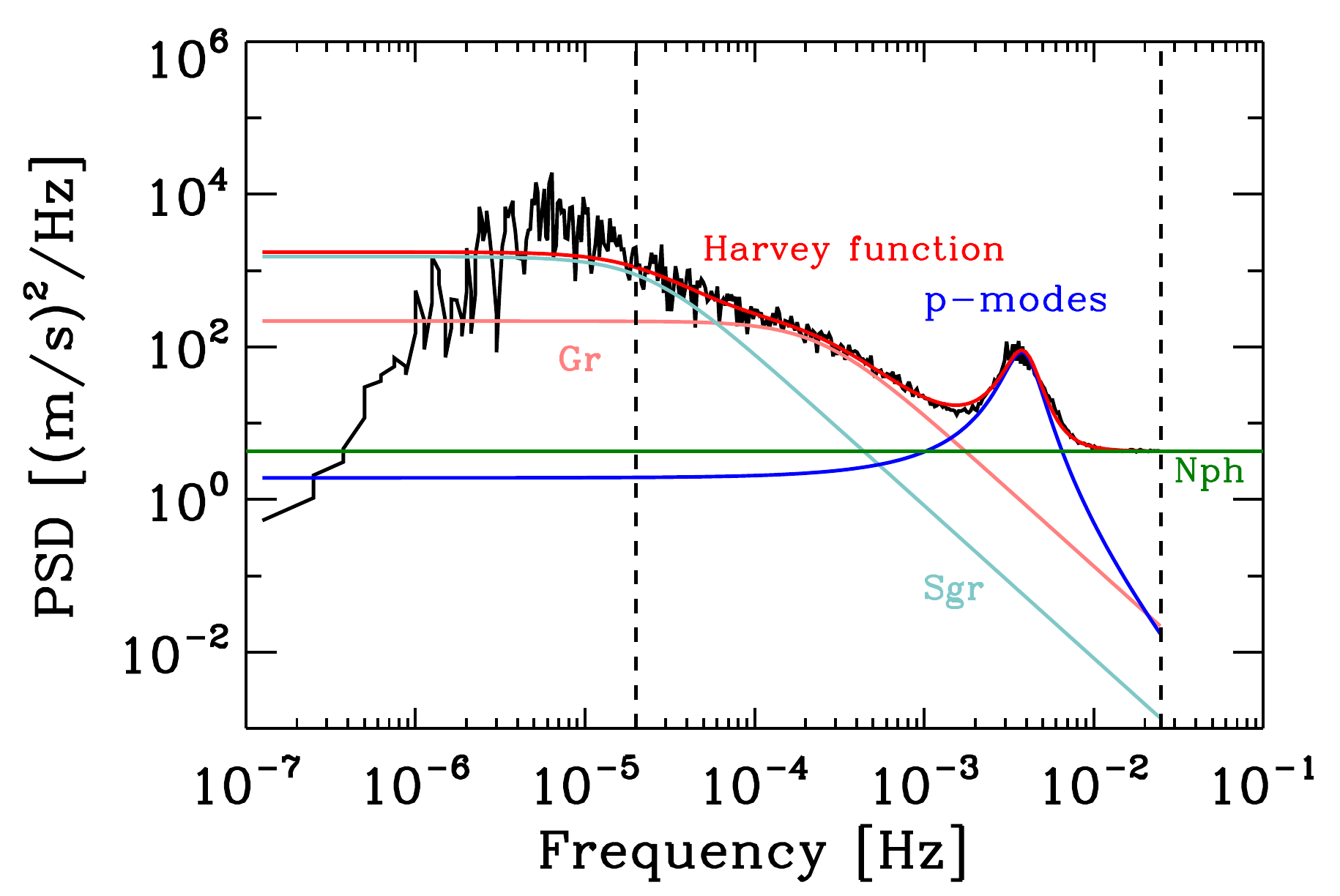} &
  	\includegraphics[width=0.46\textwidth]{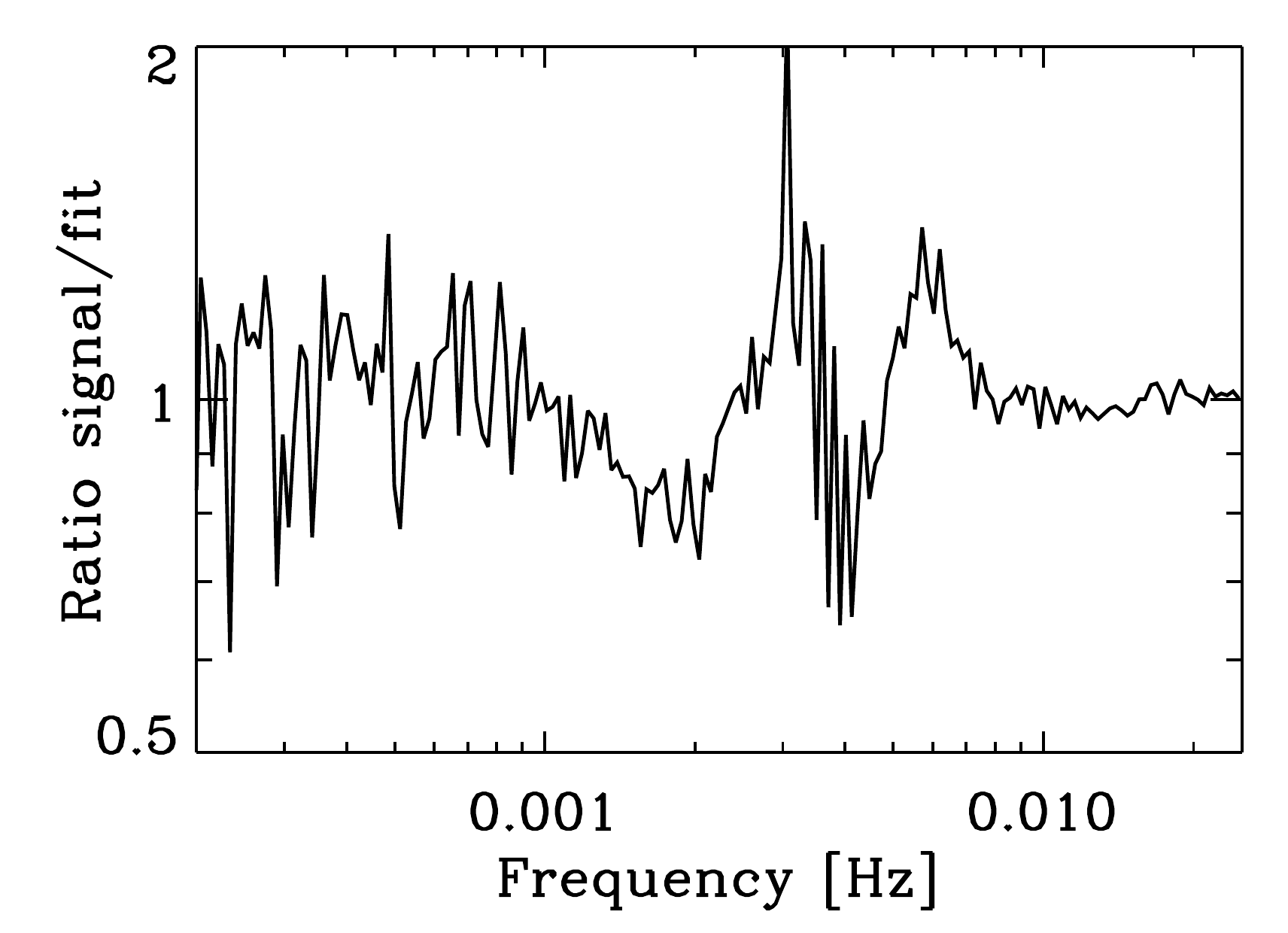} \\
	\includegraphics[width=0.5\textwidth]{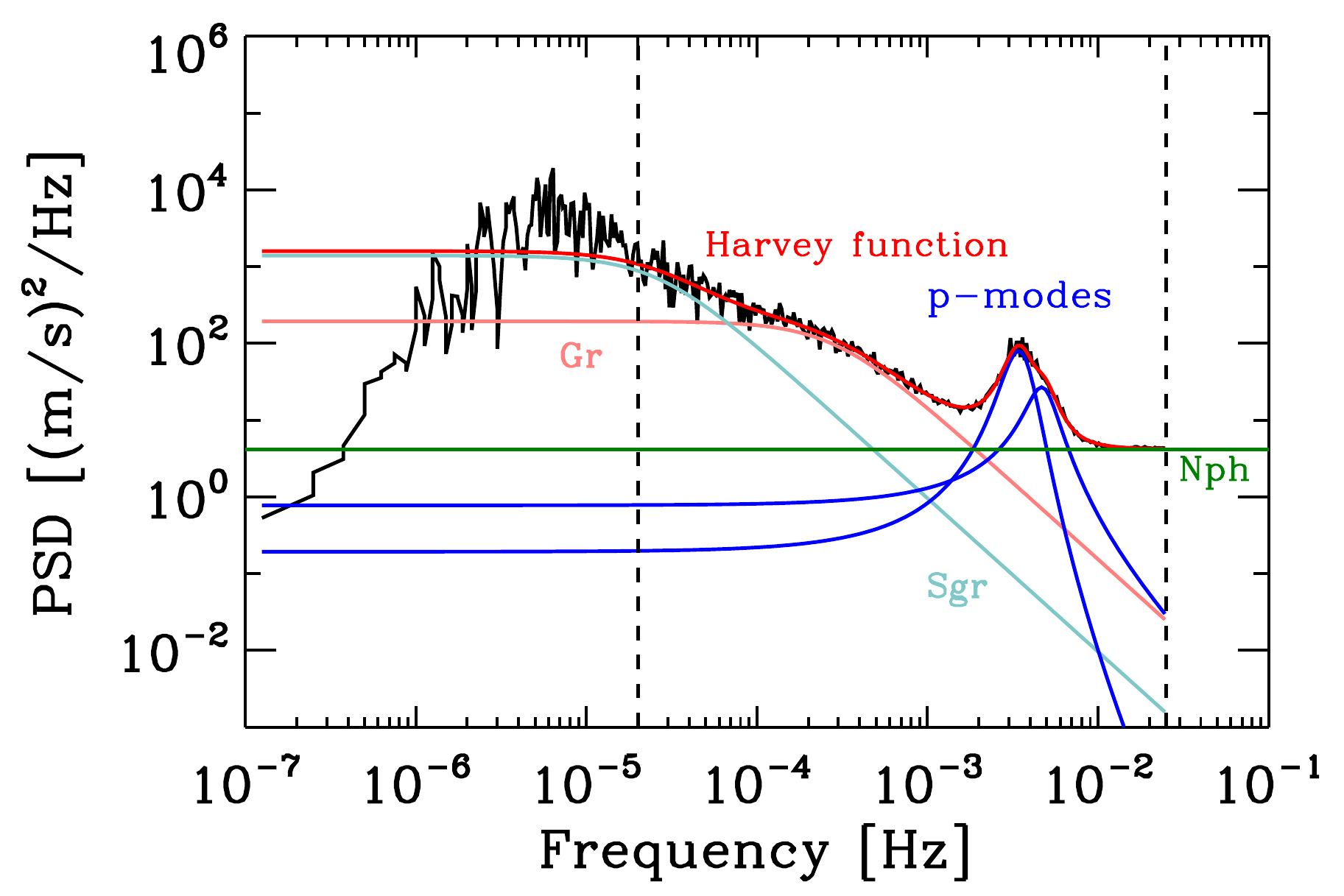} &
  	\includegraphics[width=0.47\textwidth]{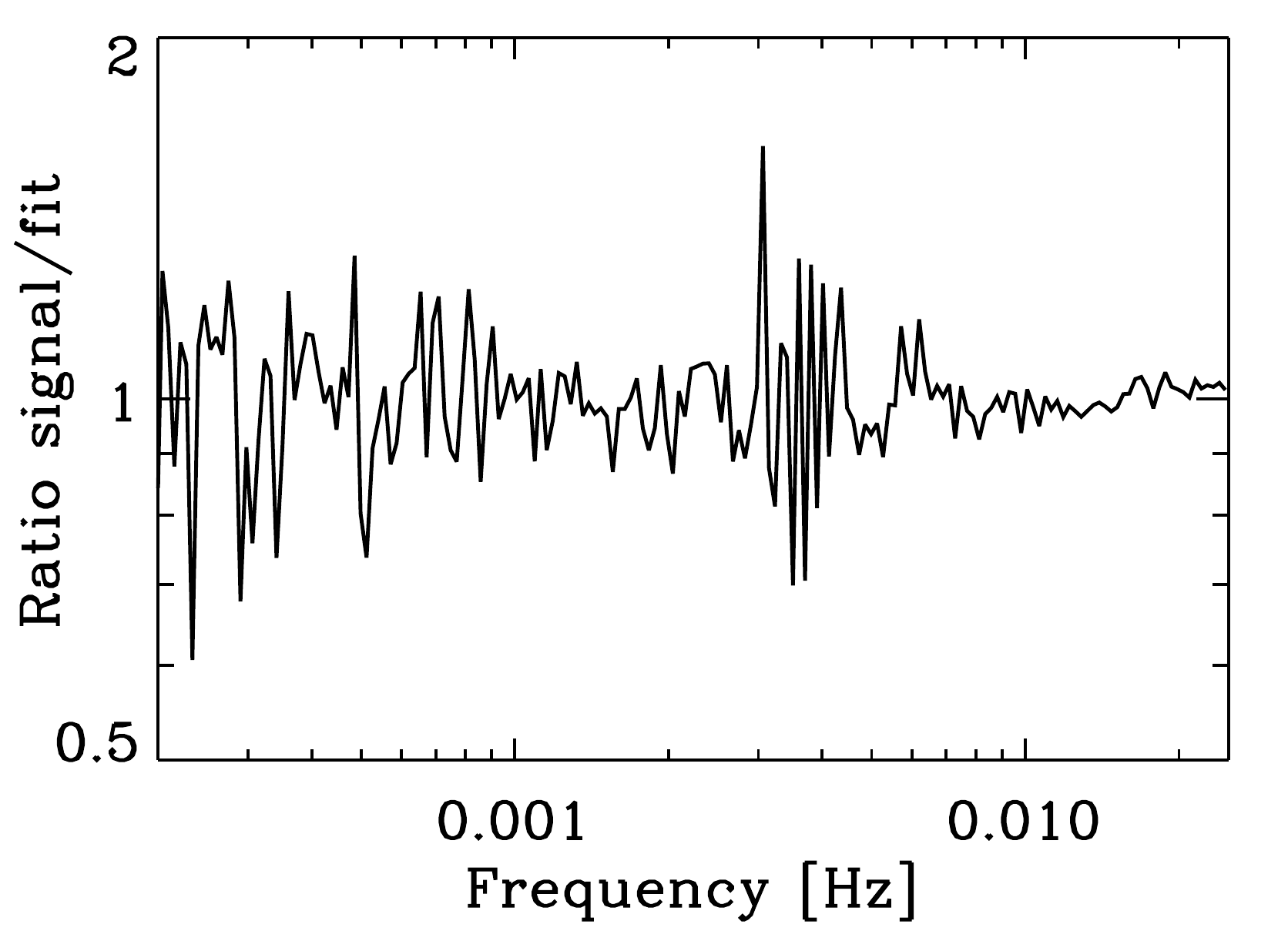} \\
\end{tabular}
  	\caption[GOLF background fitting]{Results of two different background fits applied to the GOLF spectrum of an arbitrary taken subseries of 91.25 days long. Top: Left, PSD with a fit using 8 parameters (one lorentzian) to adjust the $p$-mode envelope; Right, ratio between the PSD and the fit around the envelope of $p$-modes. Bottom: Left, PSD with a fit using 11 parameters (two lorentzians) to adjust the $p$-mode envelope; Right, ratio between the PSD and the fit around the envelope of $p$ modes.  The dashed lines represent the range in which the fit is performed. The color used for the different fits are: gray for the super granulation contribution, magenta for the granulation contribution, blue for the $p$-mode envelope, green for the noise and red for the harvey function (the sum of the granulation and supergranulation contributions).}
  	\label{Lefeb1}
\end{figure*}

An application of the background fit for the Doppler velocity solar spectrum measured by GOLF is shown in Fig.~\ref{Lefeb1}. This PSD has been modeled with two non periodic components, one for the granulation and one for the supergranulation, in both cases with the exponent $b_i=2$. To fit correctly the $p$-mode envelope measured by GOLF, it is necessary to use 2 Lorentzian profiles instead of 1 \cite[more details can be seen in][]{2008A&A...490.1143L}.  

\begin{figure}[!htb]
  \centering
   \includegraphics[width=1\textwidth]{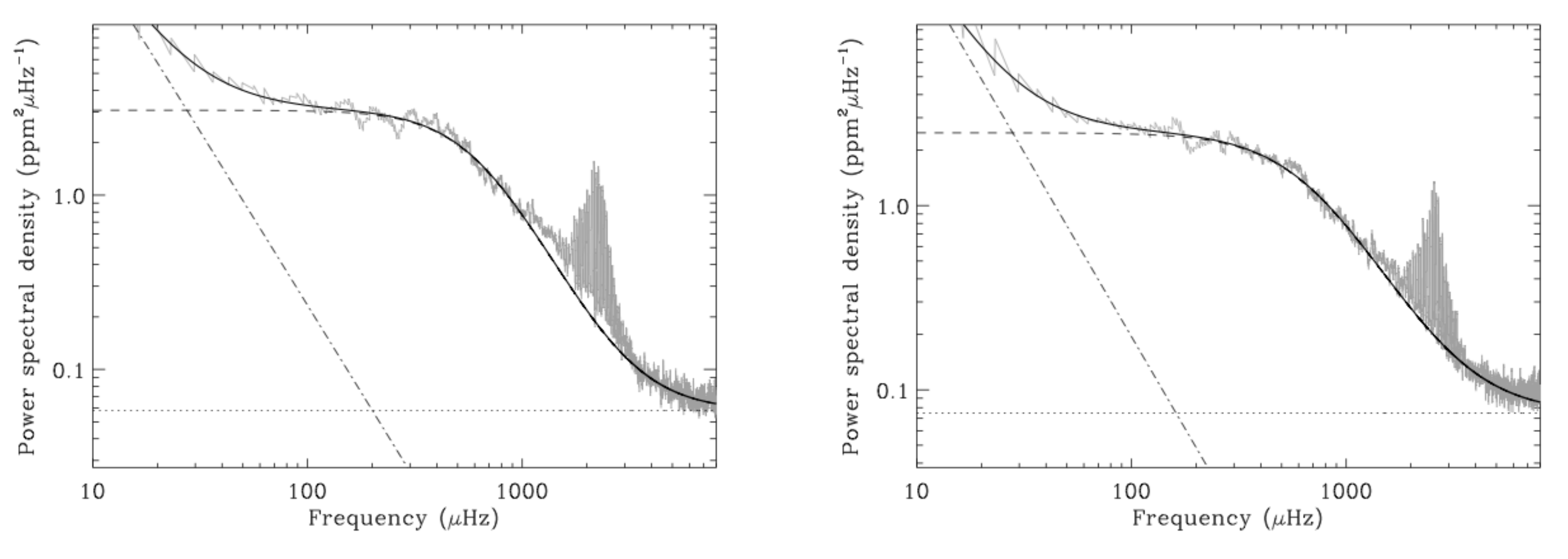}
      \caption[Background fitting of the two solar analogs 16~Cyg~A and B observed by {\it Kepler}]{Background fitting of the two solar analogs 16~Cyg~A (left) and B (right) observed by {\it Kepler}. The PSD has been smoothed by a  20 $\mu$Hz boxcar (grey), with best-fitting background components attributed to granulation (dashed lines), stellar activity and/or larger scales of granulation (dot- dashed lines) and shot noise (dotted lines), with the sum of the background components plotted as solid black lines. Figure from \citet{2012ApJ...748L..10M}. }
         \label{16CygBack}
\end{figure}

For other stars, thanks to the measurements done by CoRoT and {\it Kepler} it has been demonstrated that the same approach can be followed even for red giants \citep{2011ApJ...741..119M,2014A&A...570A..41K}. For example, in Fig.~\ref{16CygBack} we represent the background fitting of the two solar analogs 16~Cyg~A and B observed by {\it Kepler} \citep{2012ApJ...748L..10M,2015MNRAS.446.2959D}.

\begin{figure}[!htb]
  \centering
   \includegraphics[width=0.9\textwidth]{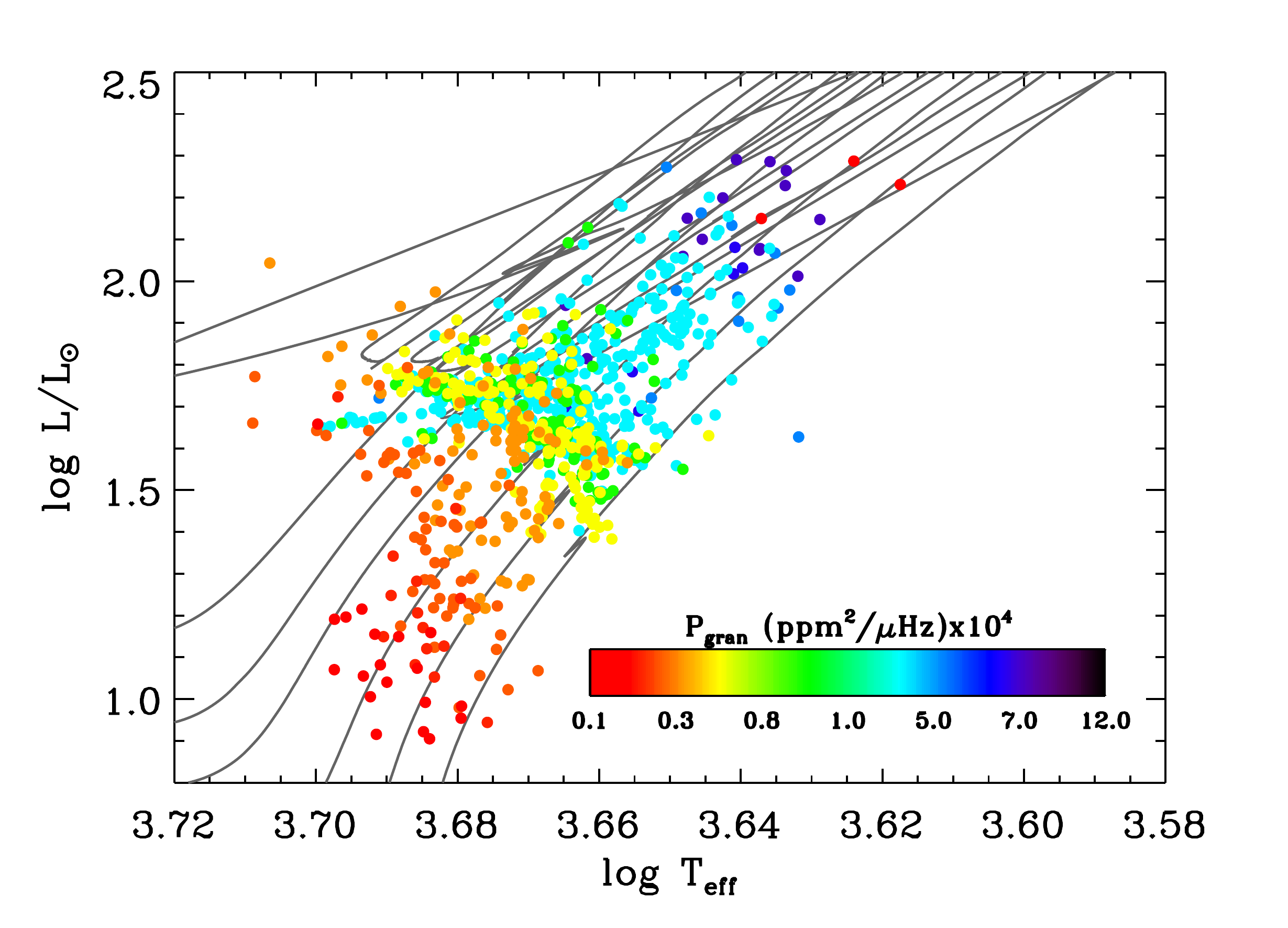}
      \caption{Distribution of  $P_{\rm gran}$  in the HR diagram (color code) for the sample of a thousand red giants observed by {\it Kepler}. The gray lines represent BaSTI evolutionary tracks computed with solar metallicity. Figure from \citet{2011ApJ...741..119M}. }
         \label{Mat_gran}
\end{figure}

The analysis of the ``ensemble'' set of a thousand pulsating red giants of {\it Kepler}  \citep{2011ApJ...741..119M} showed the existence of a relation between the convective parameters and the evolutionary state of the star (e.g. representated by $\nu_{max}$). Theoretical calculations have proven a similar relation and demonstrated the relation between the stellar granulation and the Mach number \citep{2013A&A...559A..40S}. Moreover, the observational study of the granulation power of red giants, $P_{\rm gran}$ defined as $P_{\rm gran}~=~4\sigma^2 \tau_{\rm gran}$ uncovered that larger stars present larger intensity fluctuations (see Fig.~\ref{Mat_gran}). Part of the reason for this is the smaller total number of granules covering the surface of larger stars, and hence the fluctuations are less averaged, compared to a star with many more (unresolved) granules. We also found that stars in the red clump have very similar values of granulation parameters.

Today, it is of common practice in asteroseismology to fit the background and all the p modes in one step \cite[see e.g.][]{2008A&A...488..705A,2011A&A...530A..97B,2011ApJ...733...95M}. However, in the case of the Sun, this approach is not yet employed because the number of modes --with their associated free parameters--, and the size of the time series --they are too big-- make the fitting procedure too slow and with dramatic convergency problems.

 \subsection{The solar p-mode spectrum as seen by GOLF and VIRGO/SPM}
 
 In this section we show an example of the maximum-likelihood fitting procedure (in the frequentist approach) using solar data from SoHO. The analysis presented here has been performed over  $\sim$14 years of data collected by GOLF and VIRGO/SPM. 5163 days of GOLF velocity time series from April 11, 1996 to May 30, 2010 with a duty cycle, dc=95.4~\% \citep{GarSTC2005}; and 5154 days of intensity data from the three VIRGO Sun photometers (SPM) at 402, 500, and 862~nm from April 11, 1996 to May 21, 2010 (dc = 95.2~\%).  Because the overall noise level of VIRGO/SPM is higher than in GOLF, the frequency range on which we can extract reliable estimates of the p modes at low frequency is reduced (as explained in Sect.~\ref{compGV}). However, due to the low visibility of the $\ell$=3 modes in intensity measurements, the extraction of the p-mode parameters of the $\ell$=1 modes could be done up to higher frequencies ($\sim$5000 $\mu$Hz). The mode blending at high frequency is the limiting factor in radial velocity measurements (see Fig.~\ref{fit_mp_lf}).

To characterize the p modes, we computed the power spectrum density (PSD) of the entire time series in order to maximize the frequency resolution ($\sim$ 2.24 nHz). Therefore, the obtained linewidths of the modes could be slightly overestimated because of the shift of the modes during the solar activity cycle \cite[e.g.][]{2003ApJ...595..446J}. The fits were performed by orders --fitting separately the odd and even modes-- using the methodology described in the previous sections.  

\begin{figure}[!htb]
\center
\includegraphics[width = 1.02\textwidth]{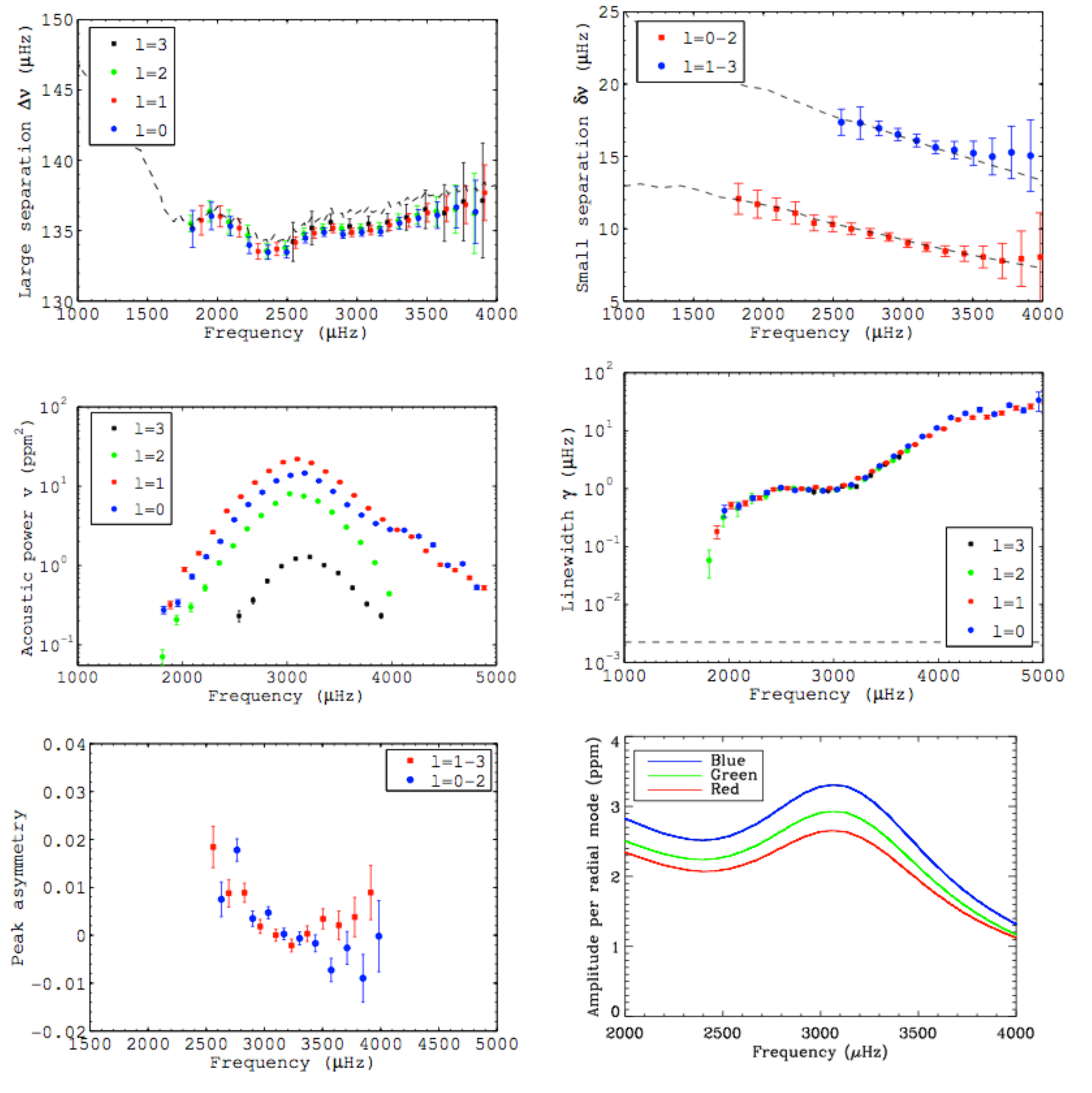}	

\caption[VIRGO/SPM p-mode characteristics]{Top left:  large separation as a function of frequency calculated from the fitted VIRGO/SPM frequencies. Top right: small separation. Middle left: Full amplitudes (in units of ppm$^2$). Middle right: Linewidths (in $\mu$Hz). Bottom left: Asymmetry. Bottom right: average maximum rms amplitudes per radial mode for the three VIRGO/SPM channels. Due to the small $\ell = 3$ visibility in the VIRGO/SPM data, these modes do not perturb the  $\ell = 1$ and the linewidth and acoustic power of the $\ell=1$ modes could be fitted up to 5000 $\mu$Hz. 
\label{fig_Virgo}}
\end{figure}

\begin{figure}[!htbp]
\includegraphics[width = 0.5\textwidth]{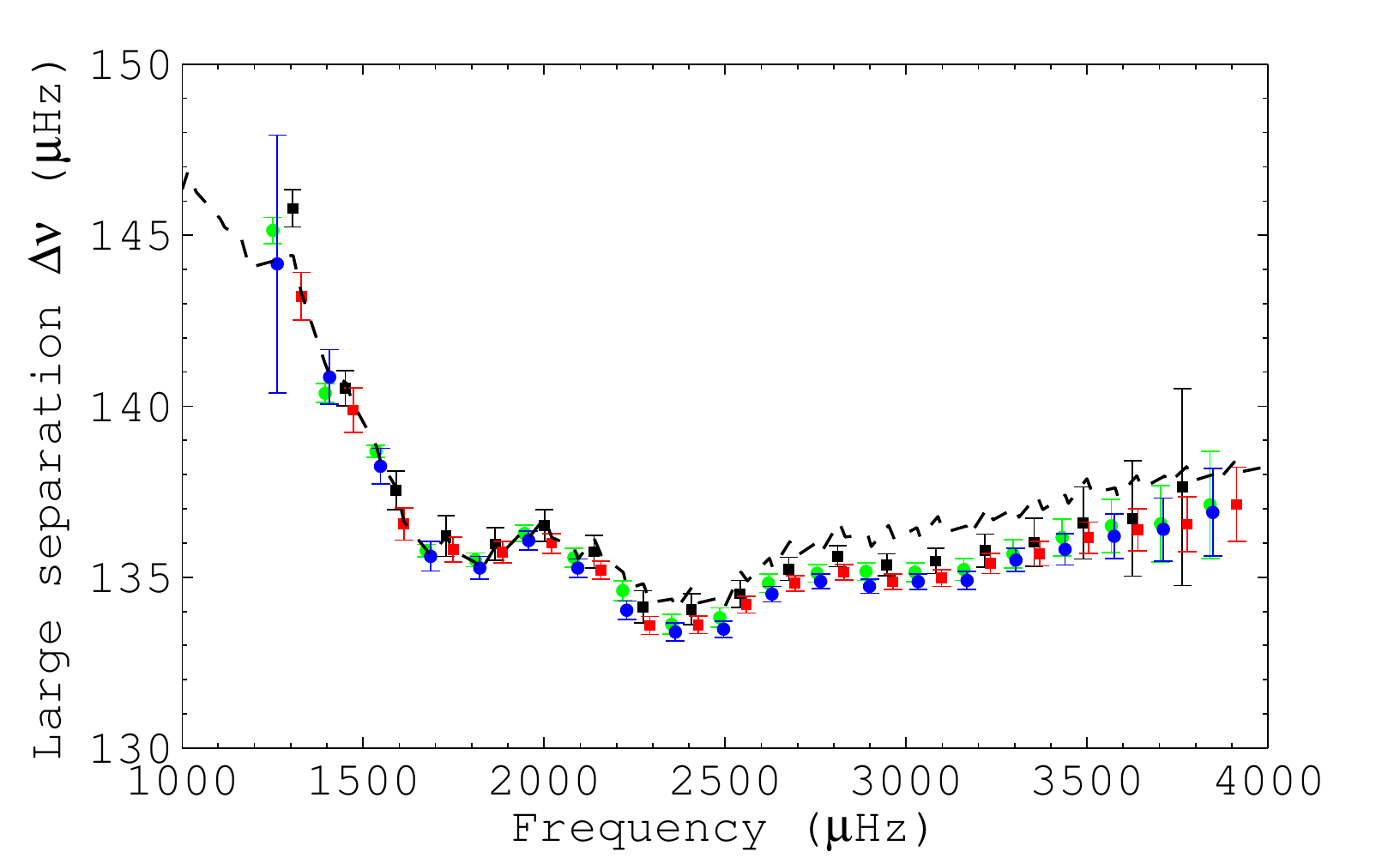}	
\includegraphics[width = 0.5\textwidth]{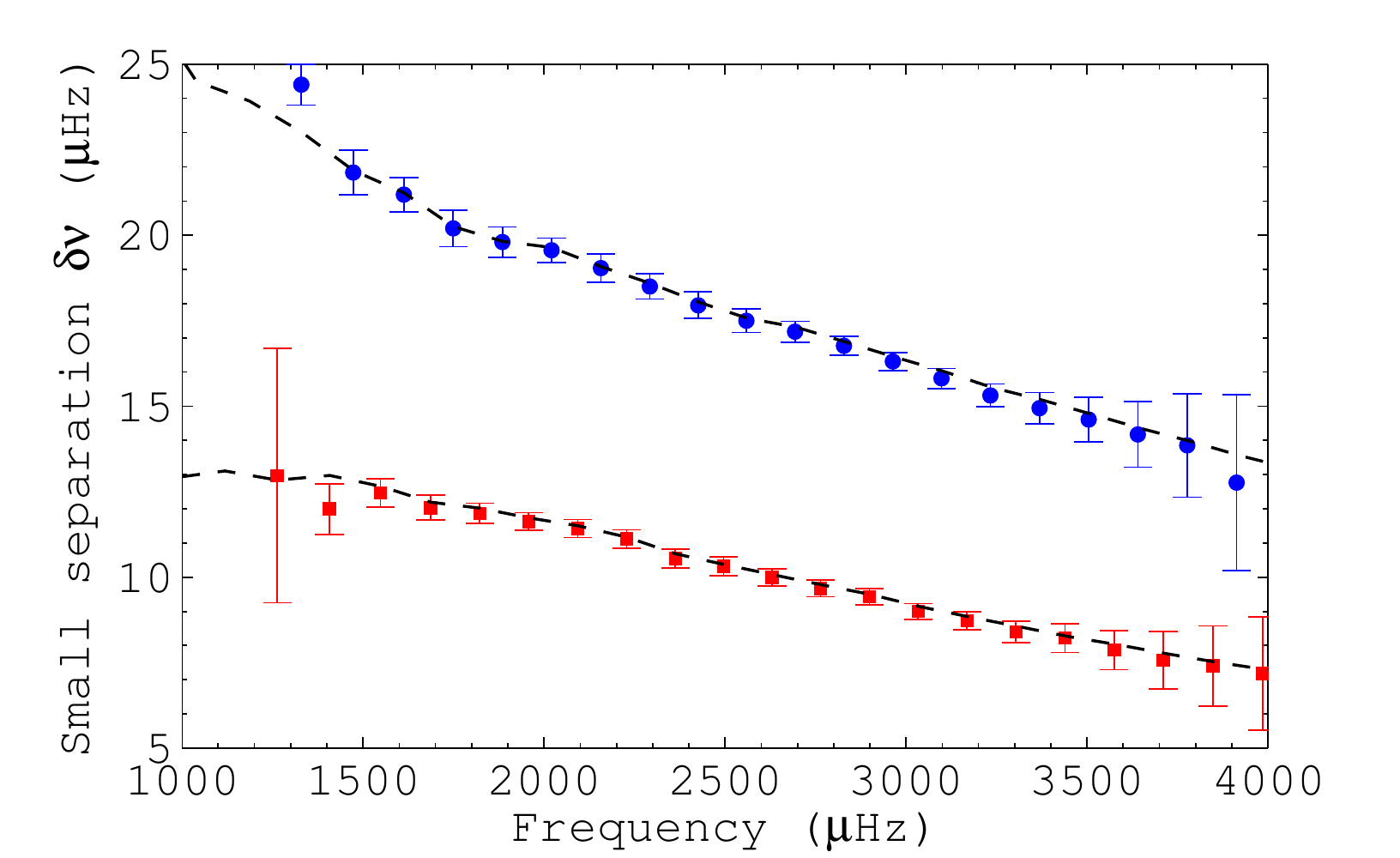}
\includegraphics[width = 0.5\textwidth]{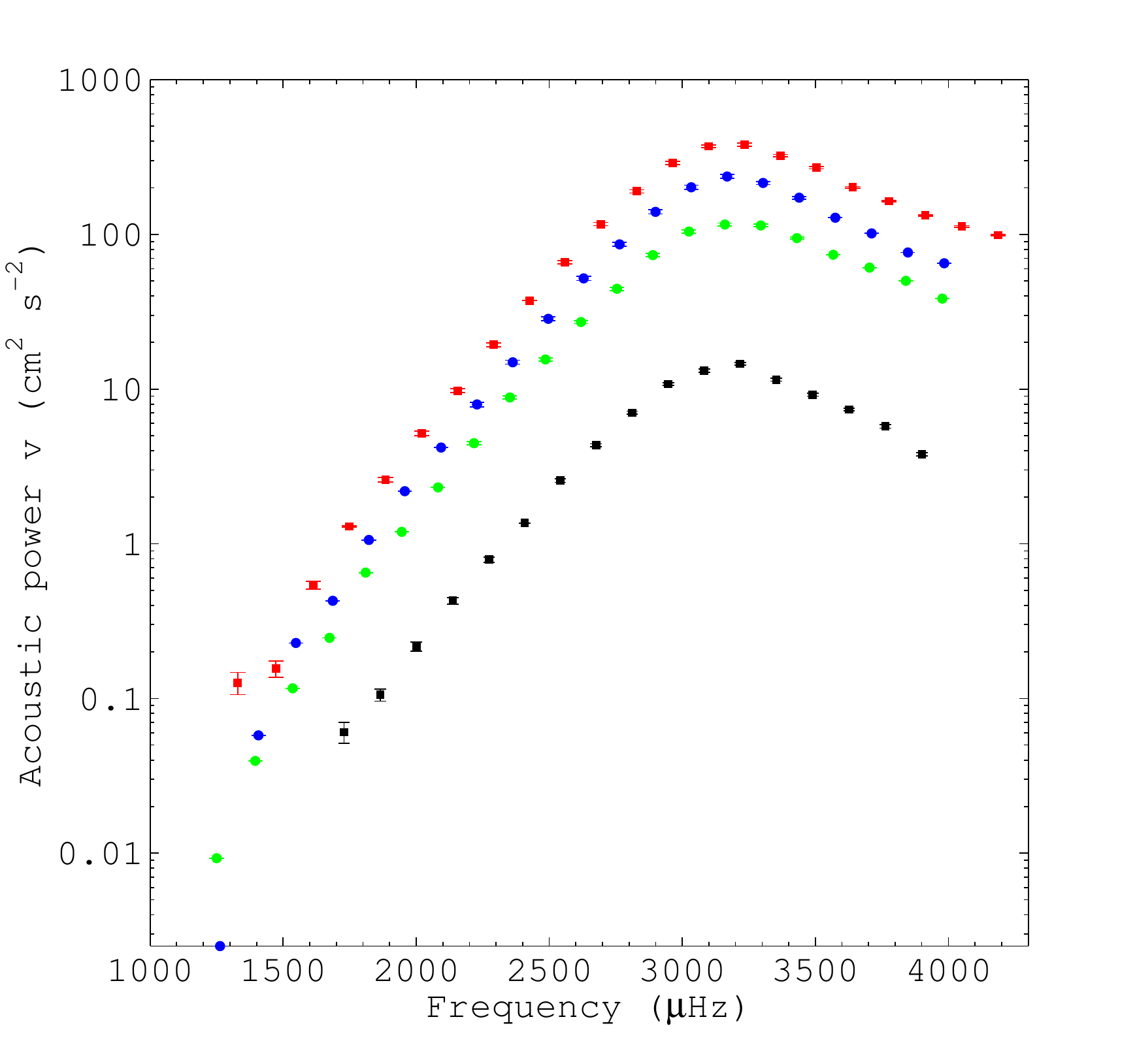}	
\includegraphics[width = 0.5\textwidth]{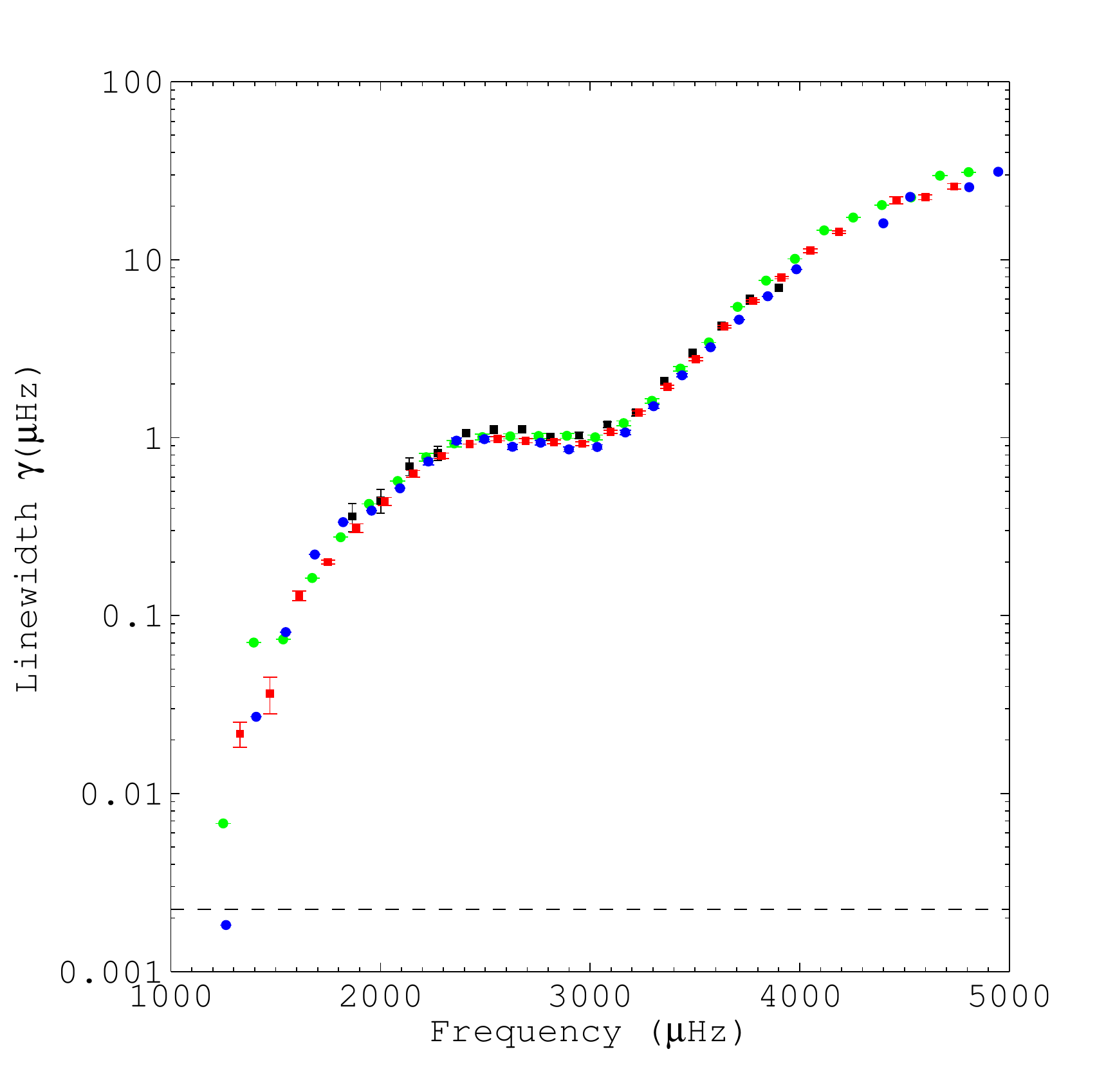}
\includegraphics[width = 0.5\textwidth]{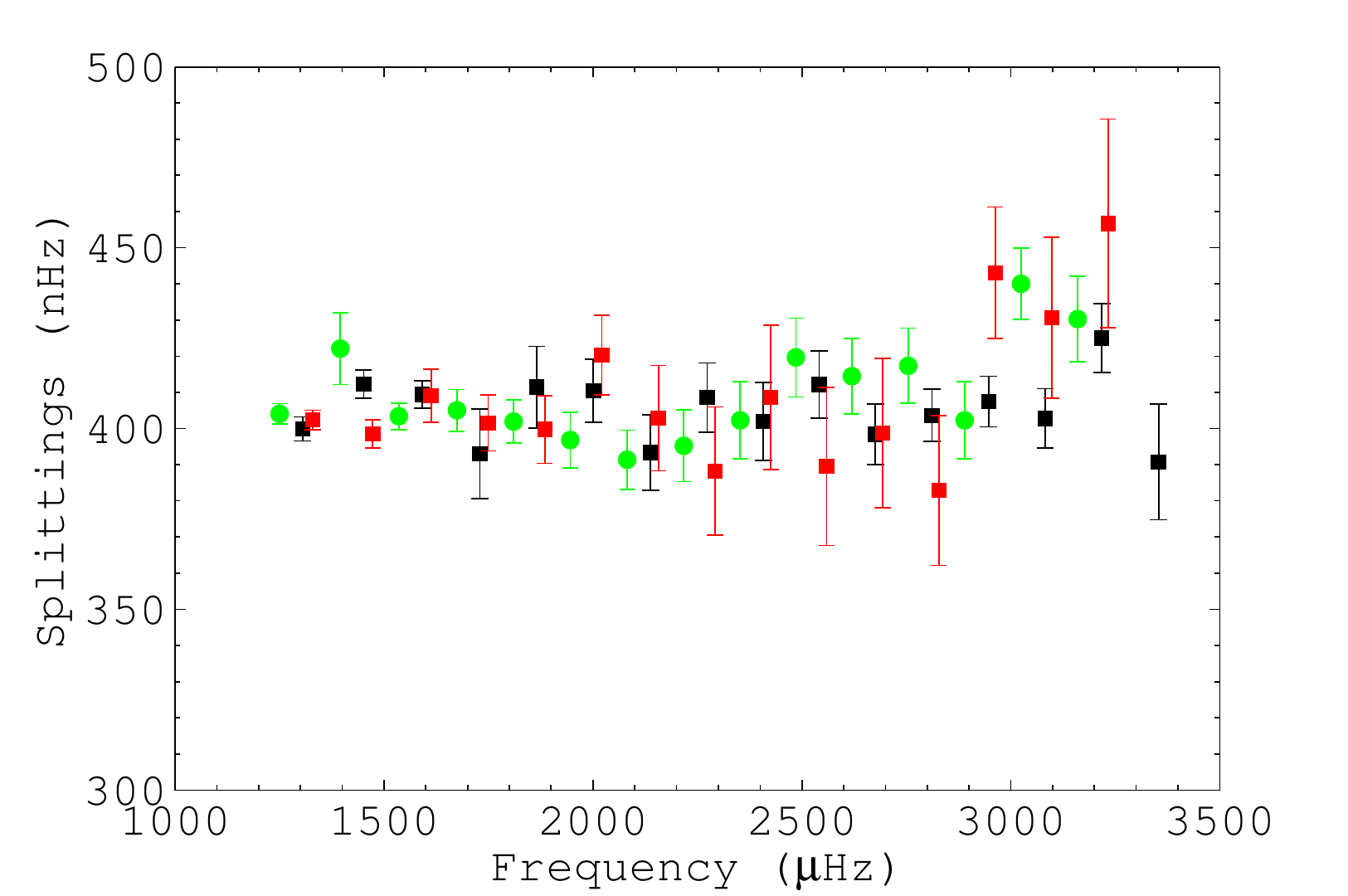}	
\includegraphics[width = 0.5\textwidth]{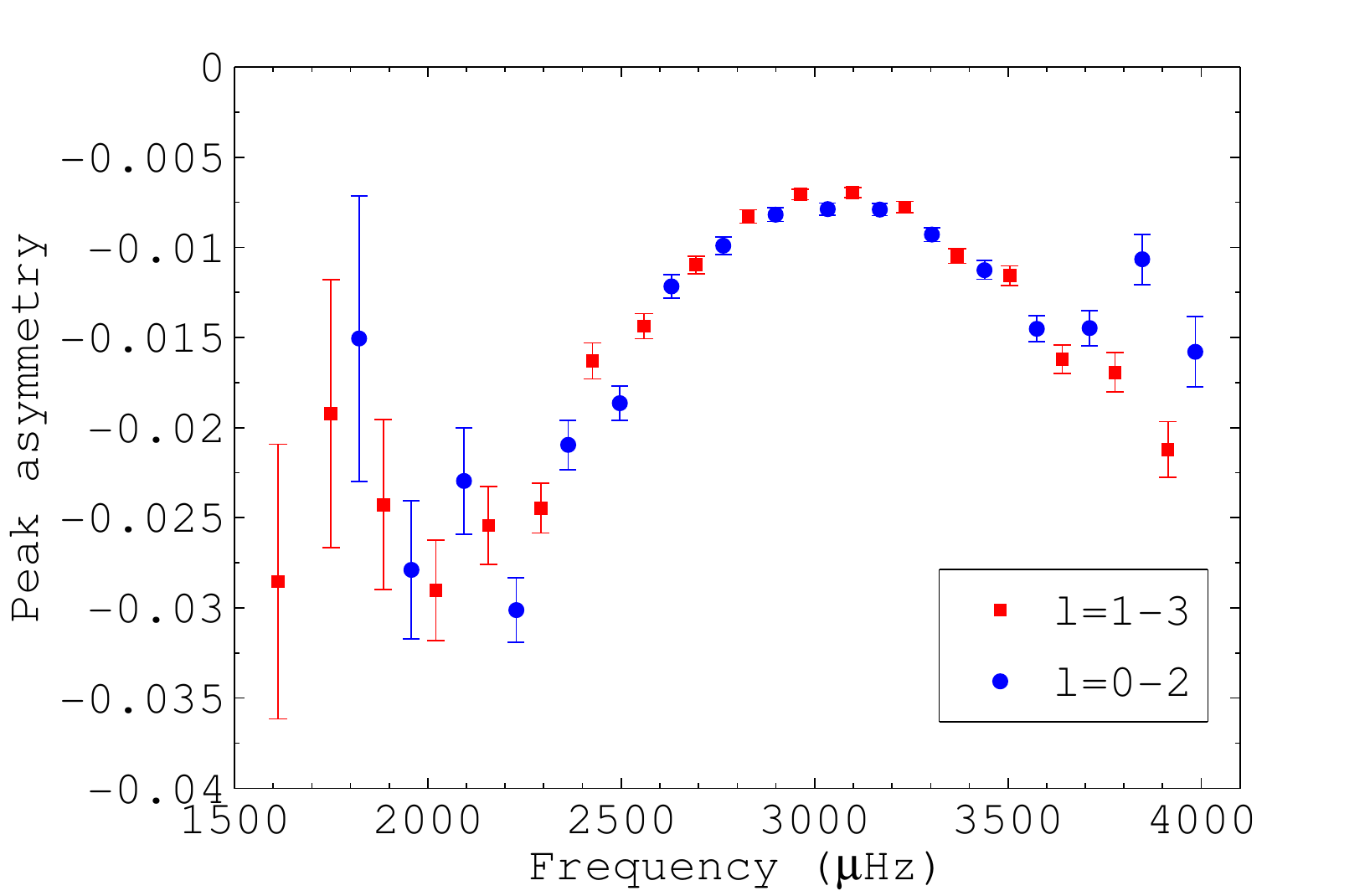}
\caption[GOLF p-mode characteristics]{Top left: large separation as a function of frequency calculated from the fitted GOLF frequencies (same legend than in previous figure (\ref{fig_Virgo}). The formal errors were multiplied by a factor 10. The dashed lines correspond to the theoretical values using the Saclay seismic model \cite{2007ApJ...668..594M}. Top right: small separation ($\ell=0-2$ modes in red, $\ell=1-3$ modes in blue). The formal errors were multiplied by a factor 10. Middle left: Full amplitudes (in units of cm$^2$ s$^{-2}$). Middle right: Linewidths (in $\mu$Hz). The horizontal dashed line corresponds to the frequency resolution. Bottom left: splittings. Above 3500 $\mu$Hz, the splittings were fixed to 400 nHz. Bottom right: Asymmetry. The increase below 2000 $\mu$Hz could not be real due to the reduction in the SNR and the fewer number of points defining the profile.
\label{fig_GOLF}}
\end{figure}

Figures~\ref{fig_Virgo} and \ref{fig_GOLF} show the p-mode parameters --large separation, small separation, acoustic power, linewidths, splittings and peak asymmetry-- as a function of frequency extracted from GOLF and VIRGO/SPM respectively between 1000 and 4000 $\mu$Hz, and up to 5000 $\mu$Hz when possible. The rotation splitting was fixed to 400 nHz for modes above 3500 $\mu$Hz. Indeed, due to the mode blending, it was impossible to obtain reliable parameters without imposing this additional condition. Thanks to that, we were able to obtain preliminary results of the linewidths extracted from GOLF up to 5000 $\mu$Hz. Also, as explained before, due to the smaller $\ell$ = 3 visibility in the VIRGO/SPM data, the linewidth and acoustic power of the $\ell$ = 1 mode could be fitted up to 5000 $\mu$Hz. Reliable estimates below 1800 $\mu$Hz could not be properly extracted in VIRGO/SPM data because of the smaller SNR.

The comparison of the large and small separation (Fig.~\ref{fig_Virgo} and \ref{fig_GOLF}, top panels) with the same quantities computed using the solar seismic model \cite[][]{STCCou2001,2007ApJ...668..594M}, shows that this model is a good reference to look for modes at low frequency. Moreover, the small separations --which are a direct probe of the core of the Sun--- are inside one-sigma error bars of those computed with the seismic model. This result was expected because the seismic model has been constructed to minimize the differences with the observations in the deepest regions of the radiative interior and produce the best estimates of the neutrino flux.

We also computed the average maximum amplitude per radial mode of the Sun (Fig~\ref{fig_Virgo} bottom right) for the three VIRGO/SPM channels, as it is commonly done in asteroseismology \cite[e.g.][]{2010A&A...511A..46M}. The maximum amplitudes were corrected by the instrumental response function using the values given by \cite{2009A&A...495..979M} for the different channels.

\subsection{Ensemble stellar p- and mixed-mode fitting}
\begin{figure}[!htb]
  \centering
   \includegraphics[width=0.8\textwidth]{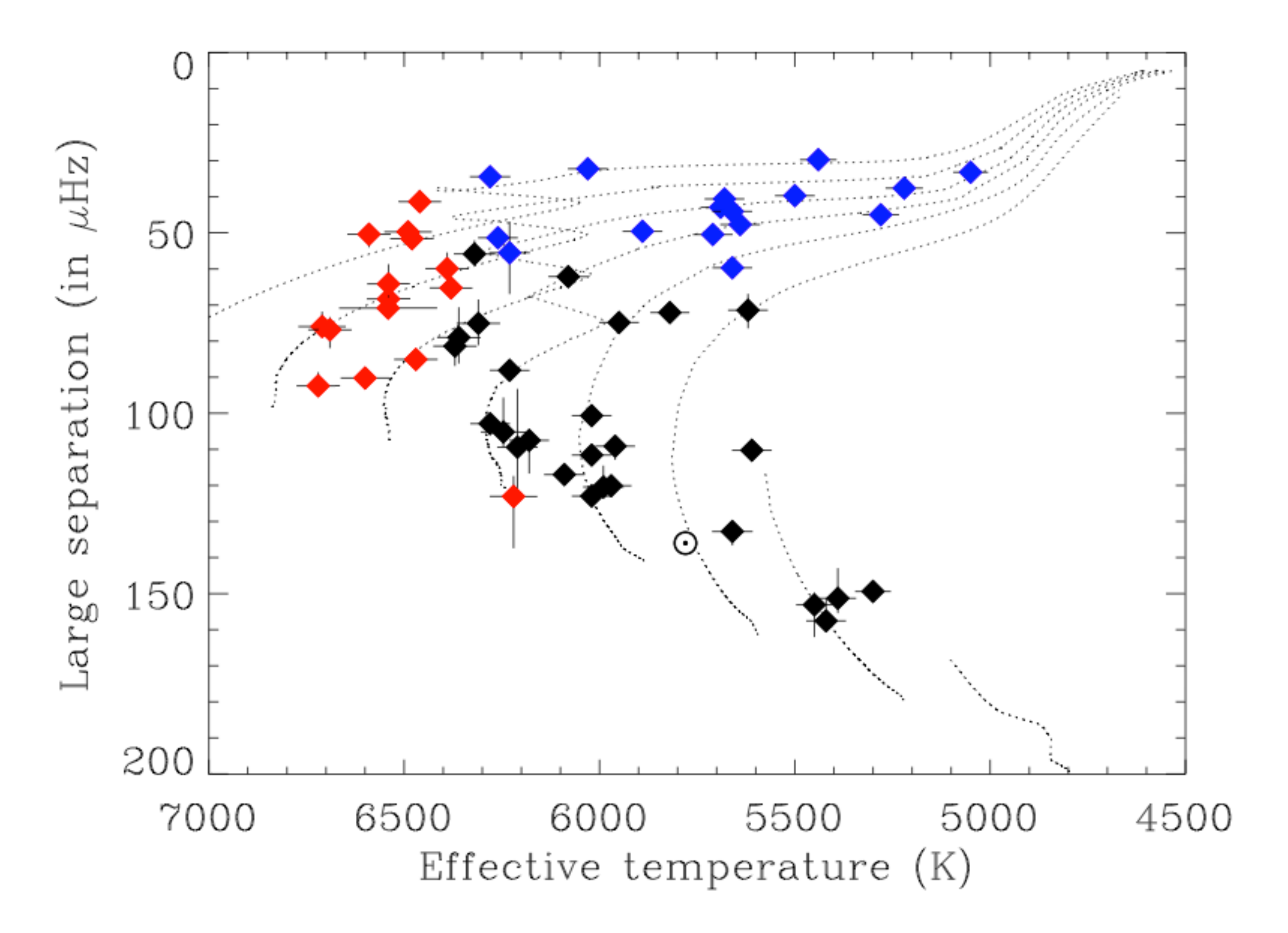}
      \caption[Seismic HR diagram in which the large separation is plotted versus the effective temperature of 61 solar-like stars observed by {\it Kepler}]{Seismic HR diagram in which the large separation is plotted versus the effective temperature of 61 solar-like stars observed by {\it Kepler}: (black) simple stars, (blue) mixed-mode stars, (red) F-like stars, ($\odot$) our Sun. The uncertainties on the large separation represent the minimum and maximum variation with respect to the median; some of these uncertainties are within the thickness of symbol. The evolutionary tracks for stars of mass 0.8 $M_\odot$ (most right) to 1.5 $M_\odot$ (most left) (by step of 0.1 $M_\odot$) are shown as dotted lines. The tracks are derived from \cite{2008A&A...482..883M}. Figure from \citet{2012A&A...543A..54A}.}
         \label{pe15_HR}
\end{figure}

{\it Kepler} allowed to do the first ``ensemble'' asteroseismic study of many solar-like stars covering a wide range of properties. 61 stars observed during Q5, Q6 and Q7 (from March 22, 2010 till December 22, 2010) were analyzed in an homogeneous way \citep{2012A&A...543A..54A}. The selected stars are plotted in a seismic HR diagram in Fig.~\ref{pe15_HR}.

\begin{figure}[!htbp]
 \centering
\includegraphics[width = 1\textwidth]{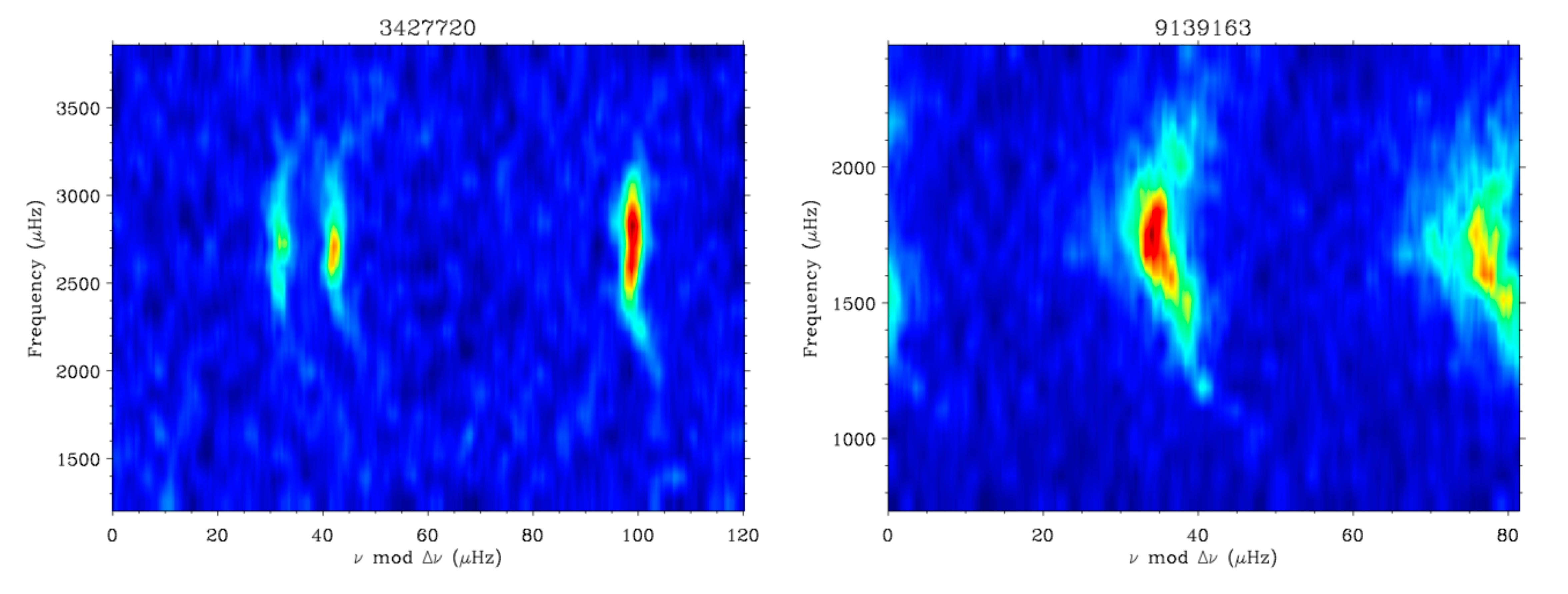}\\	
\includegraphics[width = 1\textwidth]{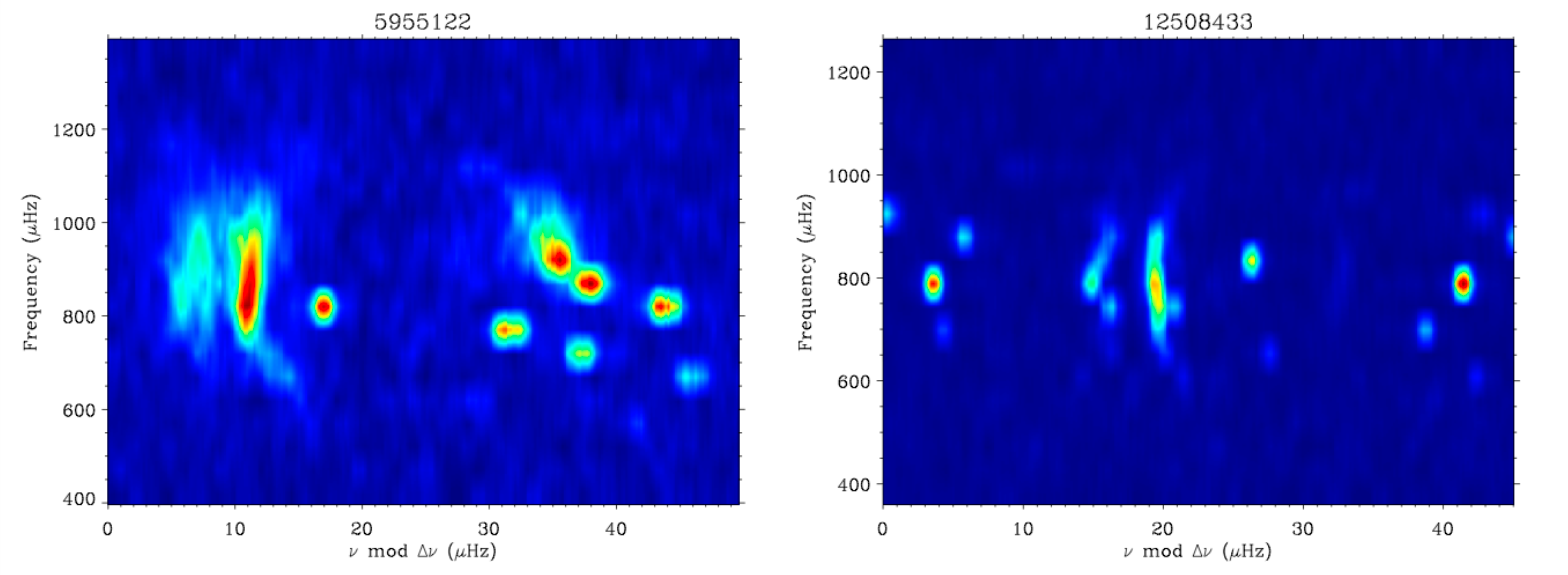}\\
\includegraphics[width = 1\textwidth]{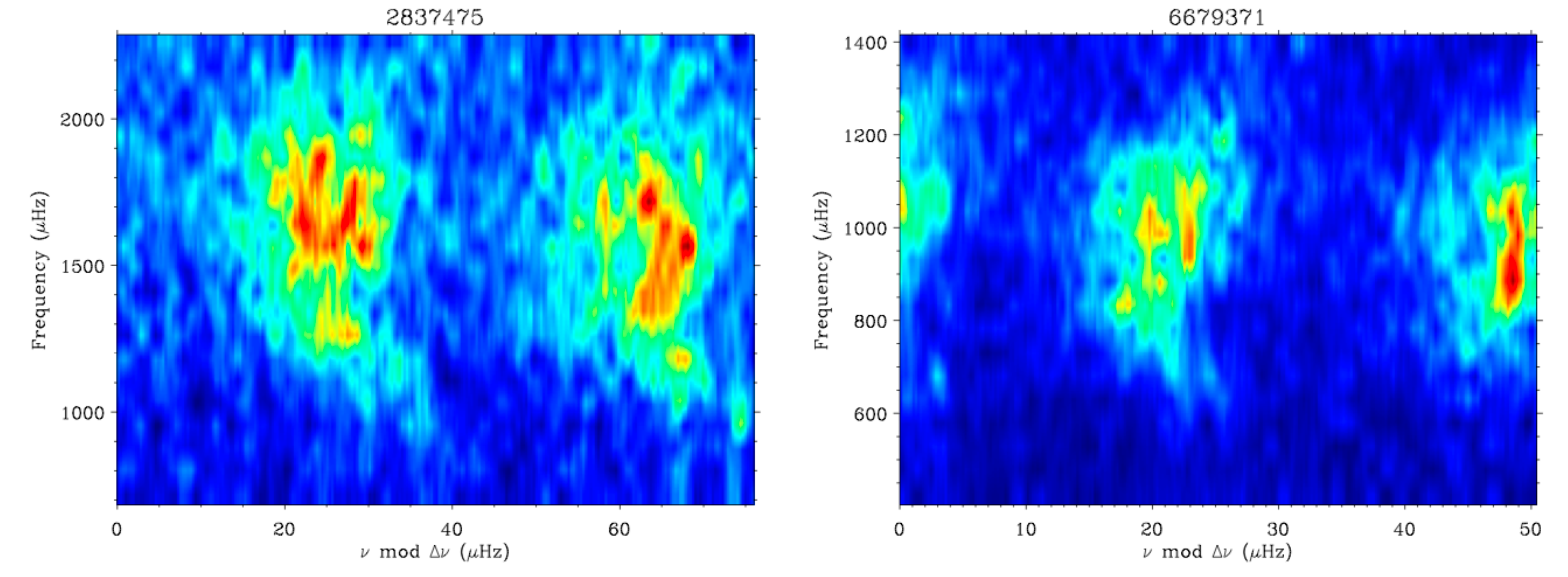}	
\caption{Example of \'echelle diagrams corresponding to the ``ensemble'' analysis of 61 solar-like {\it Kepler} stars studied by \citet{2012A&A...543A..54A}. On top two simple stars, in the middle two mixed-mode stars, and two F-like stars in the bottom panels. The power spectra are normalized by the background and smoothed over 3 $\mu$Hz. The \'echelle diagrams are smoothed over 2 orders in the vertical direction. Figures from \citet{2012A&A...543A..54A}.
\label{PE15_examples}}
\end{figure}

To properly extract the mode parameters for these stars, we separate them into three different categories: simple (sun-like), F-like, and sub giants (stars having $\ell=1$ mixed modes). Out of the 61 stars, we have 28 simple stars, 15 F-like stars, and 18 mixed-mode stars. Figure~\ref{pe15_HR} shows that the boundary between simple stars and F-like stars is about 6400 K which roughly corresponds to a linewidth at maximum mode height of about 4 $\mu$Hz \citep{2012A&A...537A.134A}. For these F-like stars, the frequency separation between the $l = 0$ and $l = 2$ modes (i.e. small separation) ranges from 4 $\mu$Hz to 12 $\mu$Hz combined with a linewidth of at least 4 $\mu$Hz justifies why the detection of the $\ell = 0, 2$ ridge is more difficult for these stars \citep[see also][]{Benomar2009a}. Examples of the three kind of stars are shown in Fig.~\ref{PE15_examples}.

Nowadays, the efforts are concentrated in the proper fitting of red giants. The first challenge is to properly identify the modes to be fitted. To do so, several semi-analytical models have been developed \citep[e.g.][]{2012A&A...540A.143M,2013A&A...549A..75G} that allow a first proper identification of the modes if the rotational splittings are not too big. Once this identification is done, bayesian techniques are more suited to this problem, and thus the first ensemble fit of 19 young red giants has been performed \citep{2015A&A...579A..83C}. The second challenge that the community will need to face in the near future will be how to do these fits in the more than 30,000 red giants measured by CoRoT and \emph{Kepler} missions.

 \section{Instrumentation}

As a consequence of the stellar oscillations, the gas is compressed and expanded in the photosphere. With each cyclic movement, the brightness of the star fluctuates. In the compression it warms up while in the expansions it cools down. As a consequence, the stellar flux is modulated and this modulation can be measured. At the same time, and due to the movement of the stellar photosphere, the absorption lines are also Doppler shifted. Stellar oscillations can then be measured as a periodic variation of the total stellar flux or by measuring the cyclic Doppler shifts induced by the pulsations. However, the magnitude of such variations are really small: around $15$ cm/s at $\nu_{\rm{max}}$ for the velocity shifts and around $3 \times 10^{-6}$ for brightness variations, corresponding to $10^{-3}$ K in temperature. It is also important to notice that solar oscillations were also looked for by measuring variations in the solar diameter using the SCLERA experiment \citep[Santa Catalina Laboratory for Experimental Relativity by Astrometry,][]{HILL1976}, and using the CNES mission PICARD \citep{2006cosp...36..170T}. Unfortunately, the complexity of this type of observations prevented the success of such measurements.

Due to this small intrinsic amplitude of the oscillations, precise and stable instrumentation are required to reach the high SNR necessary to properly record the solar and stellar pulsations.

Helio- and Astero-seismic observations can be performed either form the ground and from space. Because of the intrinsic requirement on the continuity of the measurements from weeks to months and years, it seems clear that single site observations from ground are not good (excepting the sites placed close to the Earth's Poles). The day-night cycle imposes a window function of around 8-12h which introduces daily aliases at a frequency of 11.57 $\mu$Hz in the spectral analysis. To avoid this problem, ground-based networks has been built and are currently running (e.g. GONG or BiSON). 

Apart form the magnitude used to measure the stellar oscillations, we can classify the helioseismic instruments in imaged and Sun-as-a-star ones. Due to the nature of this review, we will concentrate only on non-imaged instrumentation. I have chosen some instruments as examples of the techniques that I will explain in more details here. For a more exhaustive review on all the past, present, and future instrumentation in helio and asteroseismology, I recommend the reader the reviews by 
\citet{AppTongGar} and \citet{PalleAppJCDRAG}.

\subsection{Helioseismic Doppler velocity instruments}

Helioseismic Doppler velocity measurements are complicate in nature because the sensitivity of the measurement is not
uniform across the solar disk due to the projection effect combined with the solar rotation. A theoretical calculation of such sensitivity for the case of the GOLF instrument can be found in \citet{GarRoc1998}.

Helioseismology entered in a new era in the mid nineties with the deployment of the ground based networks around the world such as the Birmingham Solar Oscillation Network \citep[BiSON,][]{1996SoPh..168....1C}, the Global Oscillation Network Group \citep[GONG,][]{HarHil1996}, and the launch of the Solar and Heliospheric Observatory (SoHO) mission \citep{DomFle1995}. All these facilities allowed a continuous monitoring of the Sun, drastically improving the quality of the datasets available.

The SoHO mission is a three-axis, stabilized spacecraft developed by the European Space Agency (ESA) in collaboration with the National Aeronautics and Space Administration (NASA). It contains eleven scientific instruments dedicated to the study of the Sun, its heliosphere and the solar wind \citep{DomFle1995}. SoHO was one of the cornerstones of the ESA Space Science program called Horizon 2000 and it was successfully launched in December 1995.  

SoHO offers an unprecedented opportunity to study the deep interior of the Sun through helioseismology under ideal conditions at the Lagrange $L_1$ point at 1.5 10$^6$ km from Earth. At this location, no terrestrial atmospheric effects are present, continuous exposures to the Sun are possible (more than 95$\%$ duty cycle), and there is a low Sun-spacecraft line-of-sight velocity. This spacecraft carries three Helioseismic instruments, two using Doppler velocity techniques: GOLF\footnote{Global Oscillations at Low Frequency \citep{GabGre1995}} and SOI/MDI\footnote{Solar Oscillation Imager/Michelson Doppler Interferometer \citep{1995SoPh..162..129S}}, and one recording brightness variations: VIRGO\footnote{Variability of solar Irradiance and Gravity Oscillations \citep{1995SoPh..162..101F}}. 

The two main Doppler velocity instruments performing Sun-as-a-star observations, BiSON and GOLF/SoHO are based on the same technique, the spectrophotometry. 
GOLF was originally designed to measure the disk-integrated ---Sun-as-a-star--- oscillations of the Sun \citep[e.g.][]{LazBau1997,ThiBou2000,GarReg2001} and its mean magnetic field \citep{GarBou1999}. The main scientific objective of both experiments is the quantitative knowledge of the internal structure of the Sun by measuring the spectrum of its global oscillations in a wide frequency range (30 nHz to 25 mHz). In the case of GOLF, with special interest  in detecting the low-degree acoustic (p) and gravity (g) modes (located at low frequencies below 1.5 mHz). 

GOLF is an improved resonant scattering spectrophotometer that determines the line-of-sight velocity of the integrated visible solar surface of the Sun by measuring the Doppler shift of the neutral sodium doublet ($D_1$ at $\lambda$= 589.6 nm and $D_2$ at $\lambda$ = 589.0 nm). BiSON used the potassium Fraunhofer line at 770 nm instead.

\begin{figure}[htbp]
\centering
	\includegraphics[width=9cm]{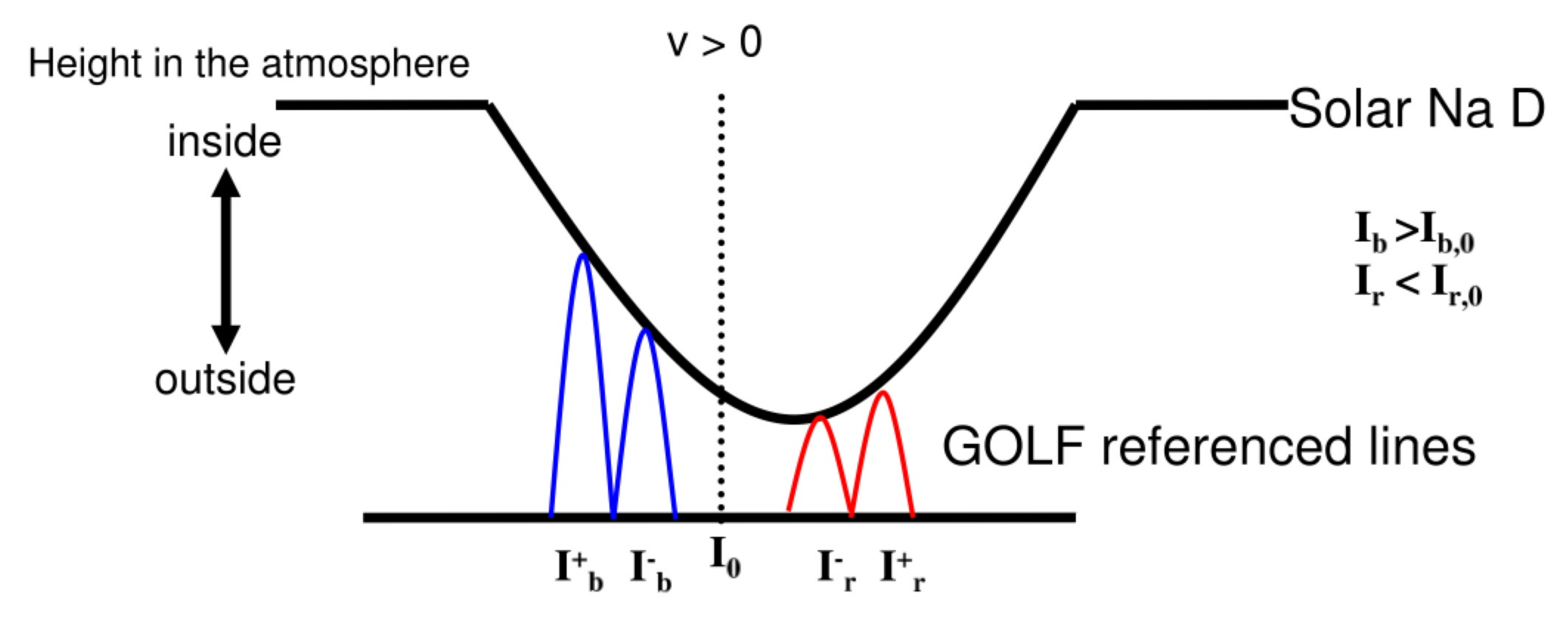}
  	\caption[Sketch of the GOLF measurements in the Sodium D1 line]{Representation of the GOLF measurements. The Solar Na D line is displaced from the rest position due to a positive velocity ($V>0$). The BW (blue wing) measures higher in the line, which means deeper (close to the photosphere) in the solar atmosphere. The RW (red wing) measures closer to the bottom of the line, i.e. upper in the solar atmosphere.}
\label{GOLF_Na}
\end{figure}

In the GOLF instrument, the light coming from the solar sodium absorption line (half-width $\sim$500 m\AA) traverses a sodium vapour cell --placed in a longitudinal magnetic field of $\sim$5000 Gauss-- which has an intrinsic (thermal) absorption line-width of the order of 25 m\AA, where it is absorbed and re-emitted in all directions. This scattered light is symmetrically split into its Zeeman components displaced around $\pm $ 108 m\AA~from the rest wavelength, allowing a measurement on either side on the wings of the solar absorption profile. The sodium cell is surrounded by a coil that changes the magnetic field $\pm$ 100 Gauss allowing the measurement of two different points on each wing (see Fig. \ref{GOLF_Na}). A polarization mechanism placed in the optical path prior to the vapor cell selects the circular polarization of the entrance light and switches every 10s between both wings. A Doppler displacement of the solar Na line will be seen as a different intensity on each wing of the instrument and thus the ratio  $(I_b-I_r)/(I_b+I_r)$ will be proportional to this velocity (see Fig. \ref{GOLF_Na}). The scattered photons are collected by 2 photomultiplier tubes. Redundancy on both, the electronics and photomultipliers guarantees the long term efficiency of the detection subsystem. A complete description of the instrument can be found in \citep{GabGre1995,GabCha1997}.

\subsection{Helioseismic photometric observations}
The closest helioseismic observations to the ones commonly done in asteroseismology are those performed by the VIRGO package aboard SoHO. It is composed of 3 types of instruments including absolute radiometers, an imager, and 3 sun photometers \citep[SPM,][]{1995SoPh..162..101F}, centered at 402 nm (blue), 500 nm (green), and 862 nm (red). 
\citet{2010ApJ...713L.155B} demonstrated that the sum of these two last channels (green and red) is a good photometric approximation of the \emph{Kepler} bandwidth.

\subsection{Stellar instrumentation}
As already said in the introduction of this chapter, asteroseismology showed its potential with the development of space instrumentation. After the pioneering measurements done with WIRE \citep[Wide-Field Infrared Explorer,][]{2000ApJ...532L.133B} and the Canadian MOST satellite \citep[Microvariability and Oscillations of Stars,][]{2000ASPC..203...74M}, space asteroseismology reached a golden age with the observations performed by CoRoT \citep[Convection, Rotation and planetary Transits,][]{2006cosp...36.3749B} and  {\it Kepler}  \citep{2010Sci...327..977B,2010ApJ...713L..79K}. The future is even more promising than the present with the on-going K2 mission \citep{2014PASP..126..398H}, and the future missions such as TESS \citep[Transiting Exoplanet Survey Satellite,][]{2014SPIE.9143E..20R} and PLATO \citep{2014ExA....38..249R}. A comparison of the asteroseismic dedicated satellites is shown in Table~\ref{tab:missions}.

 
\begin{table}[!htb]
\caption{\label{tab:missions} }
\centering
\begin{tabular}{c c c c c c c} 
\hline
Mission & Launch &$D$&FOV &$m_{\rm V}$&Number of&Total Number\\
             &           &(cm) & (deg $\times$ deg)&&stars per field&of stars\\
\hline
MOST   & 2003 & 15&0.4 $\times$ 0.4&$<$6&	1   & 150 	\\
CoRoT  & 2006 & 27&2.8 $\times$ 2.8 &$<$7     &  12,000 &150,000 \\
{\it Kepler}&2009& 95&10.5 $\times$ 10.5&$<$12& 206,000     &206,000	\\
K2		& 2014 &95& 10.5 $\times$ 10.5&$<$12 & $\sim$20,000   &$\sim$160,000	\\
Brite	     & 2013 &   3 & 24 $\times$ 24&$<$4        &  15-40   & $\sim$500 	\\
TESS    & 2017 &  10&23 $\times$ 90&$<$12	     &   20,000  &500,000\\
PLATO  & 2024 &  67&47 $\times$ 47& $<$13	      &  100,000   &1,000,000 \\
\hline
\end{tabular}
\\
\label{comparison}
\end{table}

While I am writing this chapter, the analysis of the first K2 solar-like pulsating stars is being conducted and global seismic properties have been measured in some of the targets in Campaign one. In the rest of the chapter I will give a brief overview of the two main space missions CoRoT and \emph{Kepler}.

\subsubsection{CoRoT}
The CNES-ESA mission CoRoT (Convection, Rotation and planetary Transits), launched on December 27, 2006, has been the first dedicated asteroseismic mission that has been able to perform ultra-high precision, wide-field, relative stellar photometry, for very long continuous observing runs (up to 150 days) on the same field of view. On June 17, 2014 CoRoT received the last telecommand from Earth after $\sim$~7 year in-orbit performing scientific activities, establishing the official end of the mission.

CoRoT was led by the french space agency, CNES, in association with other french laboratories and with a significant international participation. Indeed, Austria, Belgium, Germany and ESA (Science Program and RSSD/ESTEC) contributed to the payload, whereas Spain and Brazil contributed to the ground segment. CoRoT had two main scientific programs working simultaneously on adjacent regions of the sky: asteroseismology, and the search for exoplanets using the transiting method. 

\begin{figure}[!htb]
 \centering
 \includegraphics[height=.40\textheight]{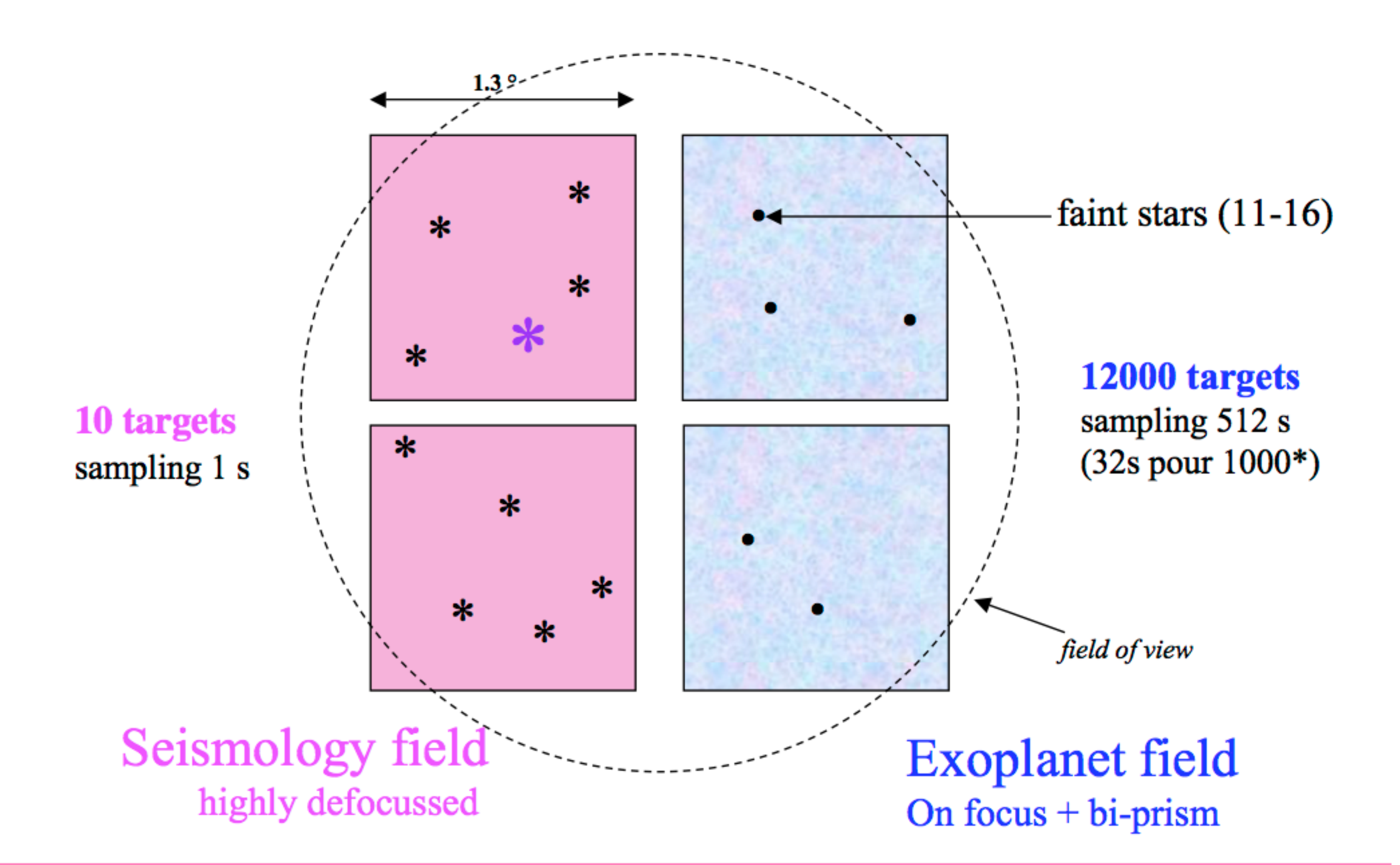}
  \caption[Focal plane of the CoRoT instrument]{\label{FP_CoRoT} Focal plane of the CoRoT instrument. At any time, around 10 targets are observed in the seismology field, while $\sim$ 12000 are monitored in the exoplanet one. }
\end{figure}

CoRoT had a 27-cm afocal telescope producing an image of the stellar field into 4 CCDs (composed by a matrix of 2048 x 4096 pixels) and installed in a proteus platform. Each CCD was working on a frame transfer mode, two optimized for asteroseismology and the other two, for exoplanet research (see Fig.~\ref{FP_CoRoT}). The images on the seismo CCDs were defocused to minimize the effects of the spacecraft jitter. Thus, stellar fluxes were measured every second with a dead time of 0.206s (sampling cycle of 79.4~\%). The data was finally integrated to a cadence of 32s in the case of the seismo filed. For the exoplanet field, the nominal integration time was also 32s but the flux of 16 read-outs was co-added on an 8.5 min cadence before being downlinked to Earth. Moreover, 500 targets per CCD preserved the nominal sampling time of 32 sec to allow a better transiting timing. These were known as ``oversampled'' light curves. At the beginning of each run, 1000 targets were selected for oversampling, but this list of targets was updated every week, thanks to a quick look analysis of the light curves.

\begin{figure}[!htb]
 \centering
 \includegraphics[height=.35\textheight]{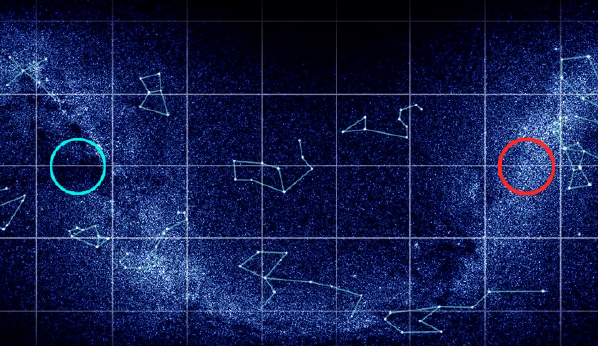}
  \caption[CoRoT eyes]{\label{corot_eyes} Regions of the sky in which CoRoT could point, usually known as the ``CoRoT eyes''. }
\end{figure}

To prevent the instrument from being blinded by the Sun and to keep the scattered light by the Earth at a minimum level, CoRoT could only observe into two 10$^\circ$ almost circular regions pointing towards the galactic center and the anticenter in the equatorial plane. These regions were called the ``CoRoT eyes'' (see Fig.~\ref{corot_eyes}). These positions were selected as a good compromise between the stellar density required for exoplanet research and the existence of seismically interesting targets. A small drift of the orbit  allowed to optimize the observing conditions for fields at the edge of the circle and slightly extend these continuous viewing zones during the lifetime of the CoRoT mission.

Every six months (in April and October), the satellite were rotated by 180$^\circ$ with respect to the polar axis and a new observation period started in the opposite direction. The longest periods of continuous observation were 150 days, the so-called Long Runs, ensuring the highest expected scientific return. Between two long runs, other fields were observed for a much shorter period of around a maximum of 25 days  (Short Run). Due to a malfunction of one module (two CCDs, one on each scientific program) in 2009, we lost half of the CoRoT field of view. Then a new strategy were designed, observing the main seismo target during roughly three months and then rotating the satellite but keeping this main target in view for another three months. In such way, we had a long run of the main target in the seismo field while the stars in the exo channel changed to maximize the detection probability. 

\subsubsection{{\it Kepler}}

Observational astroseismology reaches its maturity with the launch of {\it Kepler} on March 7, 2009 (GMT) \citep{2010Sci...327..977B,2010ApJ...713L..79K}. It was a NASA discovery mission whose primary goal was the search for and characterization of extrasolar planetary systems.  This was accomplished by time-series photometry of around 206,000 stars in a single field of view of 115 $deg^2$ --selected to provide the optimal density of stars for extrasolar planet research-- and located in the constellation of Cygnus and Lyra. The loss of two reaction wheels on the \emph{Kepler} spacecraft has ended the primary mission data collection after $\sim$~4 years of continuous operations. After a hard effort, engineers at NASA demonstrated the viability of a new mission, called K2, in which \emph{Kepler} could observe target fields along a narrow band around the ecliptic for 2 to 3 more years \citep{2014PASP..126..398H}. The first fields have already been observed and the results are very promising from the asteroseismic of solar-like stars point of view \citep{2015arXiv150507105A,2015arXiv150701827C,2015ApJ...809L...3S}.

The \emph{Kepler} main scientific objective was to measure Earth-like planets in an Earth-like orbit around Sun-like stars inside the habitable zone.  The very precise photometry required for the planet search also provided excellent data for asteroseismology. While most stars were observed at a cadence of 30 min, some of them (around 512 at any time), were observed at a cadence of 1 min \citep{2010ApJ...713L.160G,2010ApJ...713L..87J}. Combining the stars observed at both cadences it was possible to observe pulsating stars covering most of the regions in the HR diagram. Short-cadence data allowed the characterization of about 500 solar-like pulsating stars \citep{2011Sci...332..213C}, while using long cadence data, measurements of more than 13,000 red giants were possible \citep{2013ApJ...765L..41S}.

\subsection{Comparison between Doppler velocity and Intensity measurements}
\label{compGV}
\begin{figure}[!htb]
 \centering
 \includegraphics[height=.35\textheight, width=0.7\textwidth]{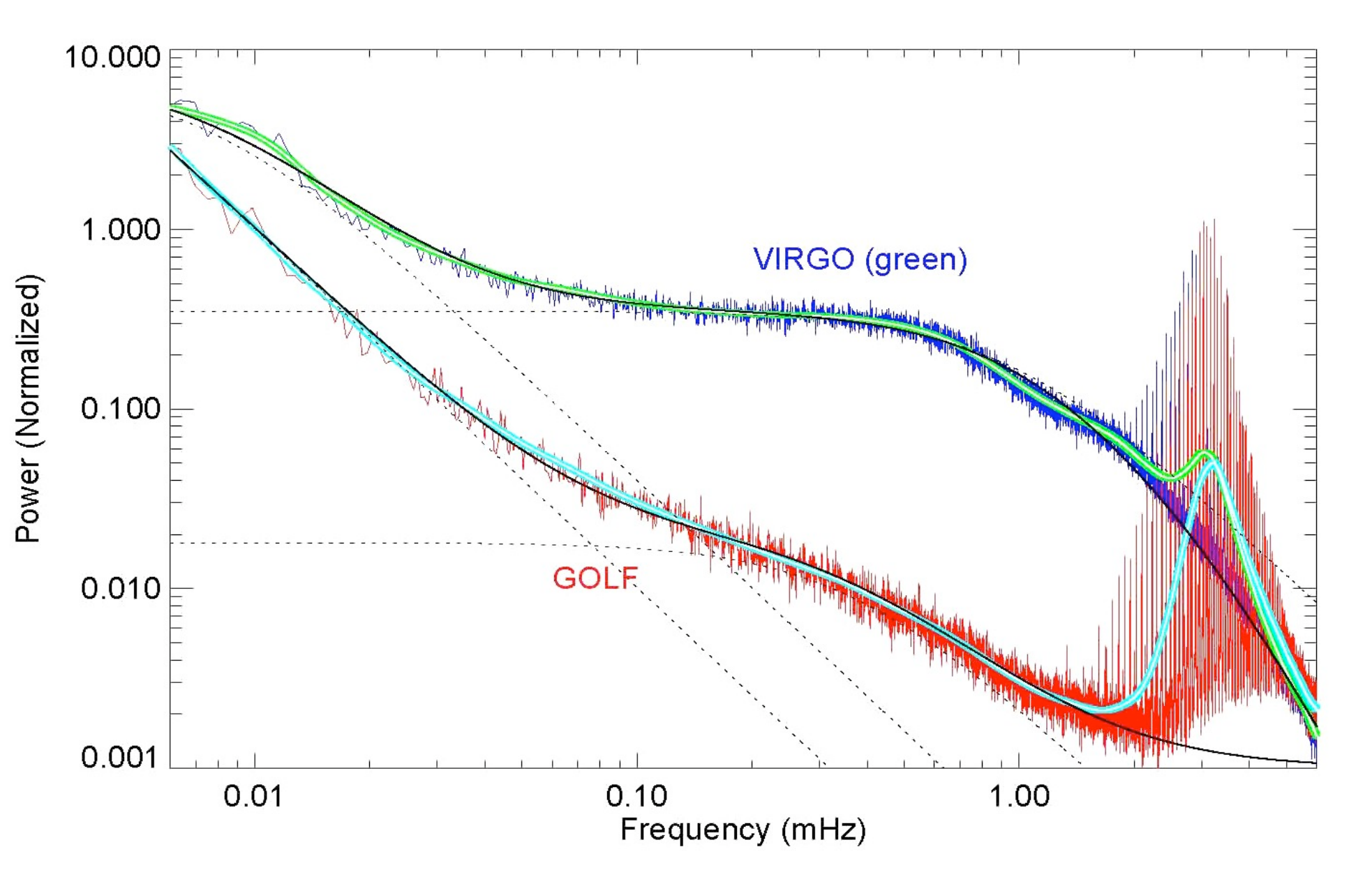}
  \caption[Comparison of the PSD obtained from velocity and intensity measurements from GOLF and VIRGO/SPM]{\label{Comp_Golf_Virgo} Comparison between the PSD extracted from Doppler velocity (GOLF) or intensity measurements (VIRGO/SPM green channel). The fitting to the convective background is also shown. (Credits, T. Bedding \& H. Kjeldsen).}
\end{figure}

Thanks to SoHO we can directly compare the resultant power spectrum obtained from Doppler velocity variations measured by GOLF and by measuring the intensity variations form the Sun spectrophotometers (SPM) of the VIRGO package. In Fig.~\ref{Comp_Golf_Virgo} the PSD obtained using both instruments is shown. We have normalized both spectrum in such a way that the maximum of the p-mode hump has the same amplitude. The convective background in intensity is higher than in velocity, which make it difficult to detect and characterize the low-order p modes below 1.8 mHz. Moreover, the signal-to-background ratio of the p modes for intensity measurements is not better than 30, while in velocity it is common to reach a level of 300.

Thus the future challenge of asteroseismology will be to observe hundred thousands of stars in Doppler velocity and from the space... But a long way is in front of us before developing such instrumentation.


\bibliographystyle{aa}
\bibliography{/Users/rgarcia/Documents/BibReader/BIBLIO}

\end{document}